\documentclass[iop]{emulateapj}
\usepackage{graphicx}
\usepackage{color}

\shorttitle{}
\shortauthors{Dud\'ik et al.}

\begin{document}

\title{Imaging and Spectroscopic Observations of a Transient Coronal Loop: Evidence for the Non-Maxwellian $\kappa$-Distributions}

\author{Jaroslav Dud\'ik\altaffilmark{1}}
    \affil{Astronomical Institute of the Academy of Sciences of the Czech Republic, Fri\v{c}ova 298, 251 65 Ond\v{r}ejov, Czech Republic}
    \email{jaroslav.dudik@asu.cas.cz}

\author{\v{S}imon Mackovjak\altaffilmark{2,}\altaffilmark{3}}
    \affil{ISDC - Data Centre for Astrophysics, Astronomy Department, University of Geneva, Chemin d'Ecogia 16, 1290 Versoix, Switzerland}

\author{Elena Dzif\v{c}\'akov\'a}
\affil{Astronomical Institute of the Academy of Sciences of the Czech Republic, Fri\v{c}ova 298, 251 65 Ond\v{r}ejov, Czech Republic}

\author{Giulio Del Zanna}
\affil{DAMTP, CMS, University of Cambridge, Wilberforce Road, Cambridge CB3 0WA, United Kingdom}

\author{David R. Williams}
\affil{University College London, Mullard Space Science Laboratory, Holmbury St Mary, Dorking, Surrey RH5 6NT, United Kingdom}

\author{Marian Karlick\'y}
\affil{Astronomical Institute of the Academy of Sciences of the Czech Republic, Fri\v{c}ova 298, 251 65 Ond\v{r}ejov, Czech Republic}

\author{Helen E. Mason}
\affil{DAMTP, CMS, University of Cambridge, Wilberforce Road, Cambridge CB3 0WA, United Kingdom}

\author{Juraj L\"{o}rin\v{c}\'ik}
\affil{DAPEM, Faculty of Mathematics Physics and Computer Science, Comenius University, Mlynsk\'a Dolina F2, 842 48 Bratislava, Slovakia}

\author{Pavel Kotr\v{c}} \author{Franti\v{s}ek F\'arn\'ik} \author{Alena Zemanov\'{a}}
\affil{Astronomical Institute of the Academy of Sciences of the Czech Republic, Fri\v{c}ova 298, 251 65 Ond\v{r}ejov, Czech Republic}

%

\altaffiltext{1}{RS Newton International Fellow, DAMTP, CMS, University of Cambridge, Wilberforce Road, Cambridge CB3 0WA, United Kingdom}
\altaffiltext{2}{DAPEM, Faculty of Mathematics Physics and Computer Science, Comenius University, Mlynsk\'a Dolina F2, 842 48 Bratislava, Slovakia}
\altaffiltext{3}{Astronomical Institute of the Academy of Sciences of the Czech Republic, Fri\v{c}ova 298, 251 65 Ond\v{r}ejov, Czech Republic}

\begin{abstract}
We report on the \textit{SDO}/AIA and \textit{Hinode}/EIS observations of a transient coronal loop. The loop brightens up in the same location after the disappearance of an arcade formed during a B8.9-class microflare three hours earlier. EIS captures this loop during its brightening phase as observed in most of the AIA filters. We use the AIA data to study the evolution of the loop, as well as to perform the DEM diagnostics as a function of $\kappa$. \ion{Fe}{11}--\ion{Fe}{13} lines observed by EIS are used to perform the diagnostics of electron density and subsequently the diagnostics of $\kappa$. Using ratios involving the \ion{Fe}{11} 257.772\AA~selfblend, we diagnose $\kappa$\,$\lesssim$\,2, i.e., an extremely non-Maxwellian distribution. Using the predicted Fe line intensities derived from the DEMs as a function of $\kappa$, we show that, with decreasing $\kappa$, all combinations of ratios of line intensities converge to the observed values, confirming the diagnosed $\kappa$\,$\lesssim$\,2. These results represent the first positive diagnostics of $\kappa$-distributions in the solar corona despite the limitations imposed by calibration uncertainties.
\end{abstract}


%
\section{Introduction}
\label{Sect:1}

The distance and nature of the stellar coronae with temperatures of up to several million Kelvin together with the proximity of the stars themselves make direct probing of these environments impossible at present. In the absence of in-situ measurements, the emitted spectrum remains the only source of information about the physical conditions in these emitting media. Most of the coronal radiation is emitted in the X-ray, extreme ultraviolet (EUV) and ultraviolet parts of the electromagnetic spectrum, requiring space-borne observatories. This poses further problems with limited availability of the data, calibration and its stability \citep{BenMoussa13,DelZanna13a}, as well as the unavoidable trade-offs between the spatial, temporal, and spectral resolutions and coverages.

In many instances, it is advantageous to combine imaging and spectroscopic observations \citep[e.g.,][]{Schmelz09,Warren12,DelZanna13b}. Imaging observations employing narrow-band EUV filters, such as those made by the Atmospheric Imaging Assembly \citep[AIA,][]{Lemen12,Boerner12} onboard the \textit{Solar Dynamics Observatory} \citep[\textit{SDO},][]{Pesnell12}, offer high spatial and temporal resolution. These are complemented with spectroscopic observations made in similar wavelength ranges, such as those performed by the EUV Imaging Spectrograph \citep[EIS,][]{Culhane07} onboard the \textit{Hinode} spacecraft \citep{Kosugi07}. Traditionally, the vast majority of both the imaging and spectroscopic data are analyzed and modelled under the assumption of the local equilibrium, i.e., Maxwellian distribution of particle energies. This is because calculation of the synthetic spectra requires integration of many individual ionization, recombination, excitation and deexcitation cross-sections over the (unknown) distribution function in order to obtain rates of these processes and finally the corresponding emissivities at individual wavelengths. The assumption of Maxwellian distribution affords easy calculation of the synthetic spectra, e.g., using the CHIANTI atomic database and software \citep{Dere97,Landi13}. 

However, this assumption is incorrect if there are correlations between the particles in the system. Such correlations can be induced by any long-range interactions in the system \citep{Collier04,Leubner04a,Livadiotis09,Livadiotis10,Livadiotis13}, e.g., particle acceleration due to magnetic reconnection \citep[e.g.,][]{Zharkova11,Petkaki11,Stanier12,Cargill12,Burge12,Burge14,Gordovskyy13,Gordovskyy14}, shocks, or wave-particle interactions \citep[e.g.,][]{Vocks08}. In such cases, the particle distribution will depart from the Maxwellian one, and will likely exhibit an enhanced high-energy tail. Furthermore, turbulence with the diffusion coefficient inversely proportional to particle velocity will also lead to the appearance of the non-Maxwellian distributions with characteristic high-energy tails \citep[e.g.,][]{Hasegawa85,Laming07,Bian14}. We note that the collision cross-section scales with $E^{-2}$, where $E$ is the particle energy \citep{Meyer-Vernet07}. This leads to the behaviour of the collision frequency as $E^{-3/2}$, i.e., the high-energy tail is difficult to equilibrate. Therefore, the currently favoured theories of nanoflare heating of the solar corona \citep[][]{Klimchuk06,Klimchuk10,Tripathi10,Bradshaw12,Winebarger12,Viall11,Viall12,Viall13} should afford situations for departures from the Maxwellian distribution. \citet{Scudder13} argue that the particle distribution in the solar and stellar coronae above 1.05$R_\odot$ is strongly non-Maxwellian.

Observational clues that the solar corona could be non-Maxwellian come from the in-situ detection of the non-Maxwellian $\kappa$-distributions in the solar wind \citep[][]{Collier96,Maksimovic97a,Maksimovic97b,Zouganelis08,LeChat11}. Spectroscopic evidence for the presence of $\kappa$-distributions in the transition region was found from active region \ion{Si}{3} spectra \citep{Dzifcakova11Si}. The $\kappa$-distributions are characterized by a high-energy power-law tail (Sect. \ref{Sect:2}), with the power-law index being given by the value of $-(\kappa$\,+\,$1/2)$. In the solar wind, the typically detected values are $\kappa$\,$\gtrsim$\,2.5, while the detection in the transition region yielded $\kappa$\,$\approx$\,7 from the active region \ion{Si}{3} spectrum. Presence of high-energy electrons in the transition region at the base of coronal loops have recently been established also by \citet{Testa14} by analyzing the transition-region \ion{Si}{4} emission observed by the IRIS instrument \citep{DePontieu14}.

Despite these detections in the solar wind and the transition region, no direct, unambiguous detection of the non-Maxwellian distribution in the coronal spectra have been made to date. Attempts at doing so were made e.g. by \citet{Feldman07}, \citet{Ralchenko07}, \citet{Hannah10}, \citet{Dzifcakova10}, and \citet{Mackovjak13}. \citet{Feldman07} assumed a bi-Maxwellian distribution, with the temperature of the second Maxwellian chosen to be 10\,MK, and argued that no such second Maxwellian is necessary to explain the observed spectra of He-like lines. This analysis was however limited and did not include the effects of $\kappa$-distributions on the spectra. \citet{Ralchenko07} showed that the quiet-Sun Si, Ca, and Ar spectra are consistent with a bi-Maxwellian distribution, where the second Maxwellian contains only a small fraction of particles, of the order of several per cent. Its temperatures range from 2.3\,MK (300\,eV) to 7.7\,MK (1000\,eV), with the fraction of particles being at most 5--7\%, and 1\% for these temperatures, respectively. No diagnostics of $\kappa$ was performed. \citet{Hannah10} used off-limb quiet-Sun observations performed by the RHESSI spacecraft \citep{Lin02} to constrain the emission measures corresponding to individual power-law and $\kappa$-distributions. However, the constraints obtained on the emission measures were rather large even for small values of $\kappa$. \citet{Dzifcakova10} and \citet{Mackovjak13} studied the possibilities of diagnosing $\kappa$ using lines observed by EIS. Despite having found several combinations of line ratios sensitive to $\kappa$, mostly involving neighbouring ionization stages, unambiguous detection were difficult due to atomic data uncertainties and possible multi-thermal effects along the line-of-sight that were not accounted for. Subsequently, \citet{Mackovjak14} showed that the techniques to obtain the differential emission measures work also for the $\kappa$-distributions, and studied the influence of $\kappa$ on such analyses.

In this paper, we use the imaging and spectroscopic observations performed by \textit{SDO}/AIA and \textit{Hinode}/EIS in conjunction with the latest instrument calibration, atomic datasets, and differential emission measure techniques to analyze a transient loop observed within an active region core. Spectral synthesis for the non-Maxwellian $\kappa$-distribution is briefly discussed in Sect. \ref{Sect:2}. Analysis of the AIA imaging observations is performed in Sect. \ref{Sect:3}. In Sect. \ref{Sect:4}, ratios of spectral lines observed by EIS are analyzed to obtain the electron density and $\kappa$ in the loop. Influence of the atomic data uncertainties on the diagnostics are discussed in Sect. \ref{Sect:5}. The results are summarized and discussed in Sect. \ref{Sect:6}.

%
   \begin{figure}[!t]
	\centering
	\includegraphics[width=8.8cm]{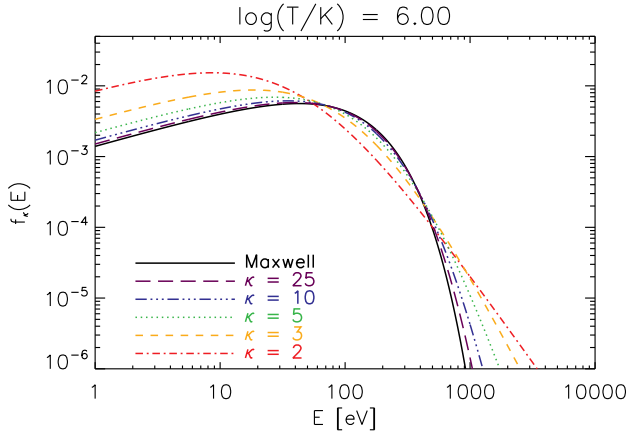}
	\caption{The $\kappa$-distributions with $\kappa$\,=\,2, 3, 5, 10, 25 and the Maxwellian distribution plotted for log($T$/K)\,=\,6.0. Colors and linestyles denote the different values of $\kappa$. 
	\\ A color version of this image is available in the online journal.}
       \label{Fig:Kappa}
   \end{figure}
%

%
   \begin{figure}[!t]
	\centering
	\includegraphics[width=8.8cm]{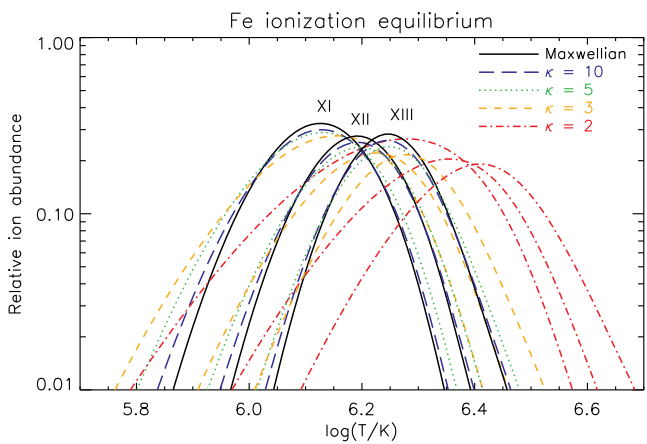}
	\caption{Ionization equilibrium for \ion{Fe}{11}--\ion{Fe}{13} for the $\kappa$-distributions, according to the calculations of \citet{Dzifcakova13}. 
		\\ A color version of this image is available in the online journal.}
       \label{Fig:Ioneq_AIA}
   \end{figure}
%

%
\section{Synthetic Spectra for the $\kappa$-distributions}
\label{Sect:2}

\subsection{The Non-Maxwellian $\kappa$-distributions}
\label{Sect:2.1}

The $\kappa$-distribution of electron energies (Fig.~\ref{Fig:Kappa}) is defined as a two-parametric distribution with parameters $T$\,$\in$\,$\left(0,+\infty\right)$ and $\kappa$\,$\in$\,$\left(3/2,+\infty\right)$ \citep[e.g.,][]{Owocki83,Livadiotis09}
        \begin{equation}
                f_\kappa(E) \mathrm{d}E = A_{\kappa} \frac{2}{\sqrt{\pi} (k_\mathrm{B}T)^{3/2}} \frac{E^{1/2}\mathrm{d}E}{\left 			  (1+ \frac{E}{(\kappa - 3/2) k_\mathrm{B}T} \right)^{\kappa+1}}\,,
                \label{Eq:Kappa}
        \end{equation}
where the $A_{\kappa}$\,=\,$\Gamma(\kappa+1)$/$\left(\Gamma(\kappa-1/2) (\kappa-3/2)^{3/2}\right)$ is the normalization constant and $k_\mathrm{B}$\,=\,1.38 $\times 10^{-16}$ erg\,s$^{-1}$ is the Boltzmann constant. The Maxwellian distribution at a given $T$ is recovered for $\kappa$\,$\to$\,$\infty$. Maximum departure from the Maxwellian occurs for $\kappa$\,$\rightarrow$\,3/2.

The mean energy $\left< E \right> = {3k_\mathrm{B}T}/{2}$ of a $\kappa$-distribution does not depend on $\kappa$. Because of this, $T$ can be defined as the temperature. \citet{Meyer-Vernet95}, \citet{Livadiotis09} and \citet{Livadiotis10} showed that $T$ indeed has the same physical meaning for the $\kappa$-distributions as the (kinetic) temperature for the Maxwellian distribution, and that it also corresponds to the definition of physical temperature in the framework of the generalized Tsallis statistical mechanics \citep{Tsallis88,Tsallis09}. This permitted these authors to generalize of the zero-th law of thermodynamics for the $\kappa$-distributions, and it also permits, e.g., the definition of electron kinetic pressure $p$\,=\,$n_\mathrm{e}k_\mathrm{B}T$ in the usual manner.

It is straightforward to see from Eq. (\ref{Eq:Kappa}) that in the high-energy limit, the $\kappa$-distribution approaches a power-law with the index of $-(\kappa+1/2)$. On the other hand, \citet{Livadiotis09} showed that, in the low-energy limit, the $\kappa$-distribution behaves as a Maxwellian with $ T_\mathrm{M} = T \left(\kappa-3/2\right) / \left(\kappa+1\right)$, and scaled to an appropriate constant \citep[see][]{Dzifcakova15}. Therefore, the $\kappa$-distribution can be thought of as a Maxwellian core (at a lower temperature) with a power-law tail.

%
\subsection{Line Intensities for the $\kappa$-Distributions}
\label{Sect:2.2}

The emissivity $\varepsilon_{ji}$ of an optically thin spectral line arising due to a transition $j \to i$, $j > i$, in a $k$-times ionized ion of the element $X$ is given by \citep[e.g.,][]{Mason94,Phillips08}
	\begin{eqnarray}
		\nonumber \varepsilon_{ji} &=& \frac{hc}{\lambda_{ji}} A_{ji} n(X_j^{+k}) = \frac{hc}{\lambda_{ji}} \frac{A_{ji}}{n_\mathrm{e}} \frac{n(X_j^{+k})}{n(X^{+k})} \frac{n(X^{+k})}{n(X)} A_X n_\mathrm{e} n_\mathrm{H}\\
		 &=& A_X G_{X,ji}(T,n_\mathrm{e},\kappa) n_\mathrm{e} n_\mathrm{H}\,,
		\label{Eq:line_emissivity}
	\end{eqnarray}
where $h$\,$\approx$\,6.62\,$\times$\,10$^{-27}$ erg\,s is the Planck constant, $c$\,$\approx$\,3\,$\times$10$^{10}$\,cm\,s$^{-1}$ represents the speed of light, $\lambda_{ji}$ is the wavelength of the resulting spectral line, $A_{ji}$ the corresponding Einstein coefficient for spontaneous emission, $n(X_j^{+k})$ is the density of the ion $+k$ with electron in the excited upper level $j$, $n(X^{+k})$ the total density of ion $+k$, $n(X)$\,$\equiv$\,$n_X$ the total density of element $X$ whose abundance is $A_X$, $n_\mathrm{H}$ the hydrogen density, and $n_\mathrm{e}$ the electron density. The function $G_{X,ji}(T,n_\mathrm{e},\kappa)$ is the contribution function for the line $\lambda_{ji}$. The $G_ {X,ji}(T,n_\mathrm{e},\kappa)$ is a function of $\kappa$ due to the dependence of the individual collisional processes of ionization, recombination, excitation and deexcitation on $\kappa$ \citep[e.g.,][]{Dzifcakova92,Dzifcakova02,Dzifcakova06,Wannawichian03,DzifcakovaMason08,Dzifcakova13,Dudik14a,Dudik14b,Dzifcakova15}.

The intensity $I_{ji}$ of the spectral line is then given by the integral of emissivity along a line of sight $l$ 
	\begin{equation}
		I_{ji} = \int A_X G_{X,ji}(T,n_\mathrm{e},\kappa) n_\mathrm{e} n_\mathrm{H} \mathrm{d}l\,,
		\label{Eq:line_intensity}
	\end{equation}
or
	\begin{equation}
		I_{ji} = \int A_X G_{X,ji}(T,n_\mathrm{e},\kappa) \mathrm{DEM}_{\kappa}(T) \mathrm{d}T\,,
		\label{Eq:line_intensity_DEM}
	\end{equation}
where the quantity EM\,=\,$\int n_\mathrm{e} n_\mathrm{H} \mathrm{d}l$ is the plasma emission measure, and DEM$_{\kappa}(T)$\,=\,$n_\mathrm{e} n_\mathrm{H} \mathrm{d}l/\mathrm{d}T$ is the differential emission measure, generalized for the $\kappa$-distributions by \citet{Mackovjak14}.

The line intensities for $\kappa$-distributions are evaluated using the ionization equilibrium calculations for $\kappa$-distributions \citep{Dzifcakova13}. Relative level population is obtained using the approximative method of \citet{Dzifcakova06b}, \citet{DzifcakovaMason08} and \citet{Dzifcakova15} based on atomic data corresponding to the CHIANTI database, version 7.1 \citep{Dere97,Landi13}. The accuracy of this method for allowed transitions is less than 10\% \citep{DzifcakovaMason08}. For the \ion{Fe}{9}--\ion{Fe}{13}, which are the ions used for diagnostics in Sect. \ref{Sect:4}, the line intensities are obtained by direct integration of the collision strengths \citep{Dudik14b}. The collision strengths for these calculations are state-of-the-art and taken from \citet{DelZanna10c}, \citet{DelZanna11}, \citet{DelZanna12a}, \citet{DelZanna12b}, \citet{DelZanna12e}, \citet{DelZanna13}, and \citet{DelZanna14}.

%
   \begin{figure*}[!t]
	\centering
	\includegraphics[width=8.8cm,clip,bb=0 40 498 385]{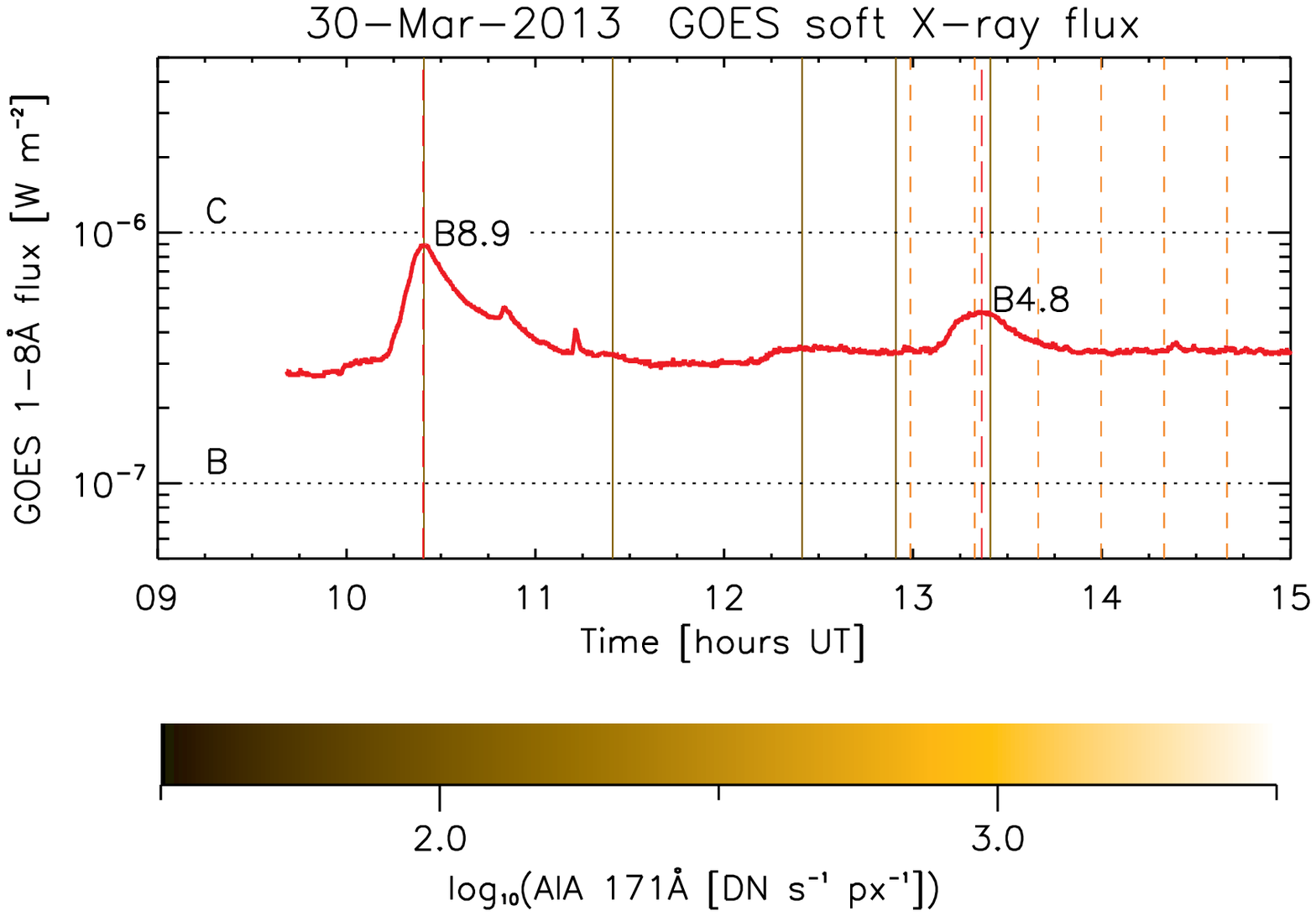}
	\includegraphics[width=8.8cm,clip,bb=0 40 498 385]{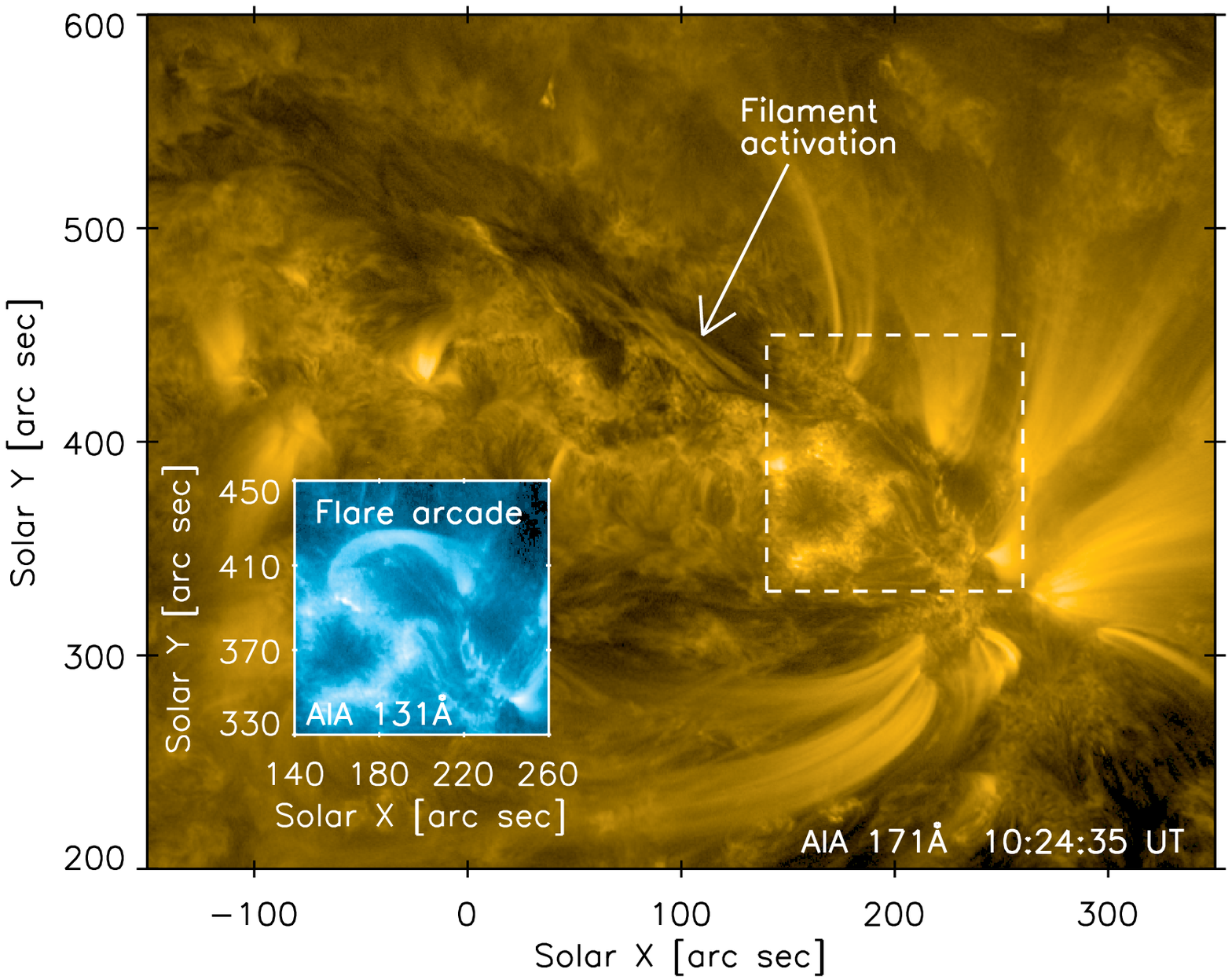}
	\includegraphics[width=8.8cm,clip,bb=0 40 498 385]{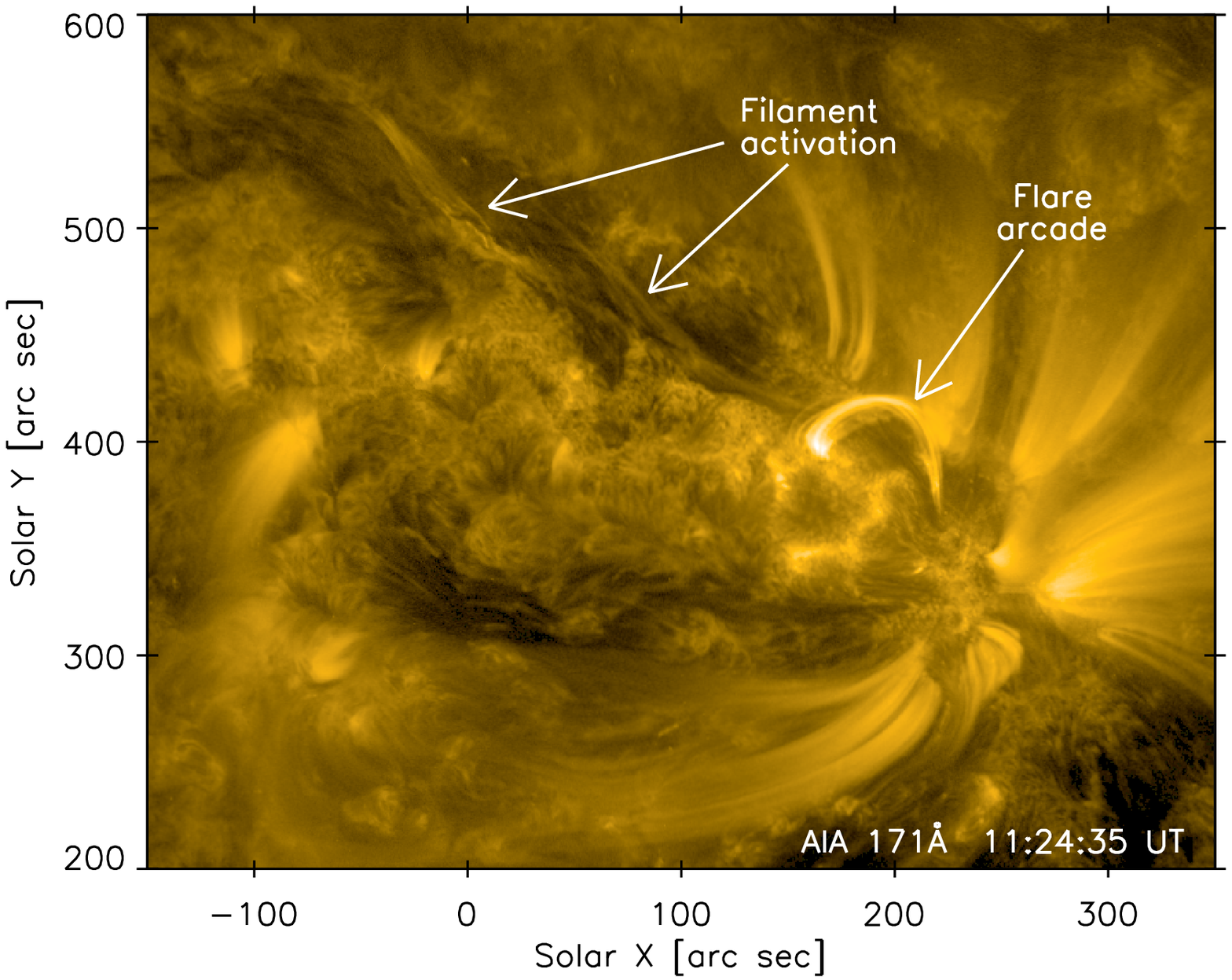}
	\includegraphics[width=8.8cm,clip,bb=0 40 498 385]{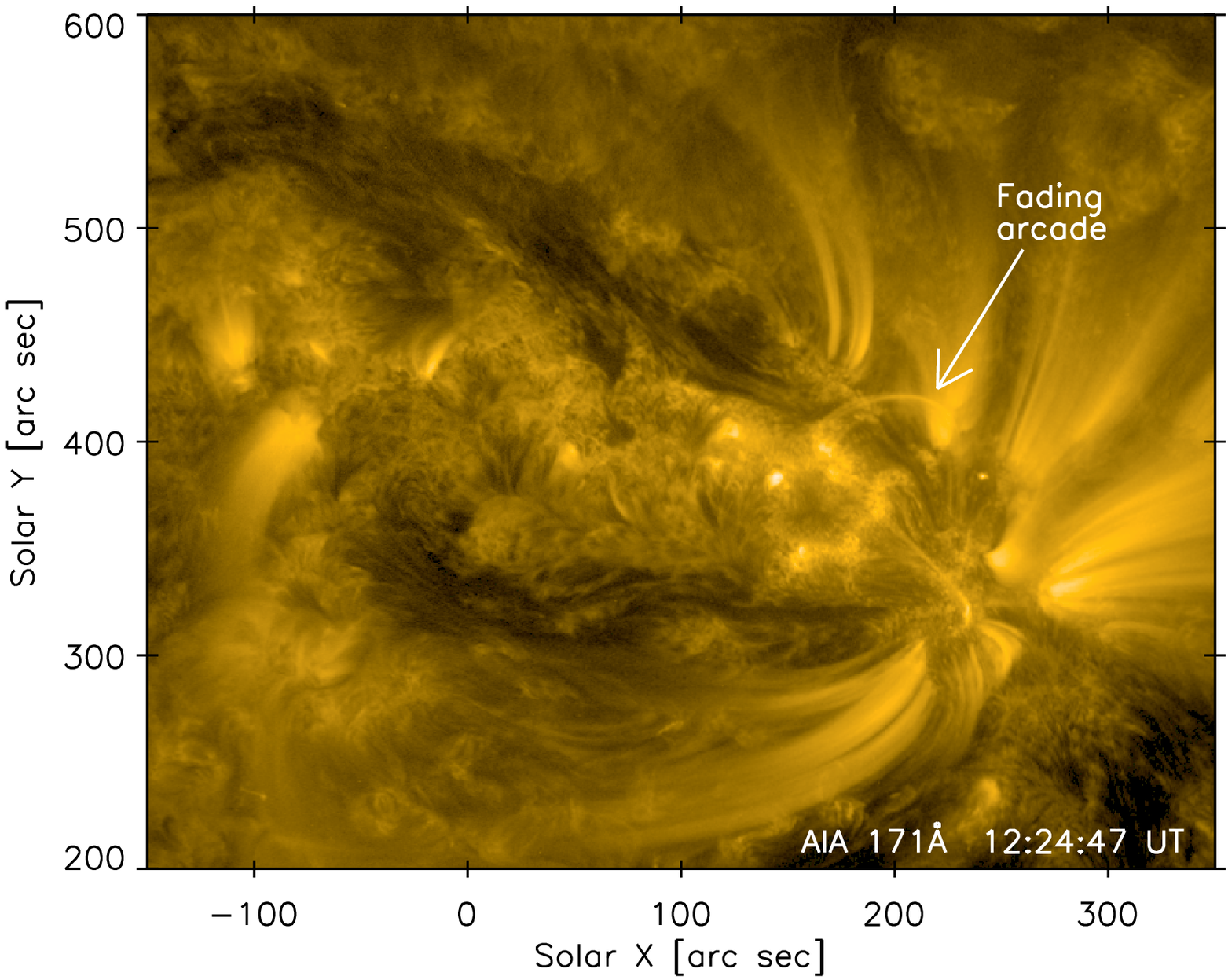}
	\includegraphics[width=8.8cm,clip,bb=0  0 498 385]{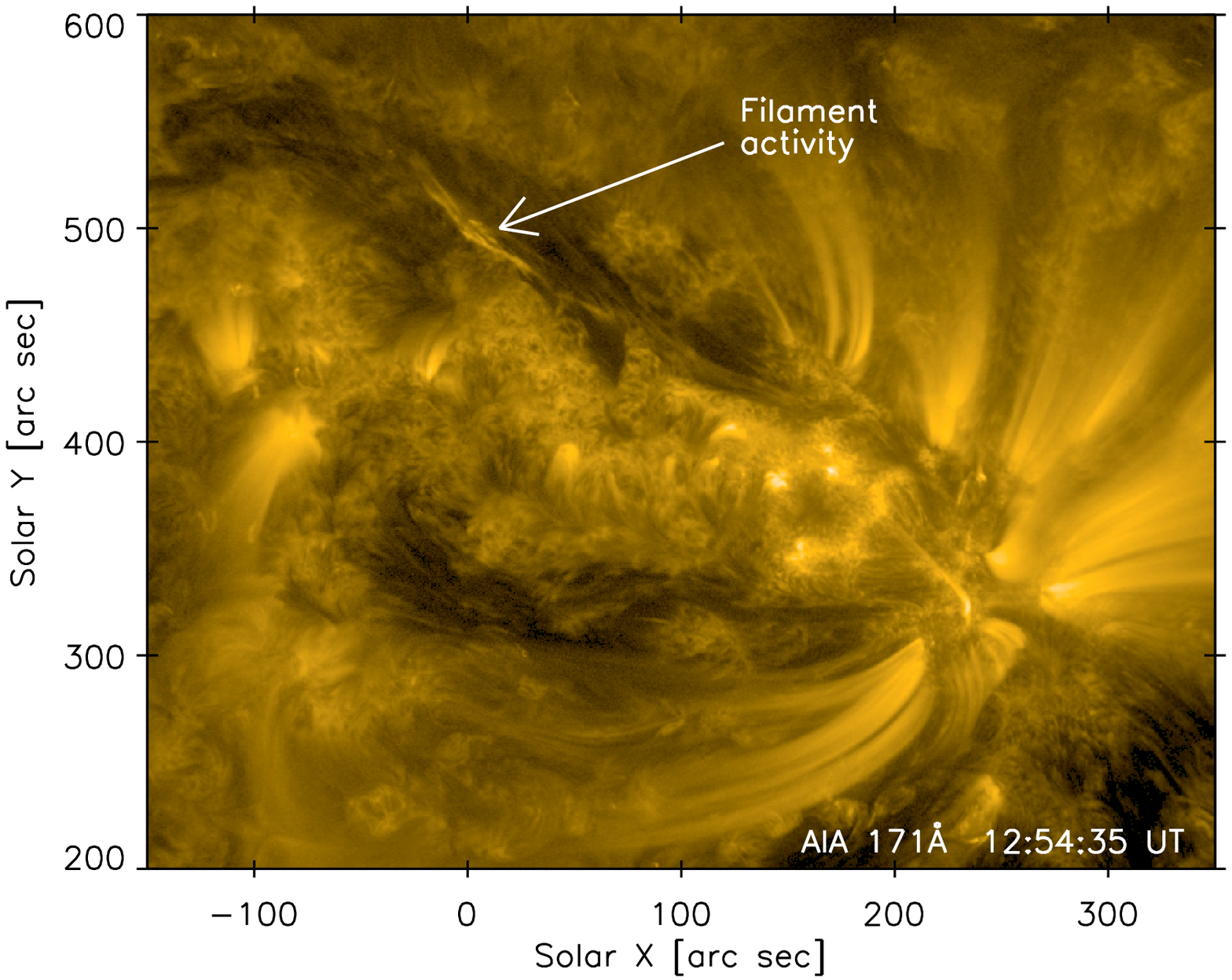}
	\includegraphics[width=8.8cm,clip,bb=0  0 498 385]{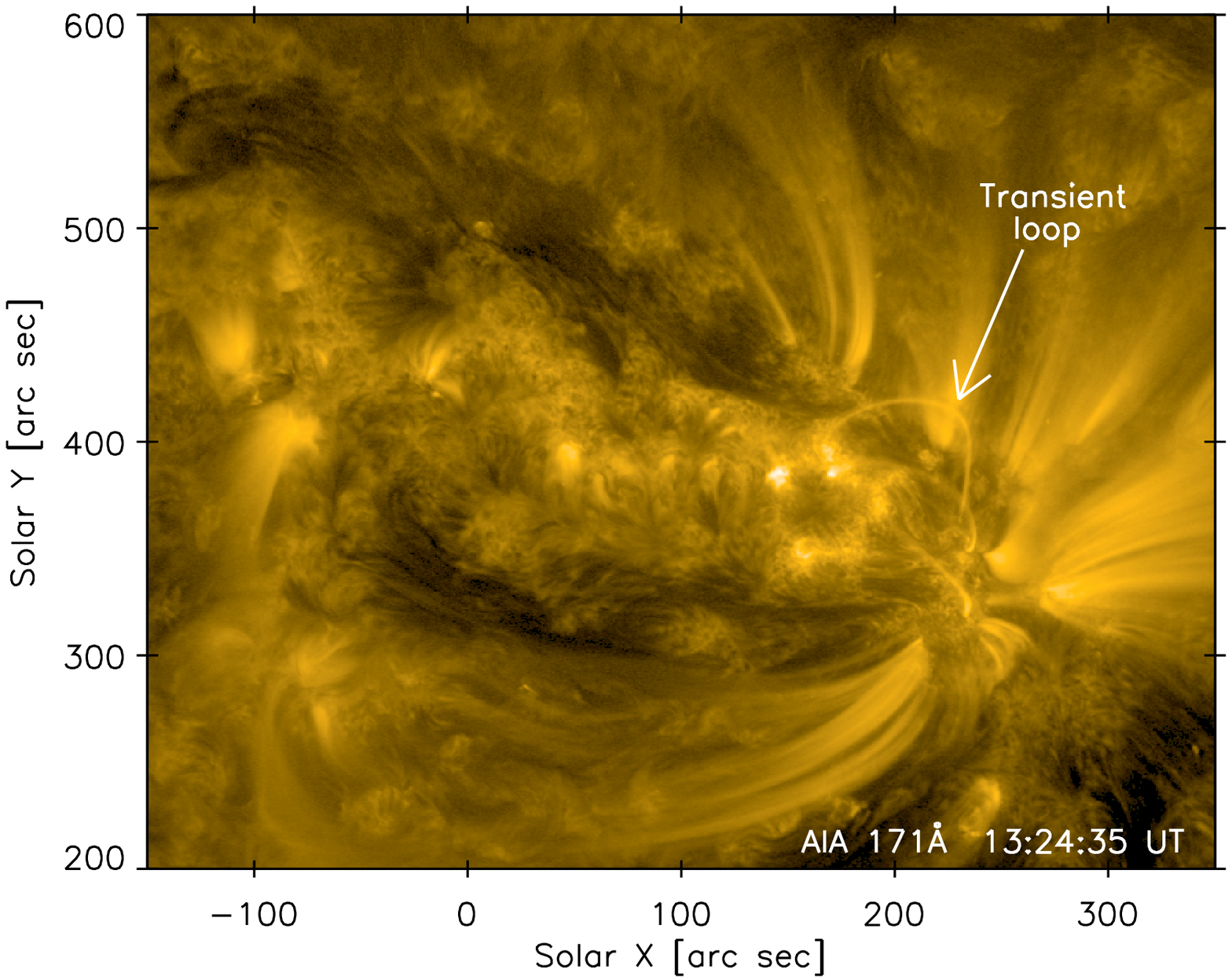}
	\caption{Evolution of the two microflares observed by AIA 171\AA. \textit{Top left}: GOES X-ray flux at 1--8\AA. Red-dashed vertical lines denote the times of the two microflares, full brown lines denote the times of AIA 171\AA~snapshots, and the orange-dashed lines denote the times of snapshots in Fig.~\ref{Fig:AIA_Loop}. \textit{Inset in top right}: AIA 131\AA~1-minute average image showing the arcade of during the B8.9 microflare. The location of the inset corresponds to the box shown. \\ A color version of this image is available in the online journal.}
        \label{Fig:Evolution}
   \end{figure*}
%
%
   \begin{figure*}[!ht]
       \centering
       \includegraphics[width=4.20cm, bb=0   0 246 56, clip]{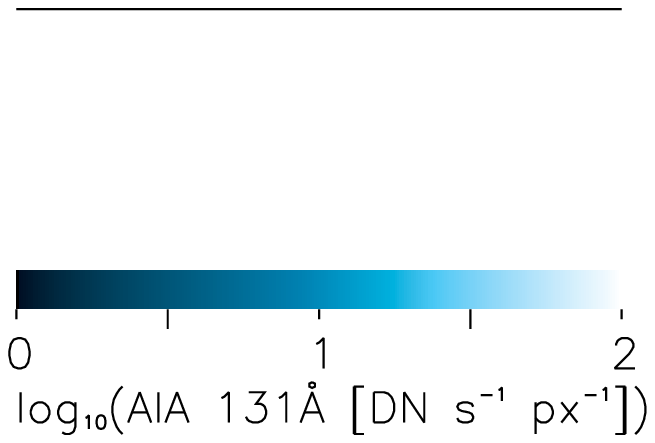}
       \includegraphics[width=3.30cm, bb=50  0 246 56, clip]{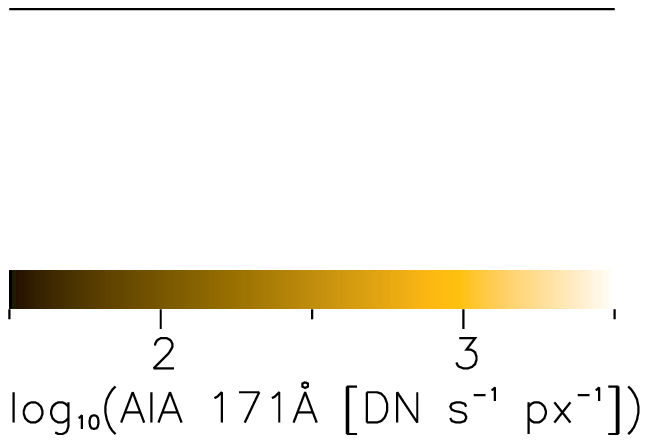}
       \includegraphics[width=3.30cm, bb=50  0 246 56, clip]{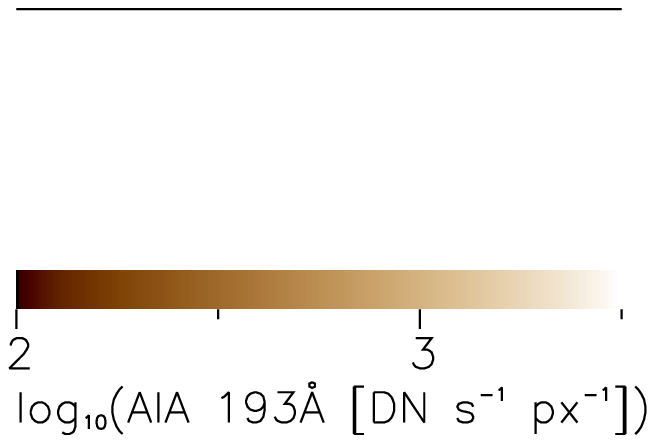}
       \includegraphics[width=3.30cm, bb=50  0 246 56, clip]{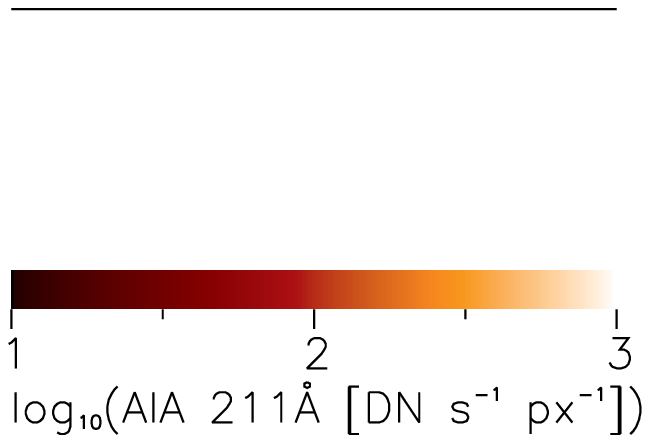}
       \includegraphics[width=3.30cm, bb=50  0 246 56, clip]{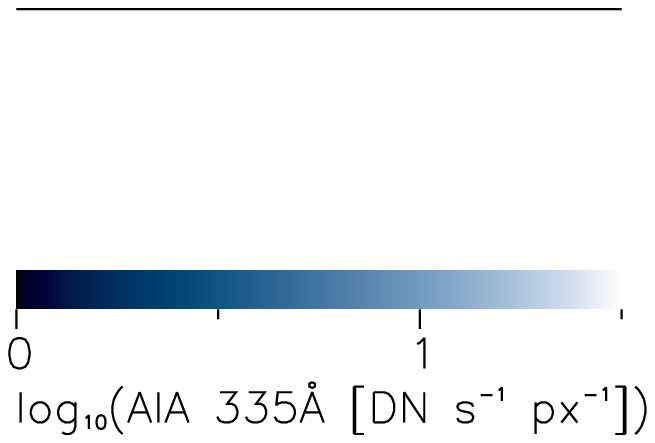}

       \includegraphics[width=4.20cm, bb=0  40 246 235, clip]{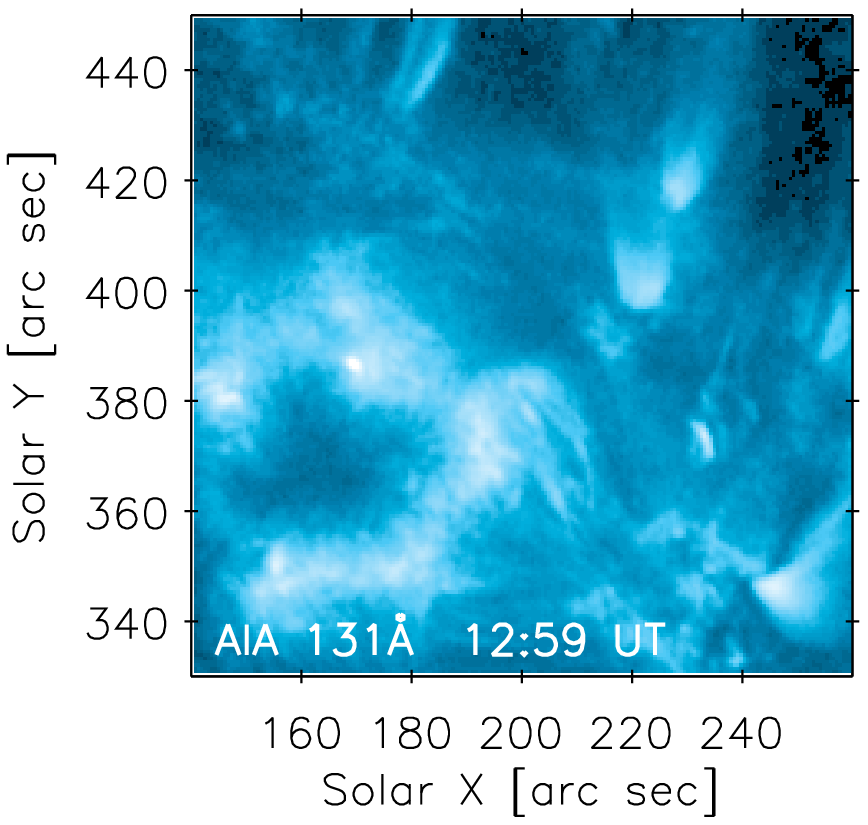}
       \includegraphics[width=3.30cm, bb=53 40 246 235, clip]{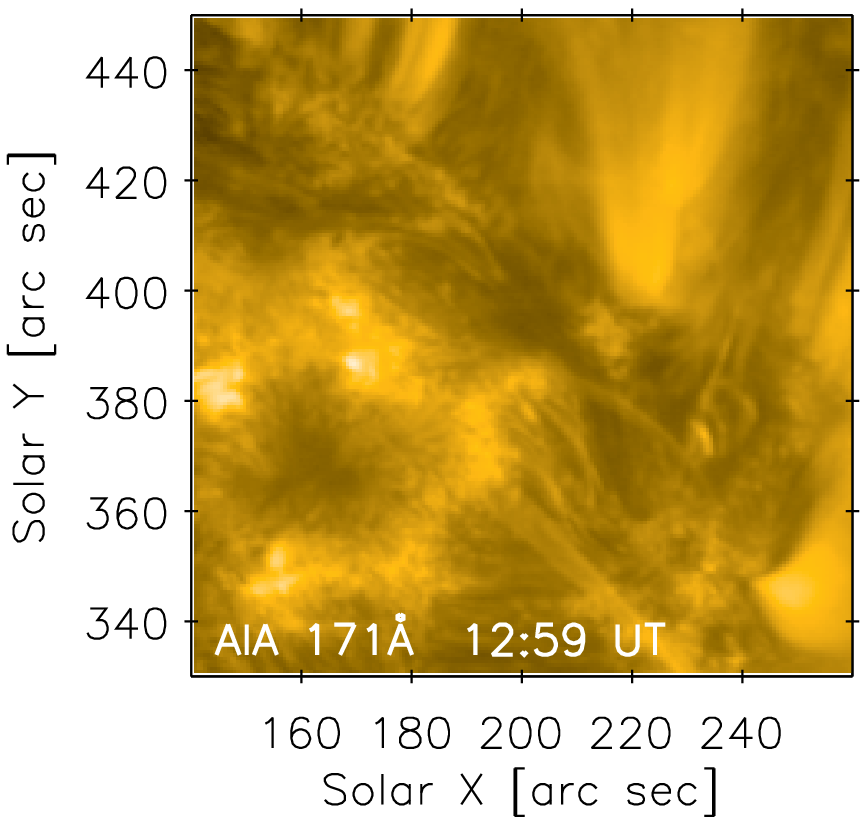}
       \includegraphics[width=3.30cm, bb=53 40 246 235, clip]{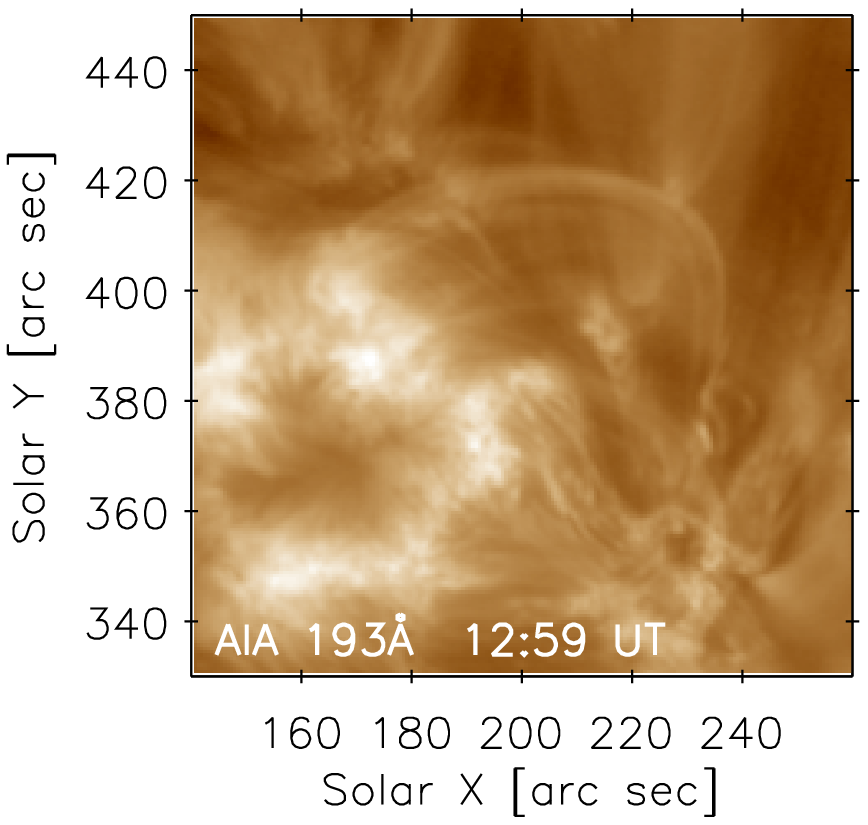}
       \includegraphics[width=3.30cm, bb=53 40 246 235, clip]{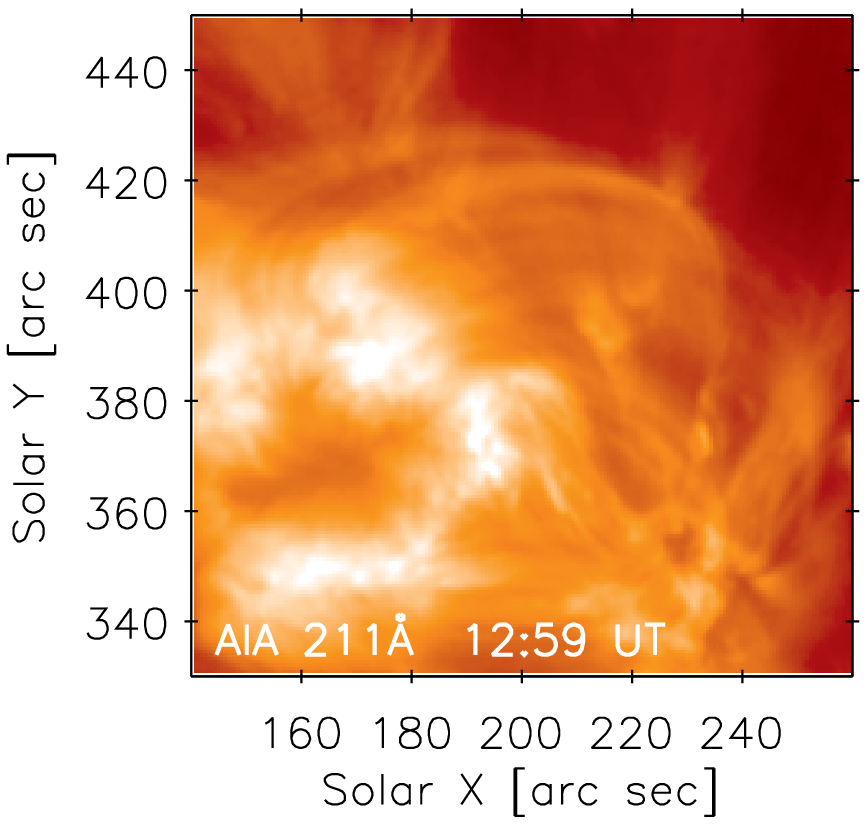}
       \includegraphics[width=3.30cm, bb=53 40 246 235, clip]{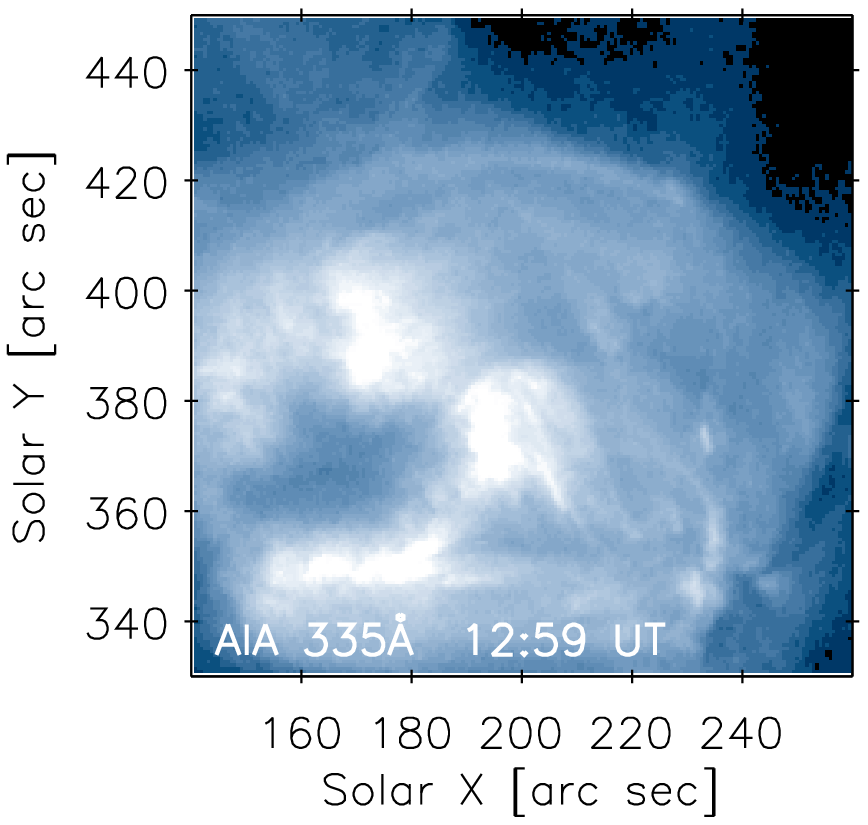}

       \includegraphics[width=4.20cm, bb=0  40 246 235, clip]{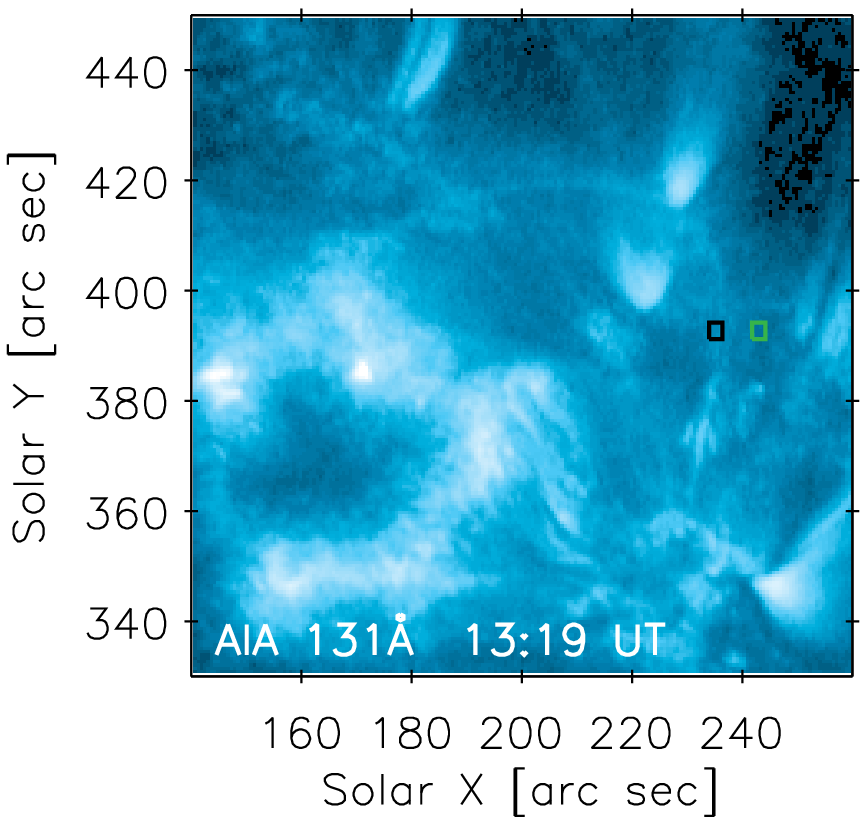}
       \includegraphics[width=3.30cm, bb=53 40 246 235, clip]{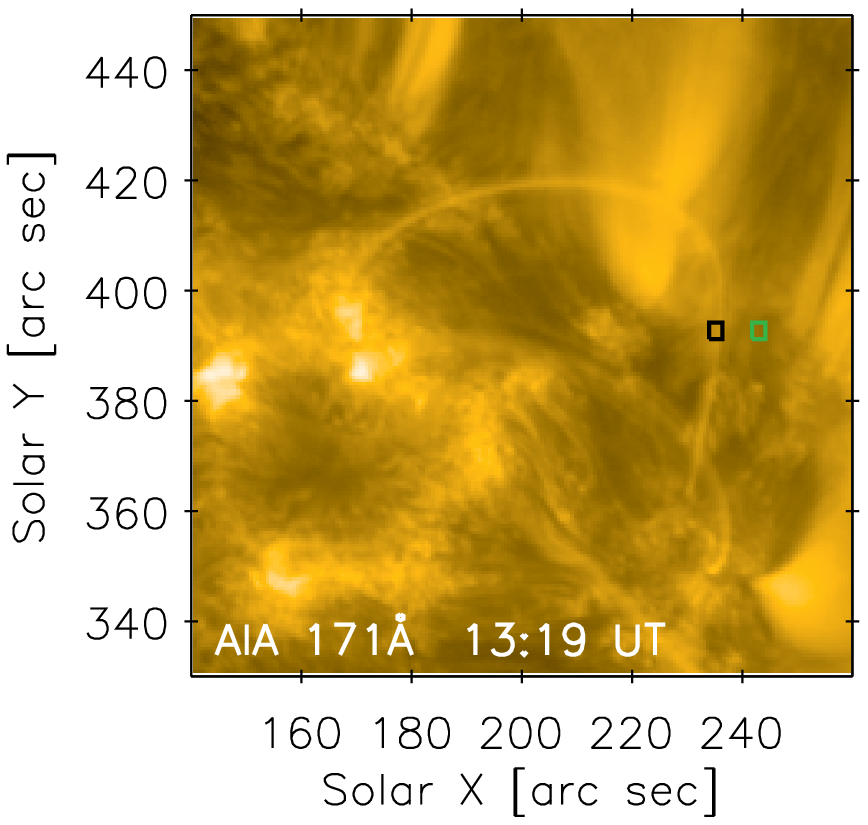}
       \includegraphics[width=3.30cm, bb=53 40 246 235, clip]{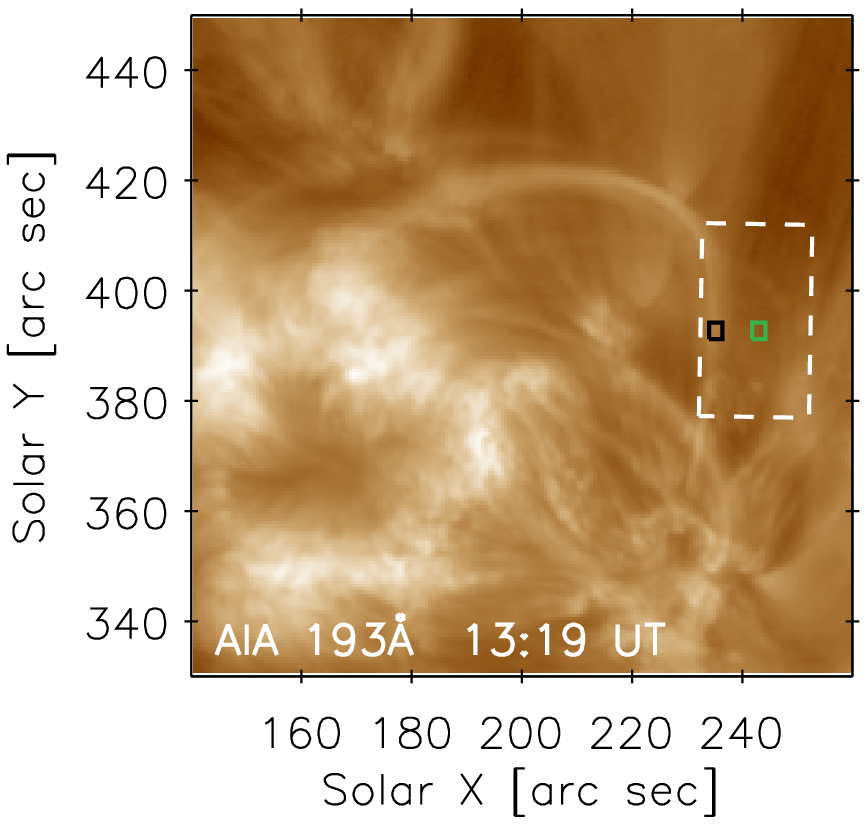}
       \includegraphics[width=3.30cm, bb=53 40 246 235, clip]{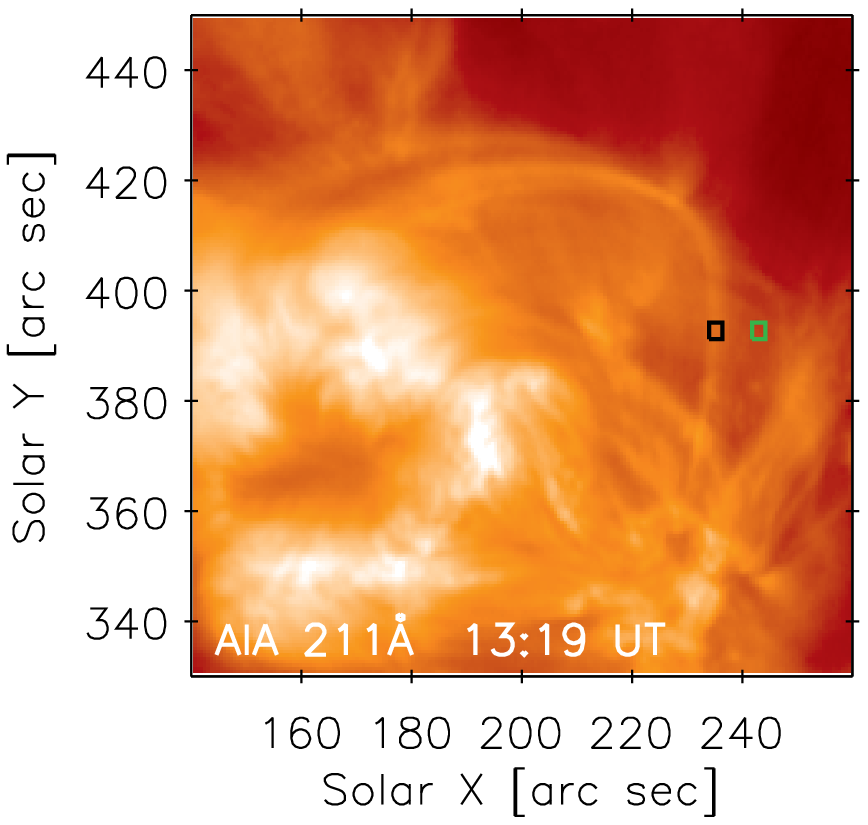}
       \includegraphics[width=3.30cm, bb=53 40 246 235, clip]{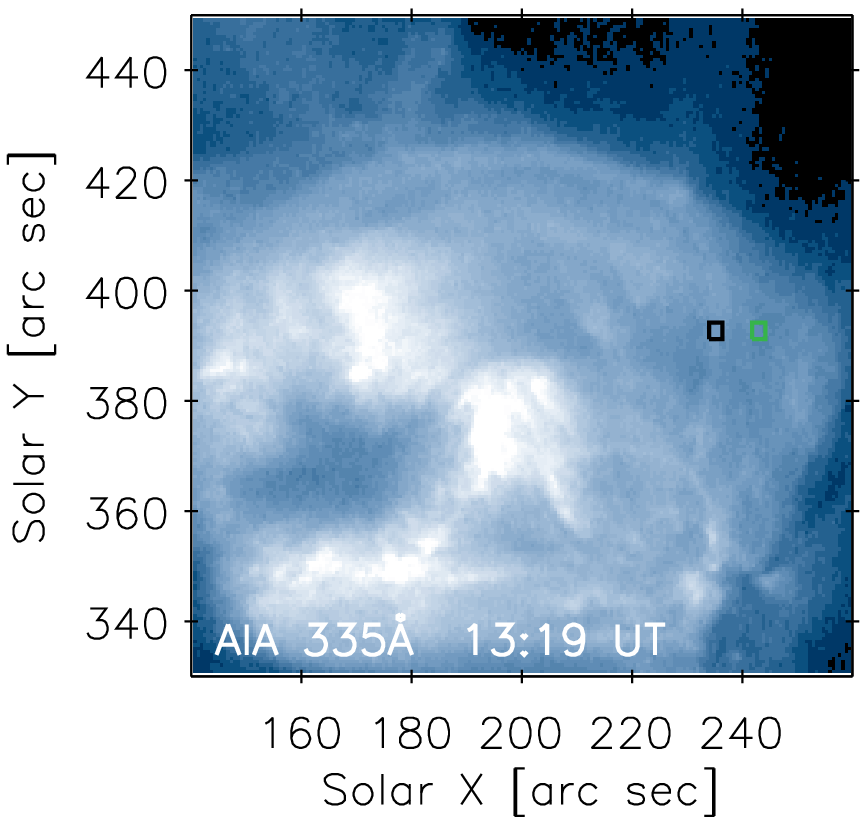}

       \includegraphics[width=4.20cm, bb=0  40 246 235, clip]{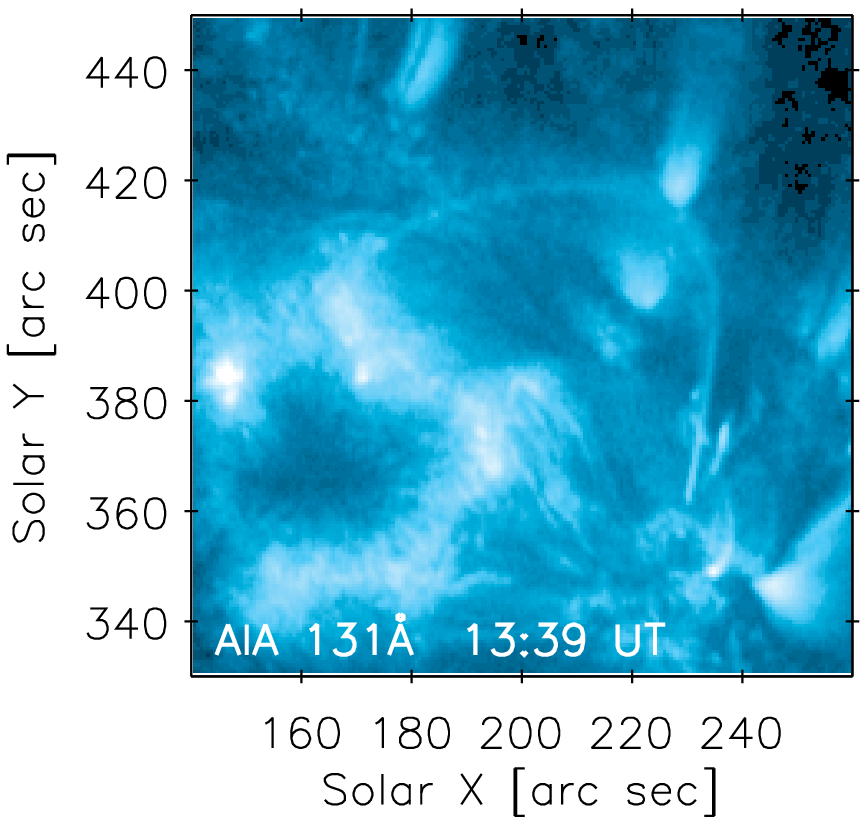}
       \includegraphics[width=3.30cm, bb=53 40 246 235, clip]{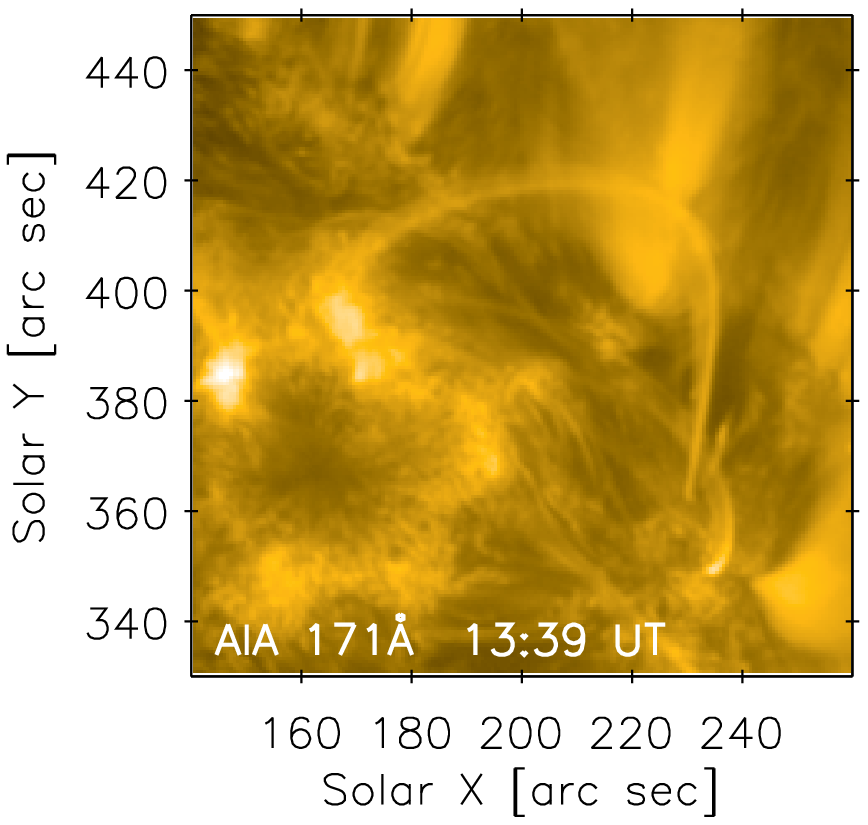}
       \includegraphics[width=3.30cm, bb=53 40 246 235, clip]{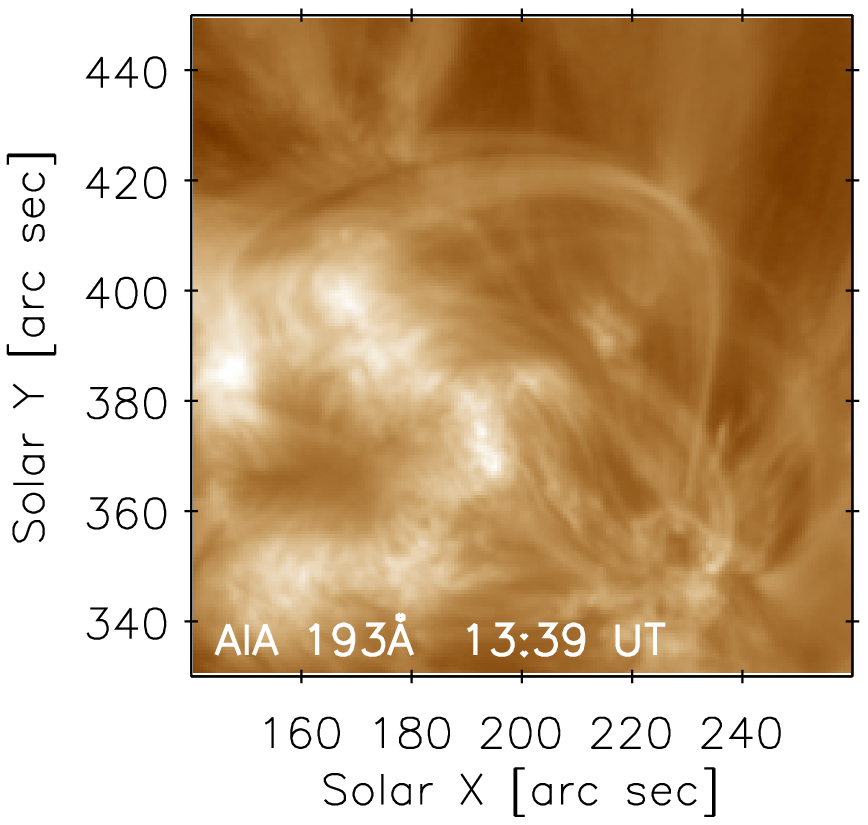}
       \includegraphics[width=3.30cm, bb=53 40 246 235, clip]{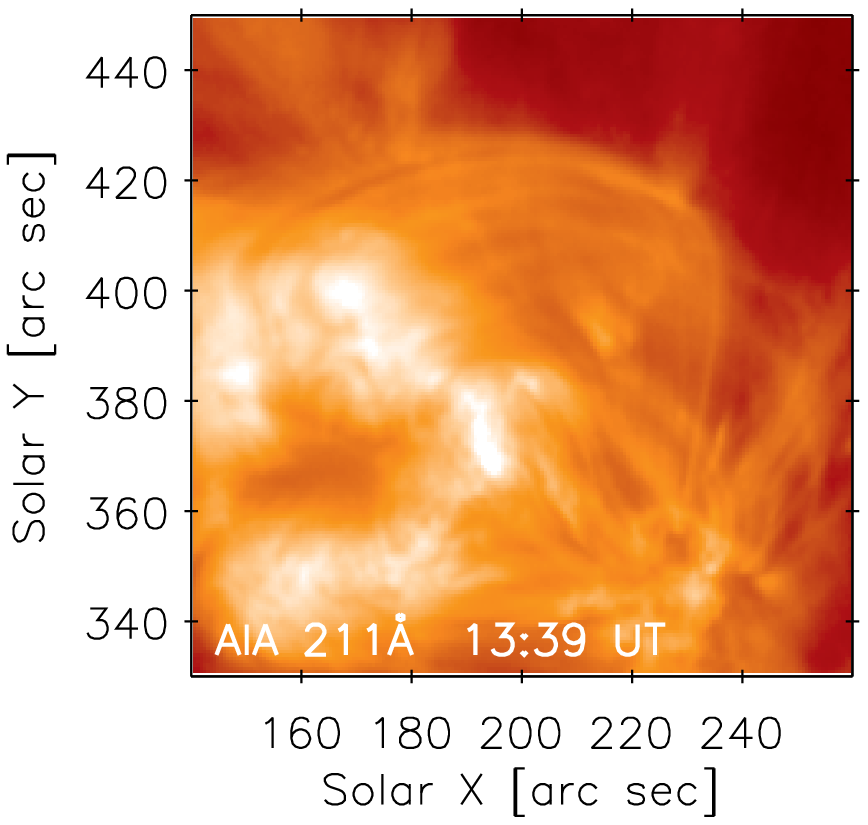}
       \includegraphics[width=3.30cm, bb=53 40 246 235, clip]{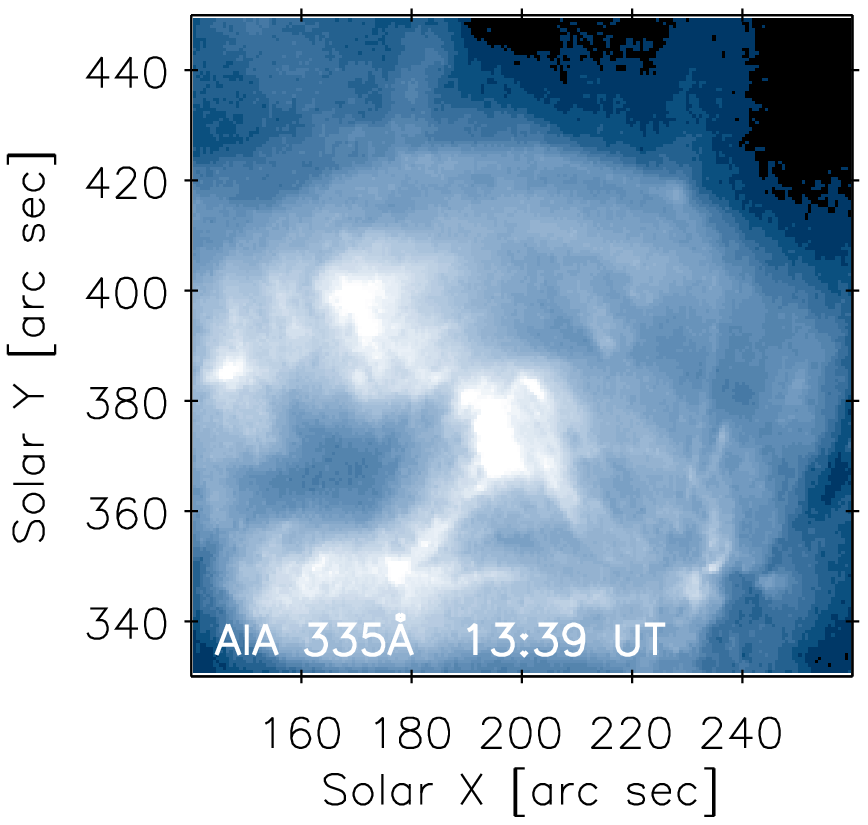}

       \includegraphics[width=4.20cm, bb=0  40 246 235, clip]{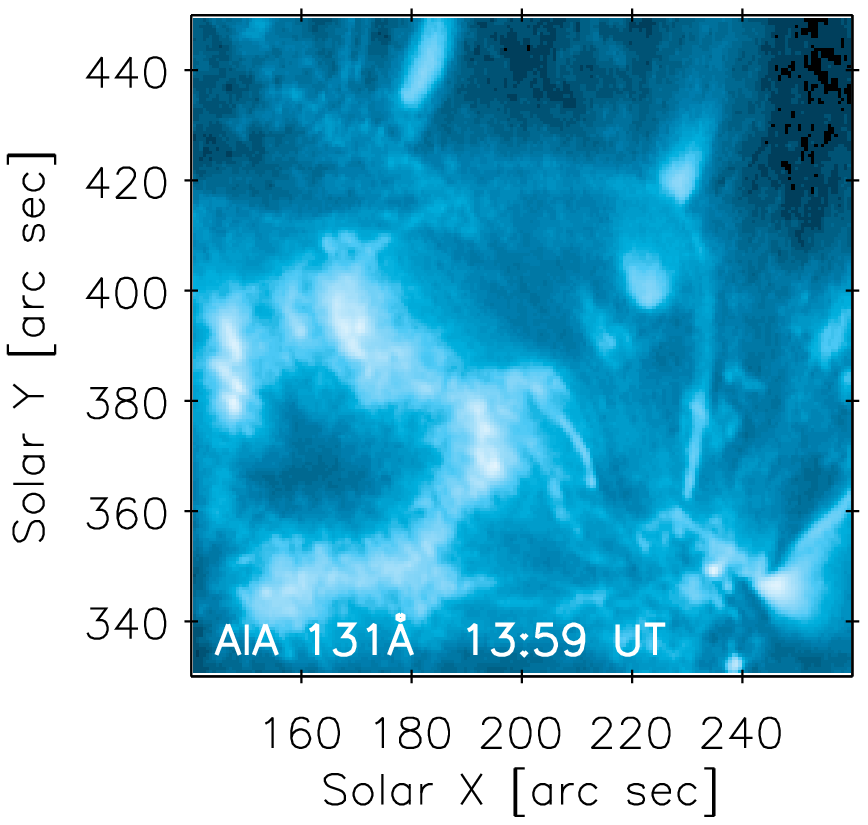}
       \includegraphics[width=3.30cm, bb=53 40 246 235, clip]{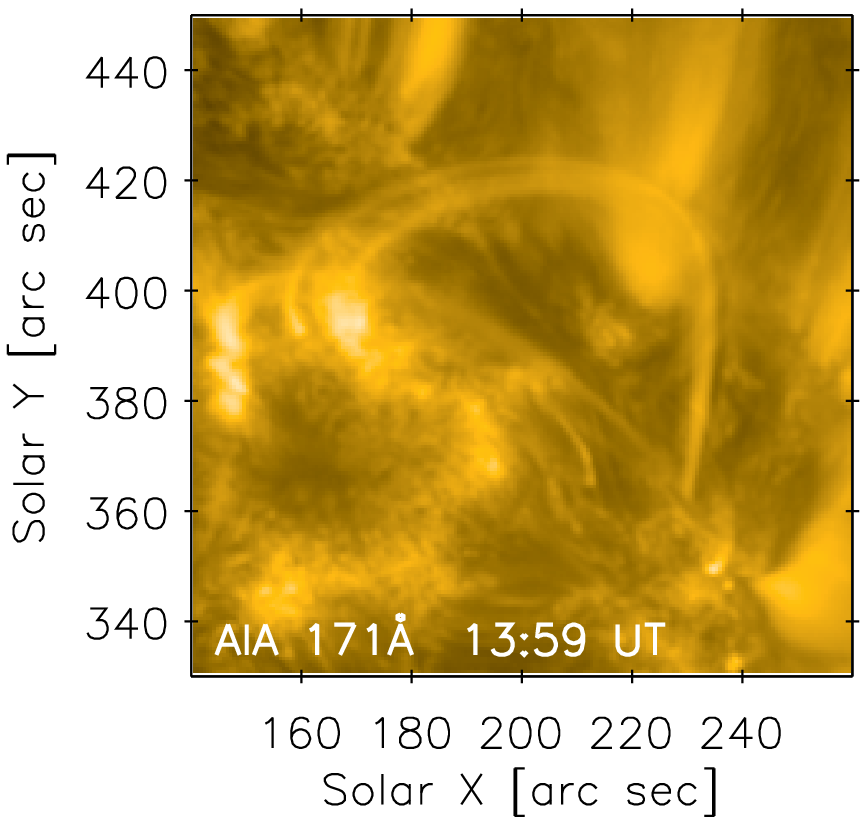}
       \includegraphics[width=3.30cm, bb=53 40 246 235, clip]{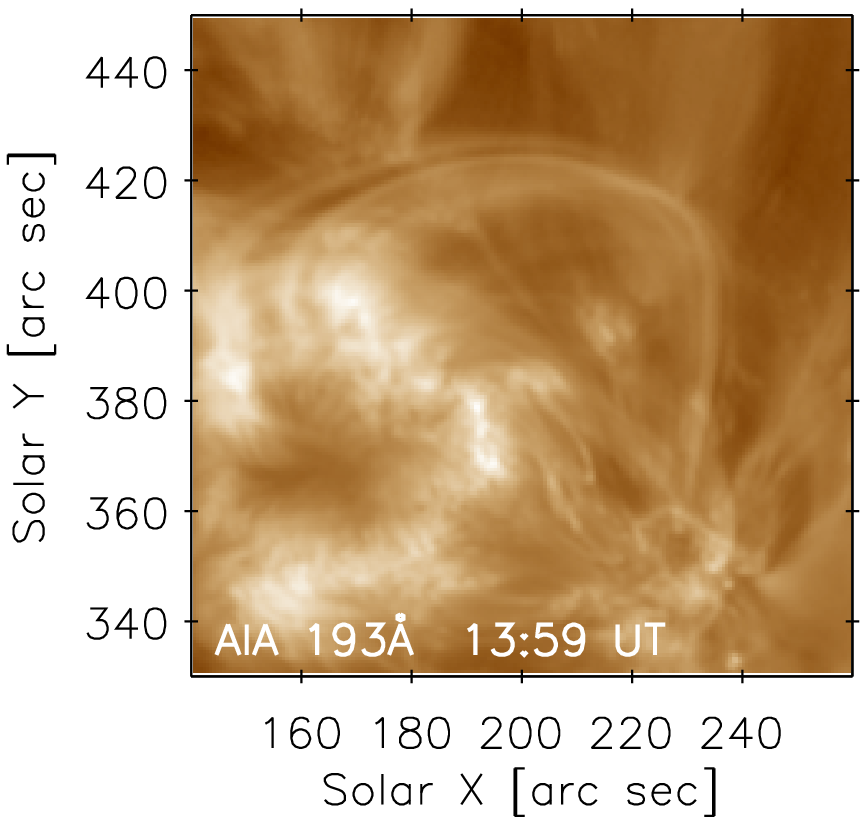}
       \includegraphics[width=3.30cm, bb=53 40 246 235, clip]{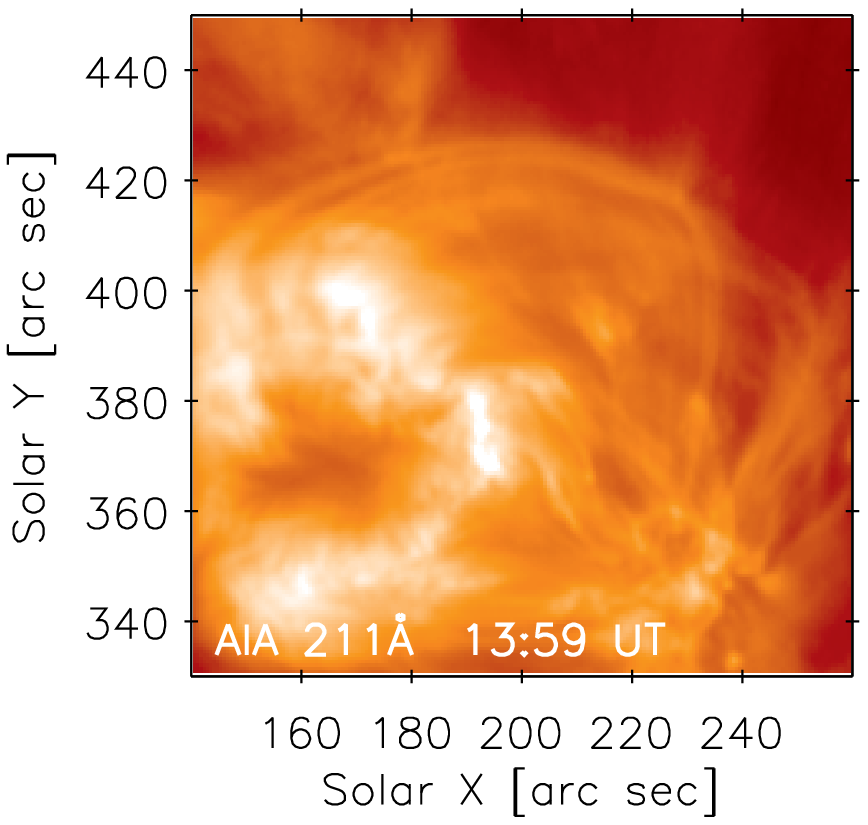}
       \includegraphics[width=3.30cm, bb=53 40 246 235, clip]{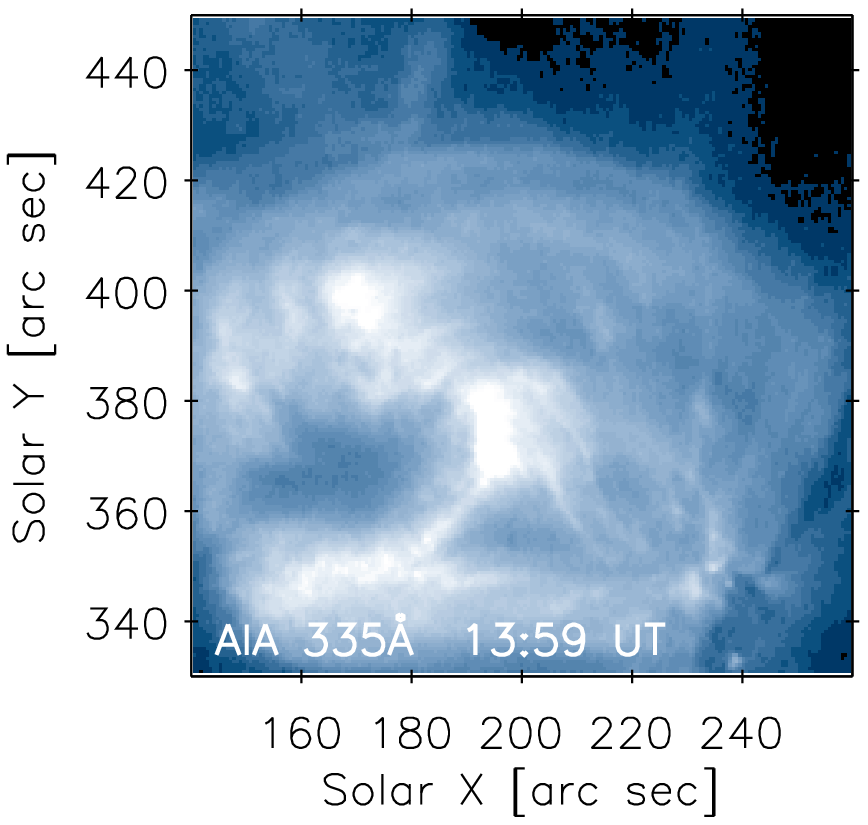}

       \includegraphics[width=4.20cm, bb=0  40 246 235, clip]{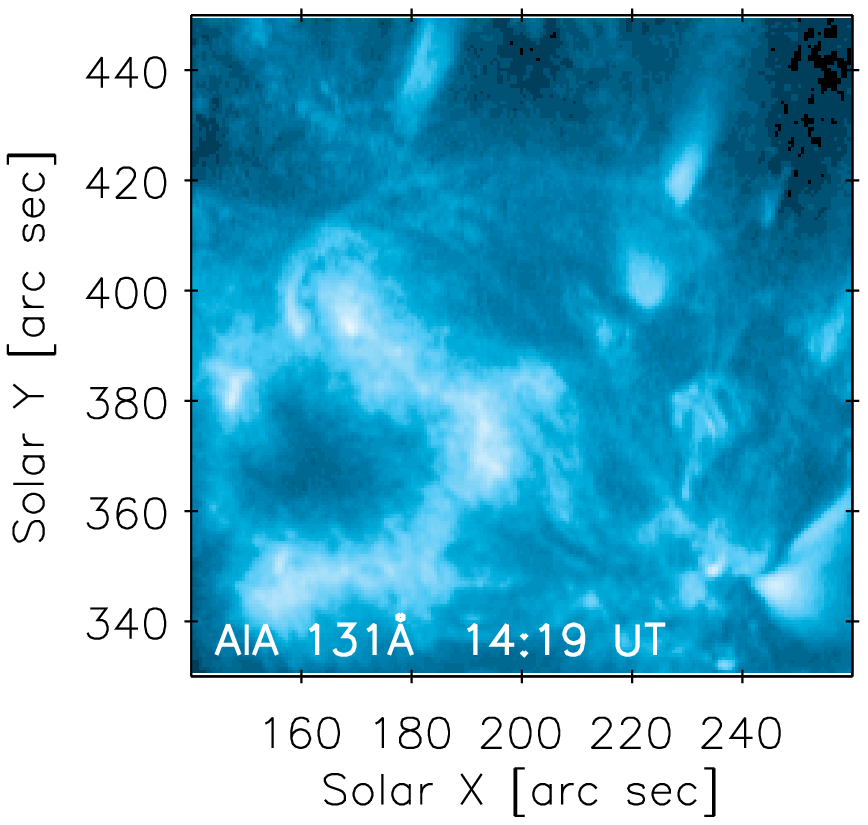}
       \includegraphics[width=3.30cm, bb=53 40 246 235, clip]{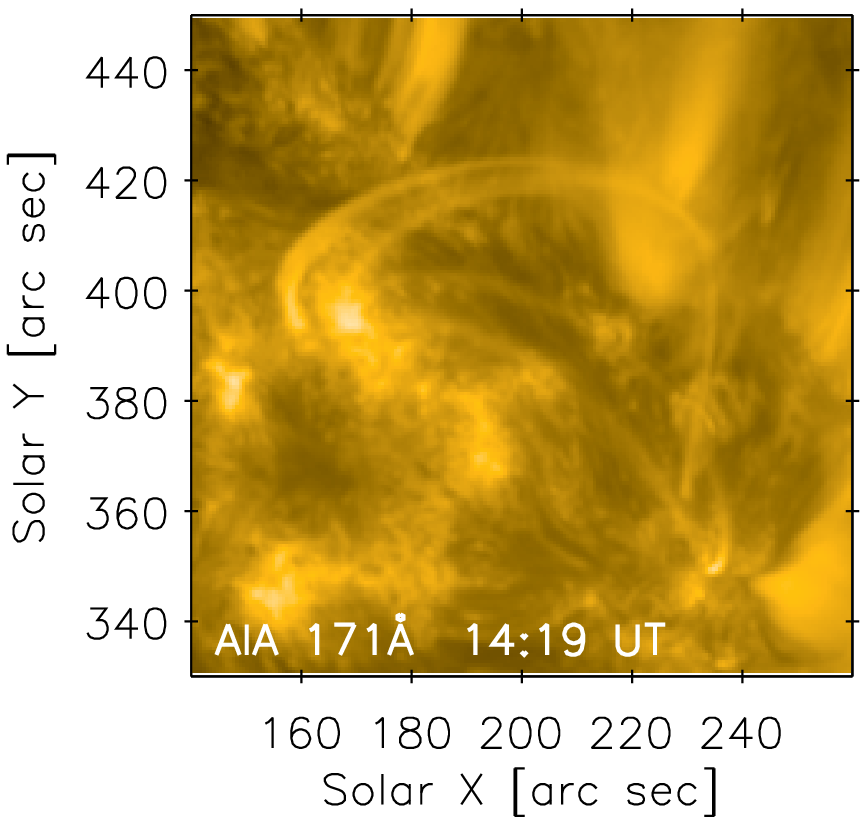}
       \includegraphics[width=3.30cm, bb=53 40 246 235, clip]{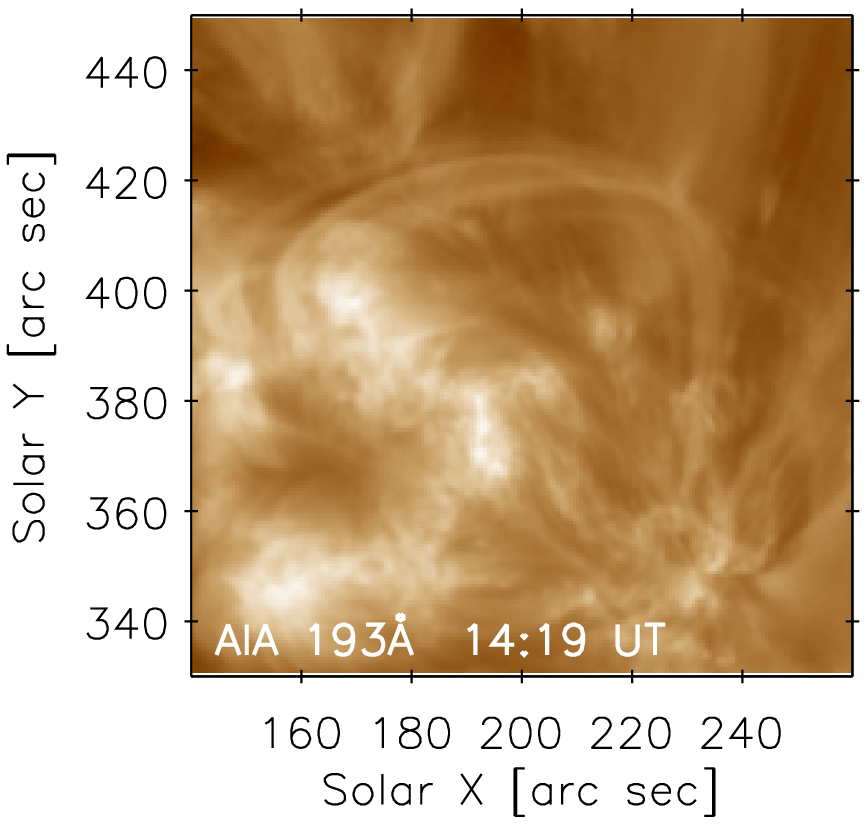}
       \includegraphics[width=3.30cm, bb=53 40 246 235, clip]{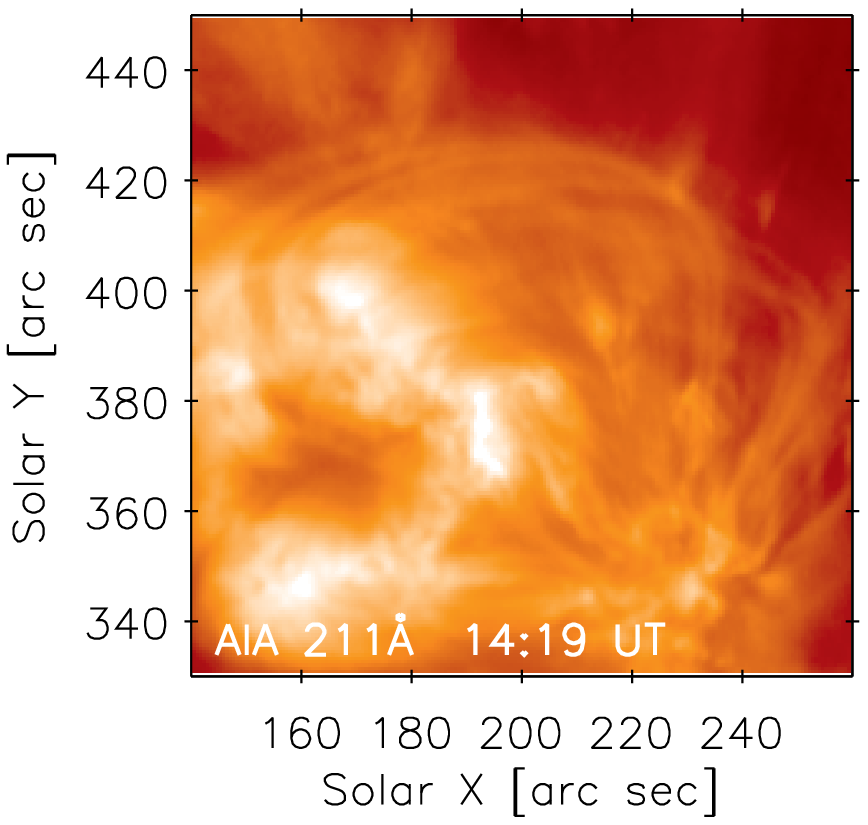}
       \includegraphics[width=3.30cm, bb=53 40 246 235, clip]{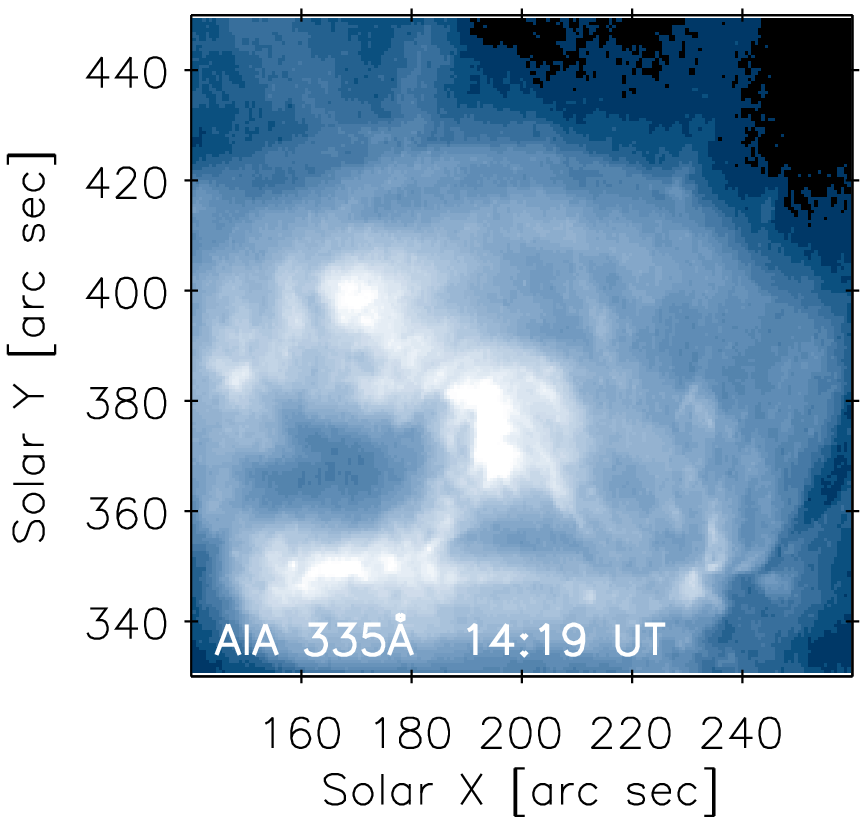}

       \includegraphics[width=4.20cm, bb=0   0 246 235, clip]{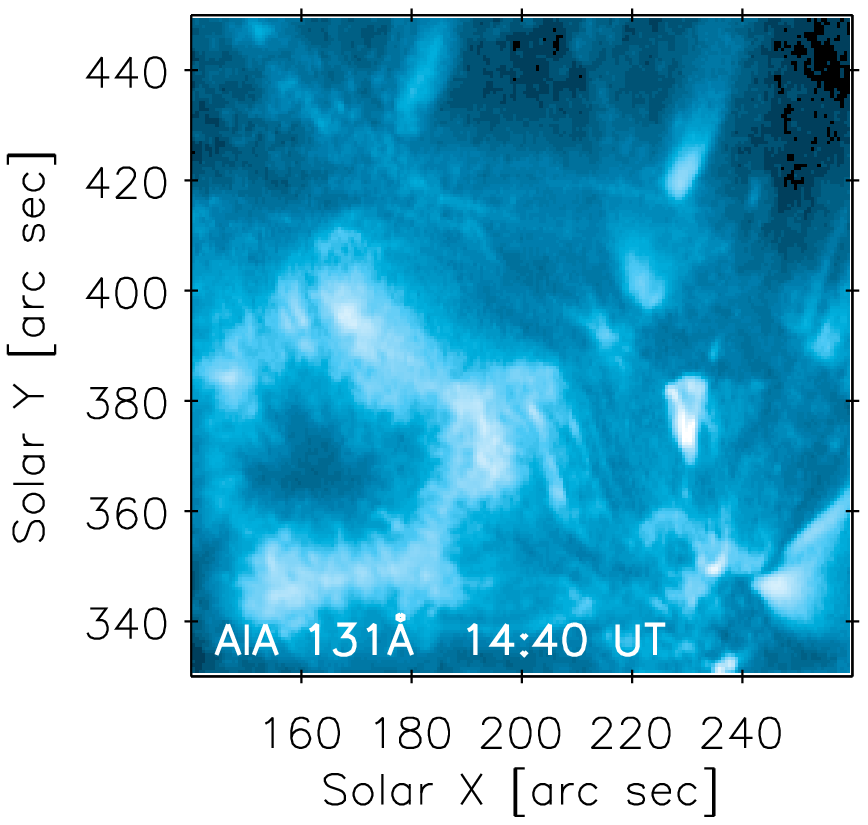}
       \includegraphics[width=3.30cm, bb=53  0 246 235, clip]{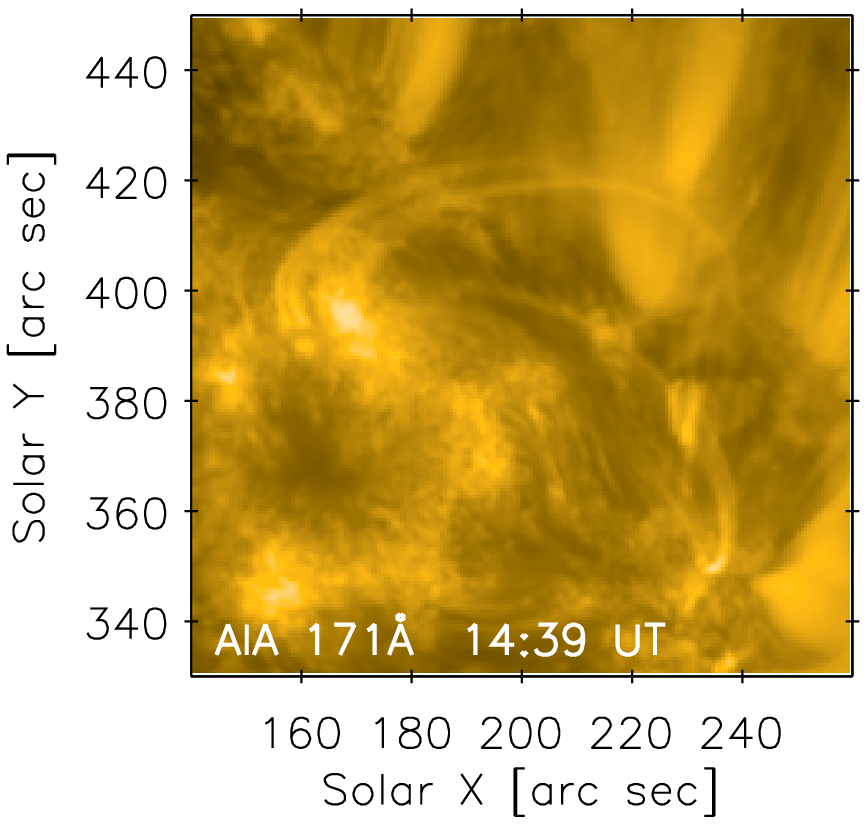}
       \includegraphics[width=3.30cm, bb=53  0 246 235, clip]{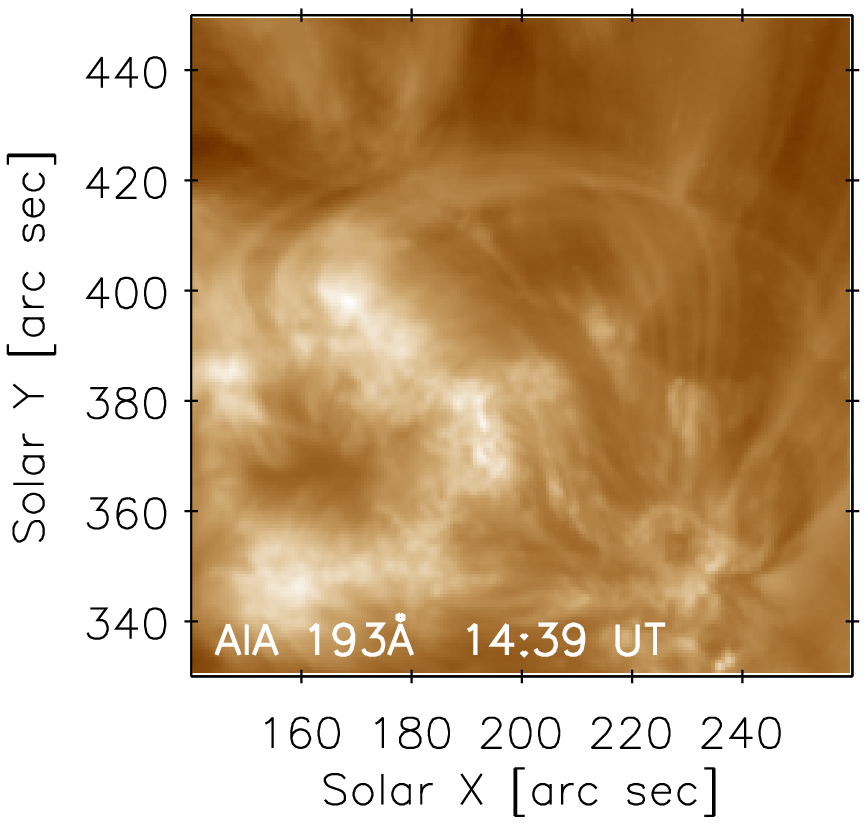}
       \includegraphics[width=3.30cm, bb=53  0 246 235, clip]{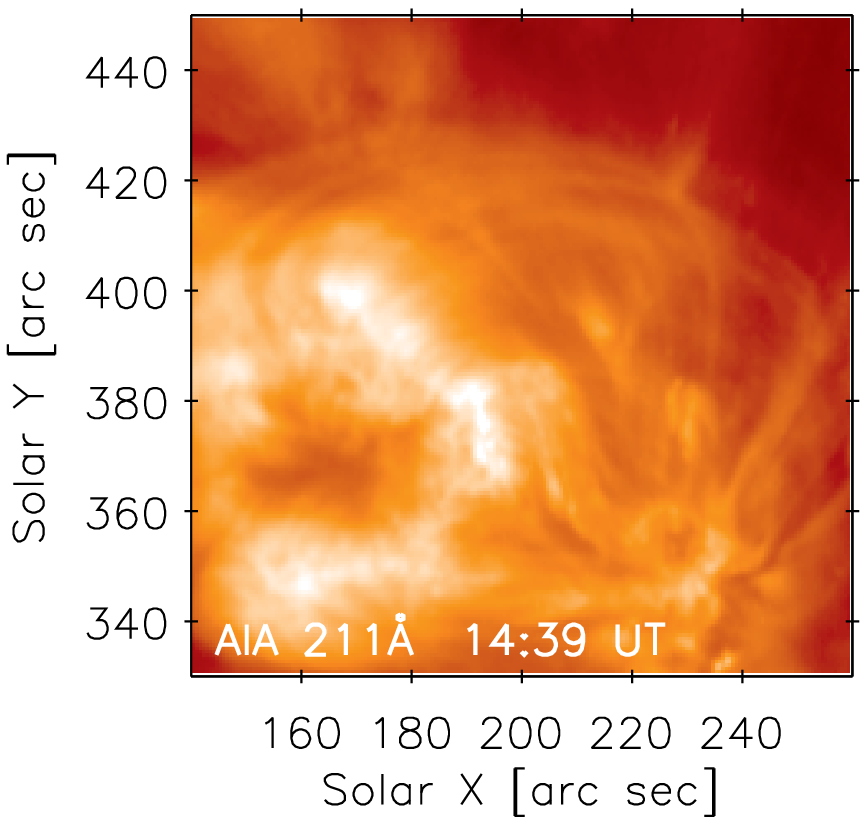}
       \includegraphics[width=3.30cm, bb=53  0 246 235, clip]{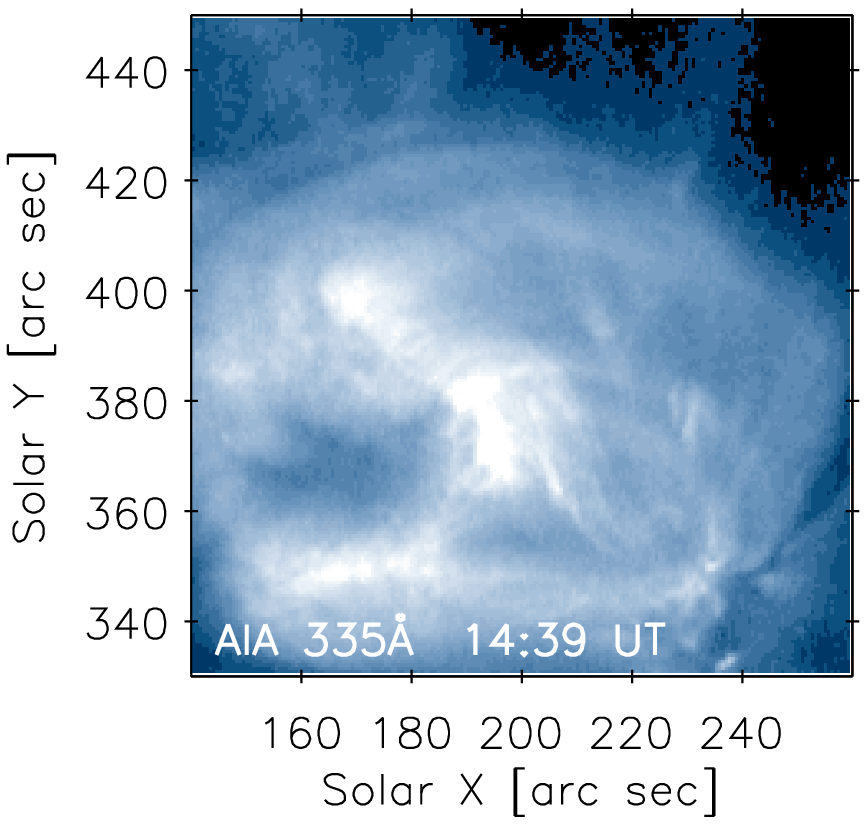}

       \caption{Multi-wavelength imaging observations of the loop evolution made by \textit{SDO}/AIA. The time cadence shown is 20\,min. AIA data are averaged over 1-minute intervals to increase the signal-to-noise especially in the 335\AA~and 131\AA~bands. The images are scaled logarithmically. The field of view shown corresponds to the box in Fig.~\ref{Fig:Evolution}. Small black and green boxes in the \textit{second row} denote the selected portion of the loop and background, respectively. The white box shows the field of view of a portion of the EIS raster shown in Fig.~\ref{Fig:EIS_raster}. \\ A color version of this image is available in the online journal.}
       \label{Fig:AIA_Loop}
	\vspace{0.4cm}
   \end{figure*}
%
%
   \begin{figure}[!t]
	\centering
	\includegraphics[width=8.8cm,clip]{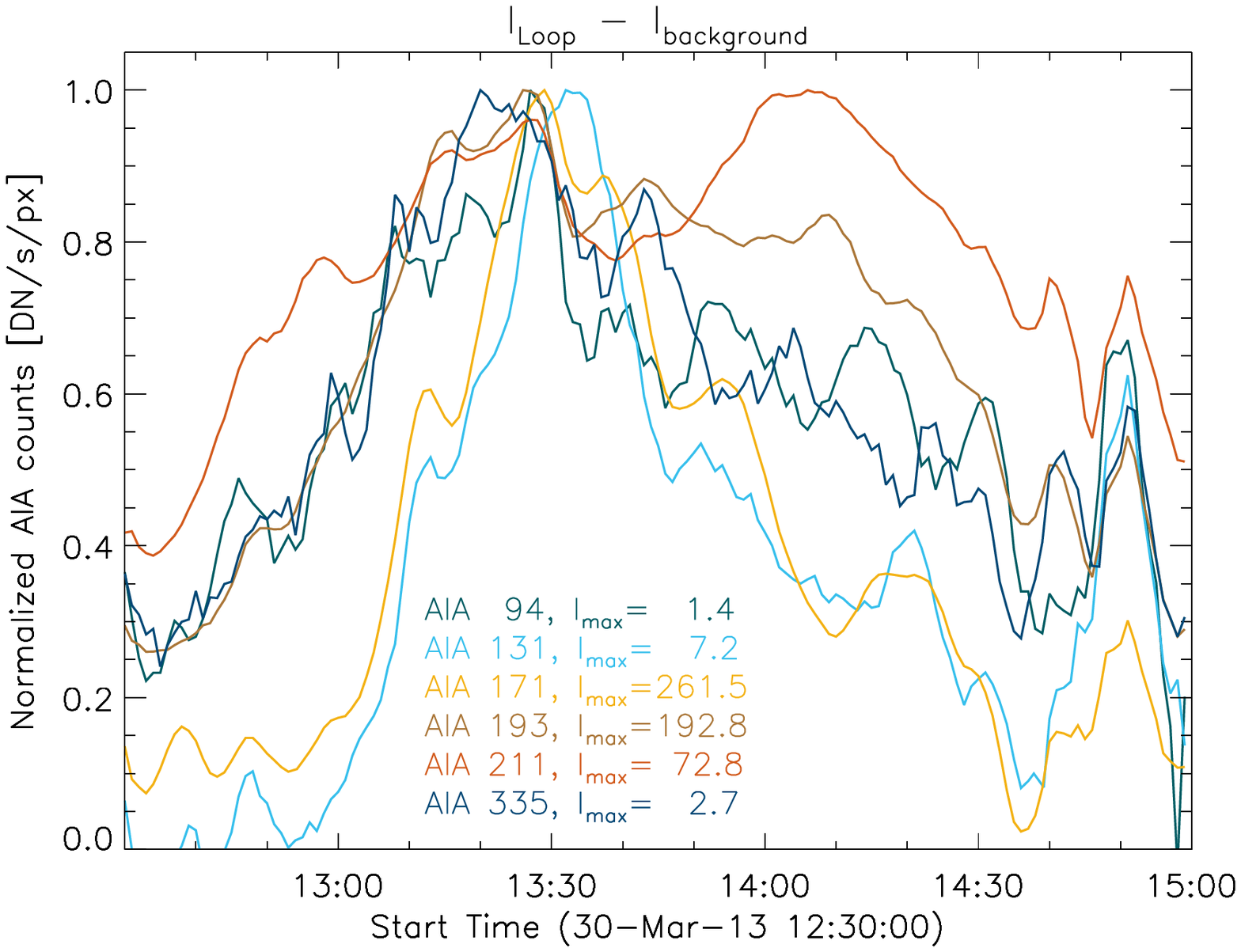}
	\includegraphics[width=8.8cm,clip,bb=0 40 498 195]{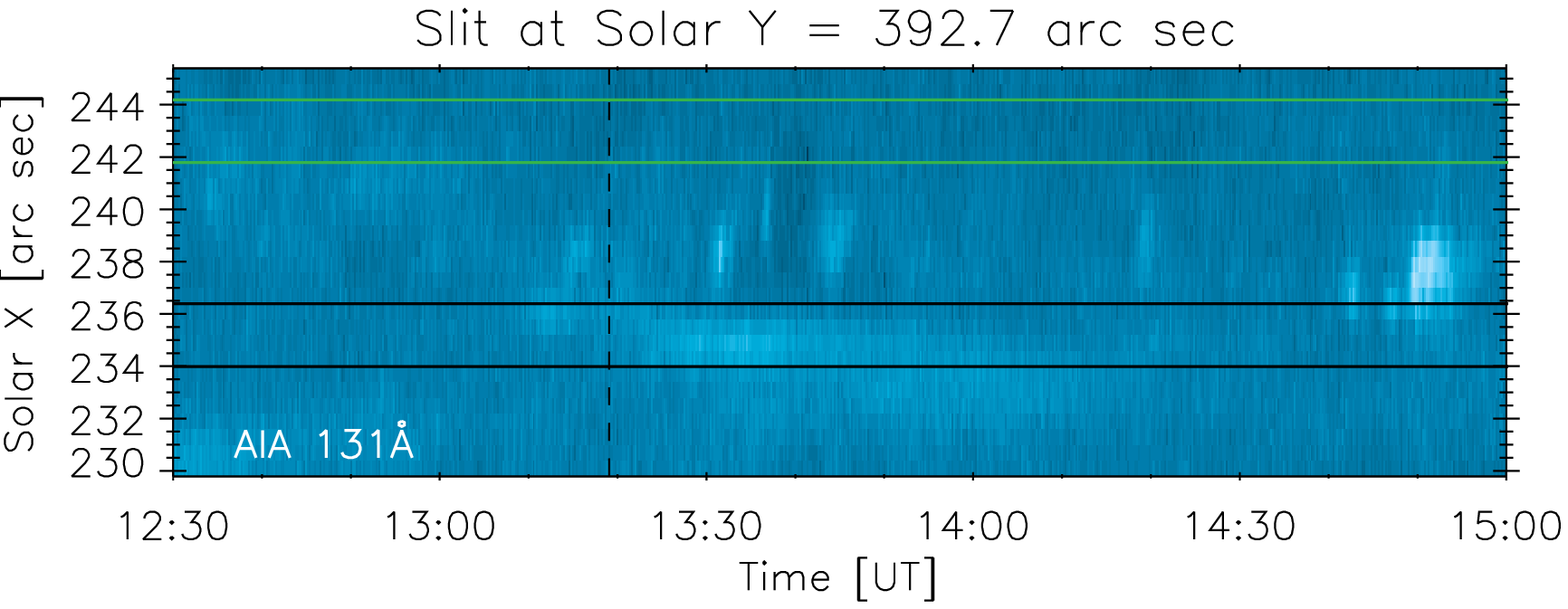}
	\includegraphics[width=8.8cm,clip,bb=0 40 498 175]{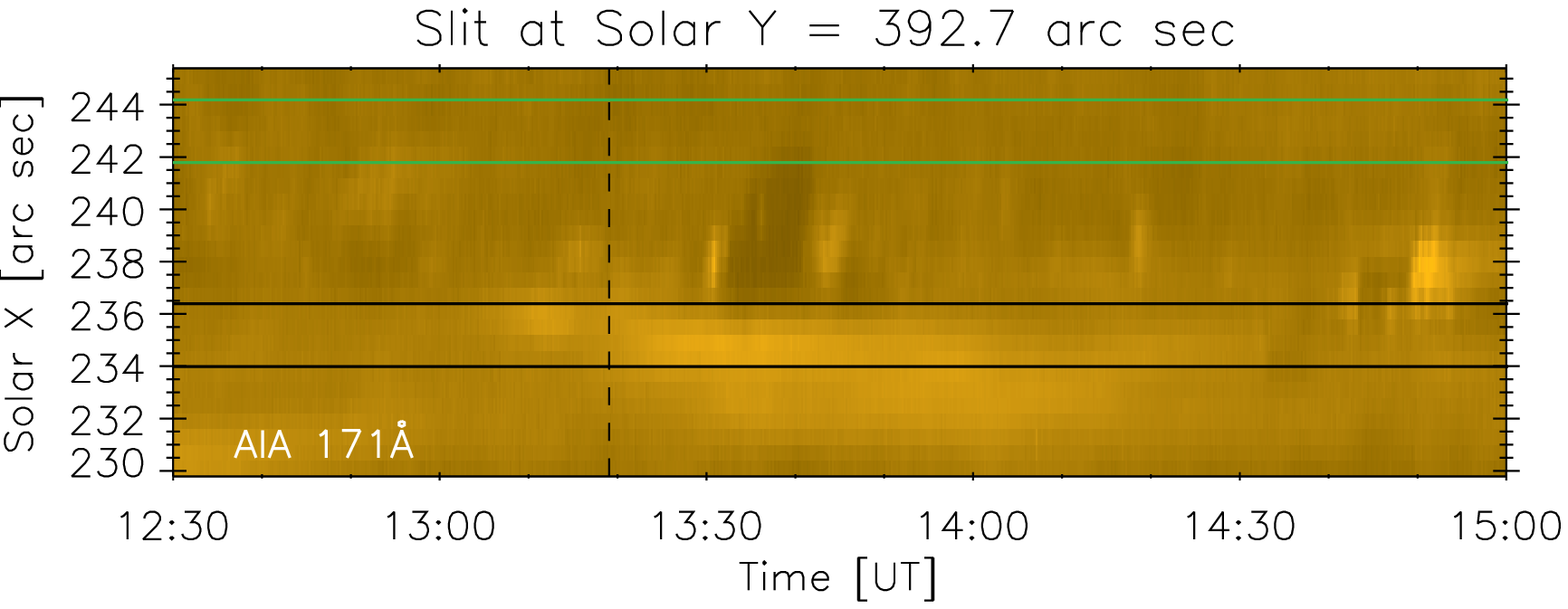}
	\includegraphics[width=8.8cm,clip,bb=0  0 498 175]{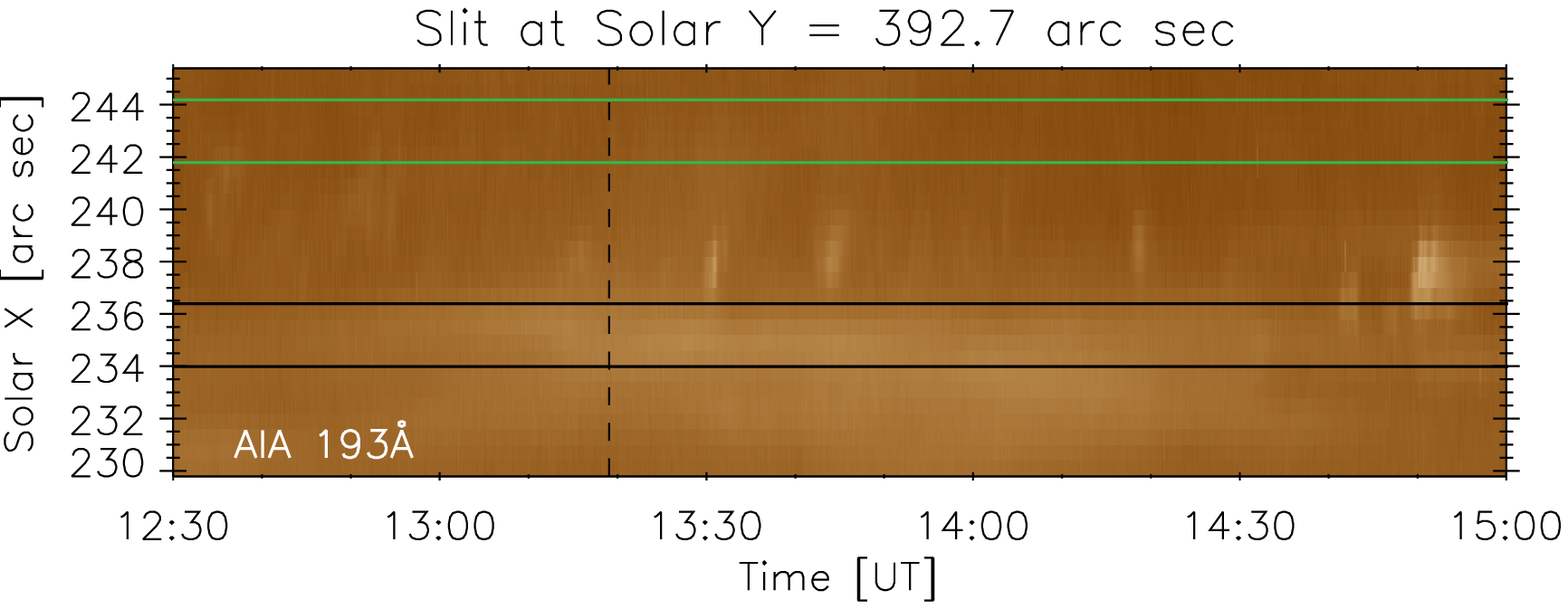}
	\caption{\textit{Top}: Background-subtracted intensity lightcurves for the small black box shown in Fig.~\ref{Fig:AIA_Loop}. The lightcurves are normalized to maximum intensity in the time window shown. The maximum intensities are listed for each filter. Individual colors stand for individual AIA filters. \textit{Bottom}: Time-distance plots in AIA 131\AA, 171\AA, and 193\AA~along a slit placed through the black and green box at Solar\,$Y$\,=\,392.7$\arcsec$. Black and green horizontal lines correspond to the Solar $X$ extensions of the respective boxes shown in Fig.~\ref{Fig:AIA_Loop}, where black stands for the loop and green for the background. The vertical dashed black line corresponds to 13:19\,UT (Sect. \ref{Sect:4}). \\ A color version of this image is available in the online journal.}
       \label{Fig:AIA_Lightcurves}
   \end{figure}
%
   \begin{figure*}[!t]
	\centering
	\includegraphics[width=8.8cm,clip]{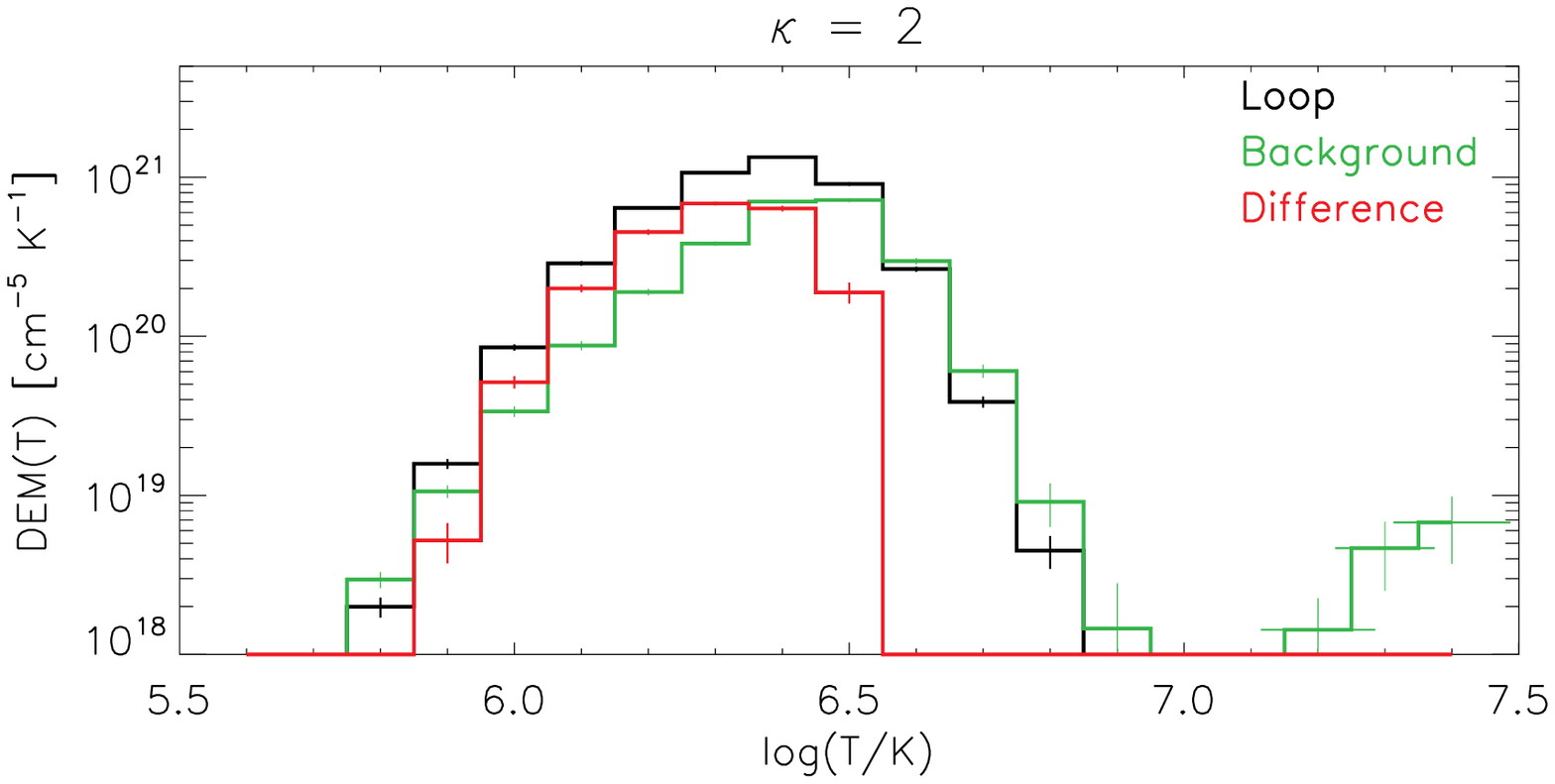}
	\includegraphics[width=8.8cm,clip]{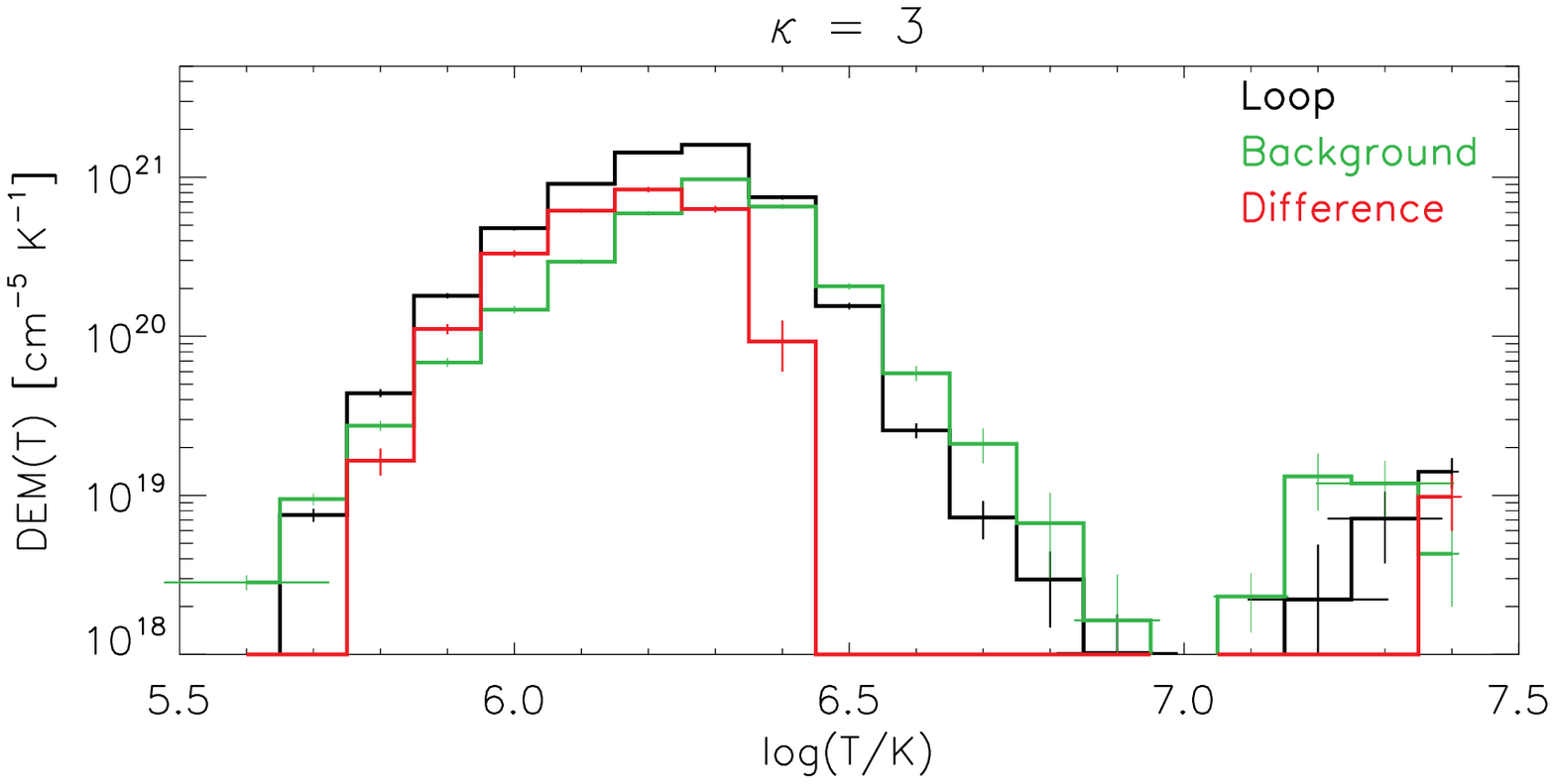}
	\includegraphics[width=8.8cm,clip]{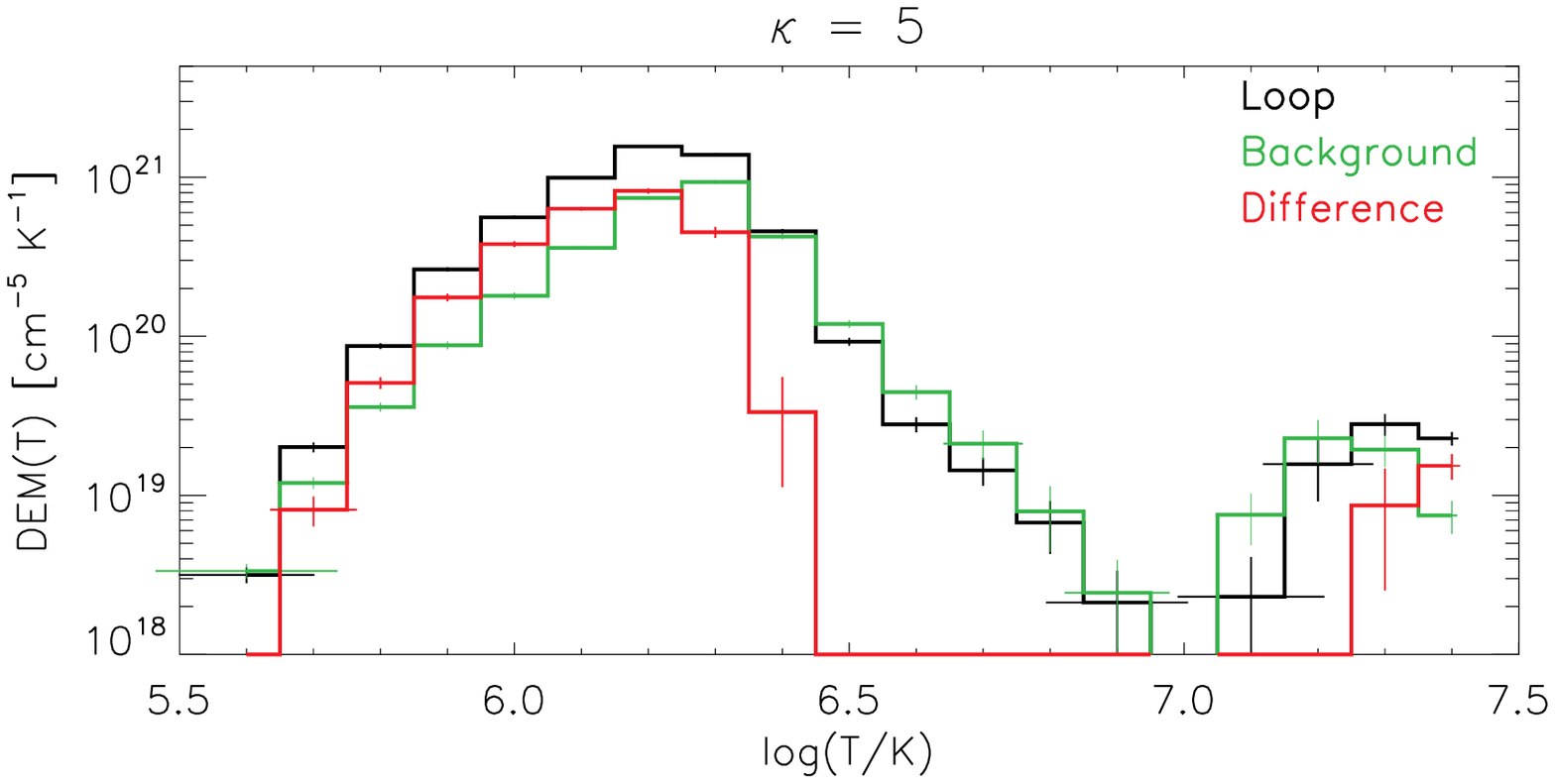}
	\includegraphics[width=8.8cm,clip]{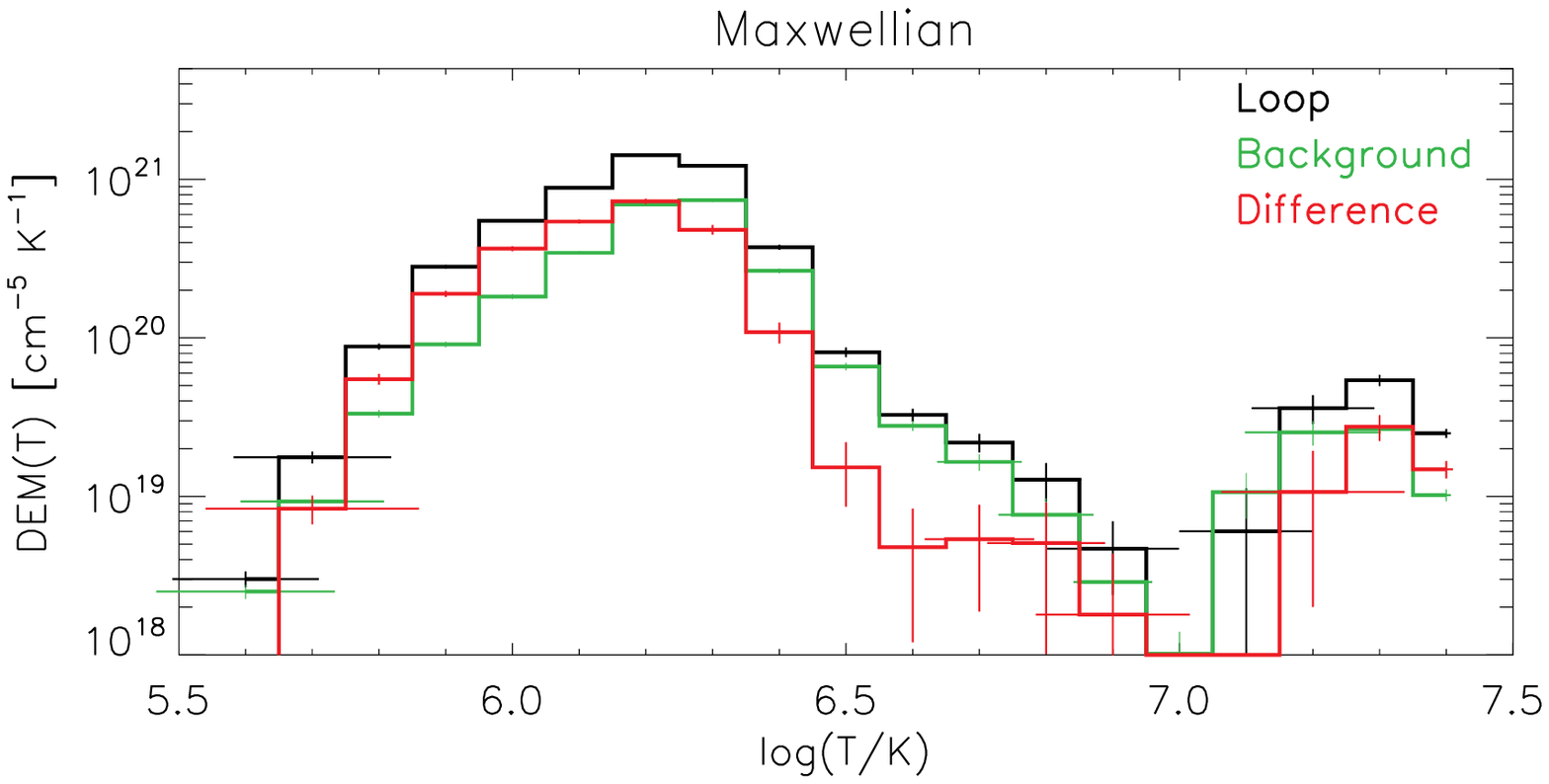}
	\caption{DEM diagnostics from AIA data using the regularized inversion method. The distribution assumed is indicated at each panel. Black and green lines stand for the average DEMs within the black and green boxes in Fig.~\ref{Fig:AIA_Loop}. Red stands for their difference. Uncertainties of the DEMs in each temperature bin are given by the respective horizontal and vertical error bars. \\ A color version of this image is available in the online journal.}
        \label{Fig:AIA_DEM}
   \end{figure*}
%

   \begin{figure*}[!ht]
       \centering
       \includegraphics[width=17.60cm,bb=13 0 993 63, clip]{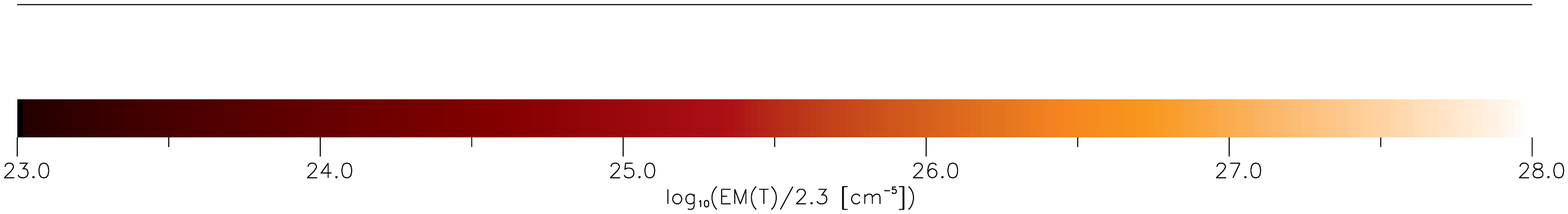}

       \includegraphics[width=4.20cm, bb=0  40 246 235, clip]{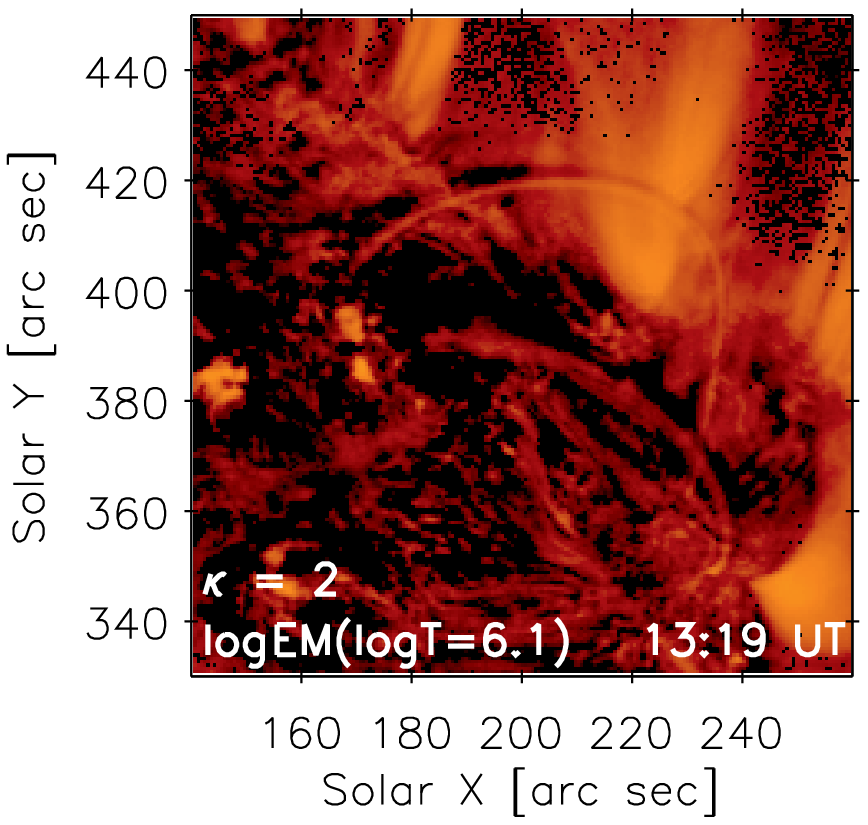}
       \includegraphics[width=3.30cm, bb=53 40 246 235, clip]{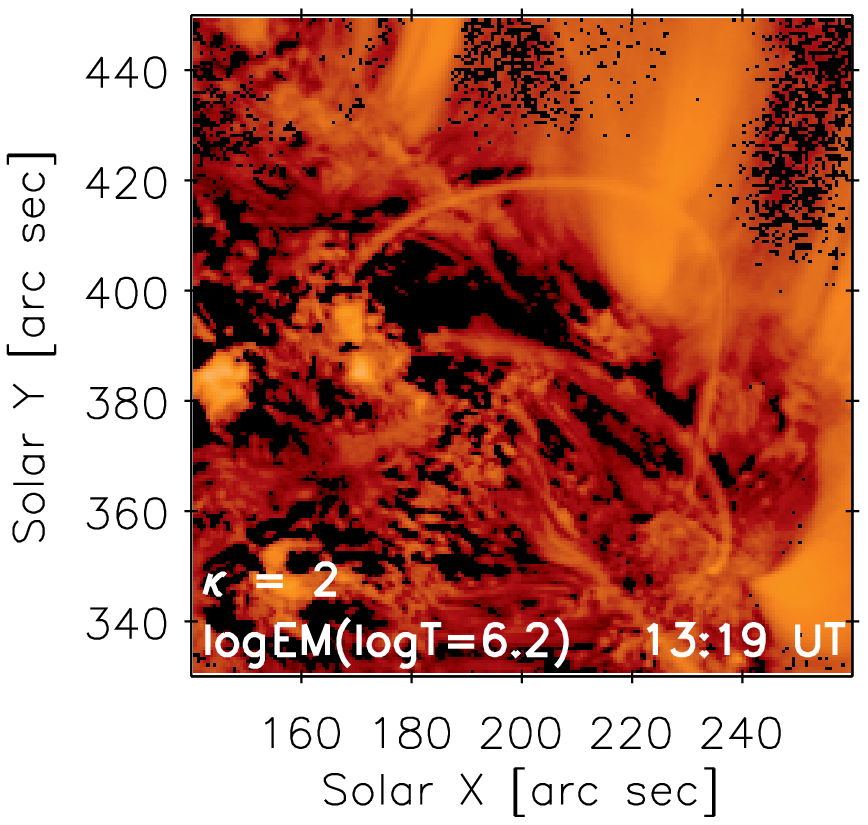}
       \includegraphics[width=3.30cm, bb=53 40 246 235, clip]{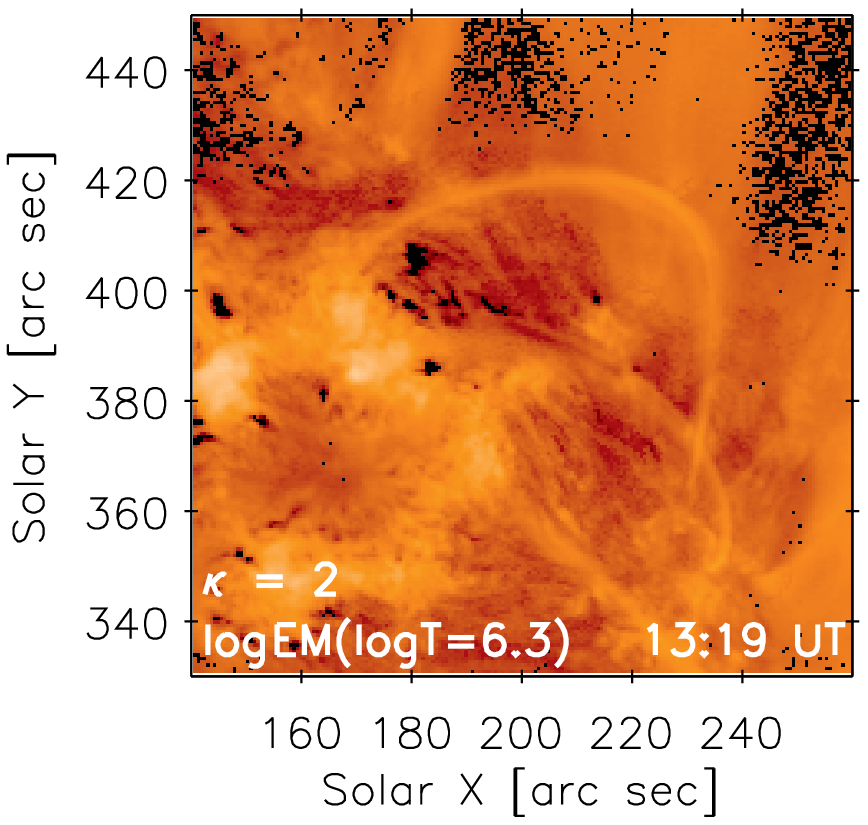}
       \includegraphics[width=3.30cm, bb=53 40 246 235, clip]{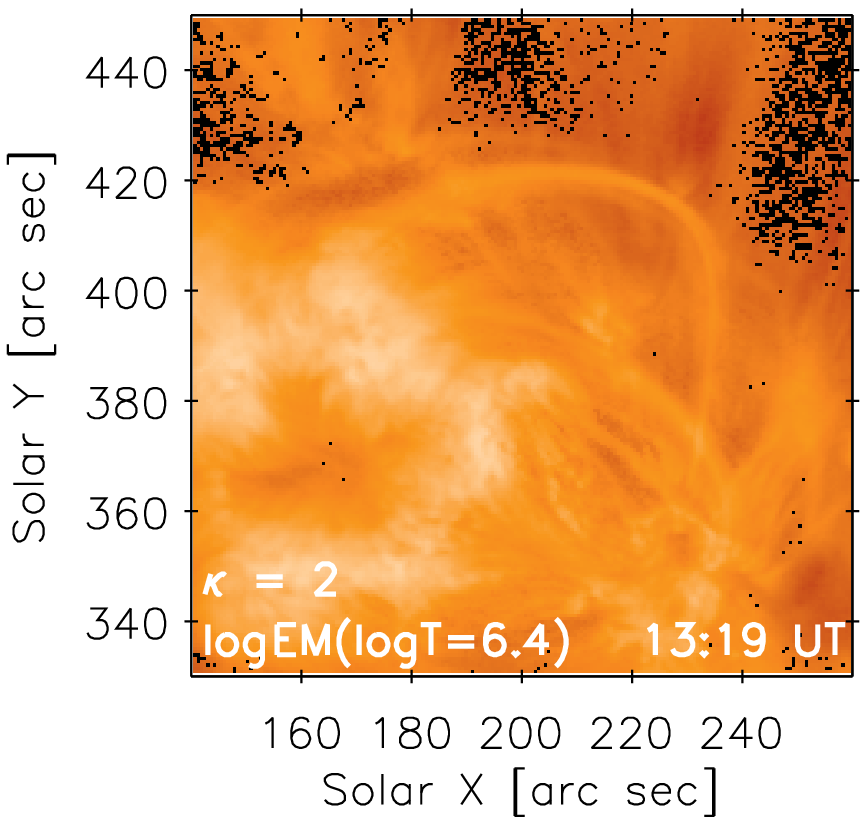}
       \includegraphics[width=3.30cm, bb=53 40 246 235, clip]{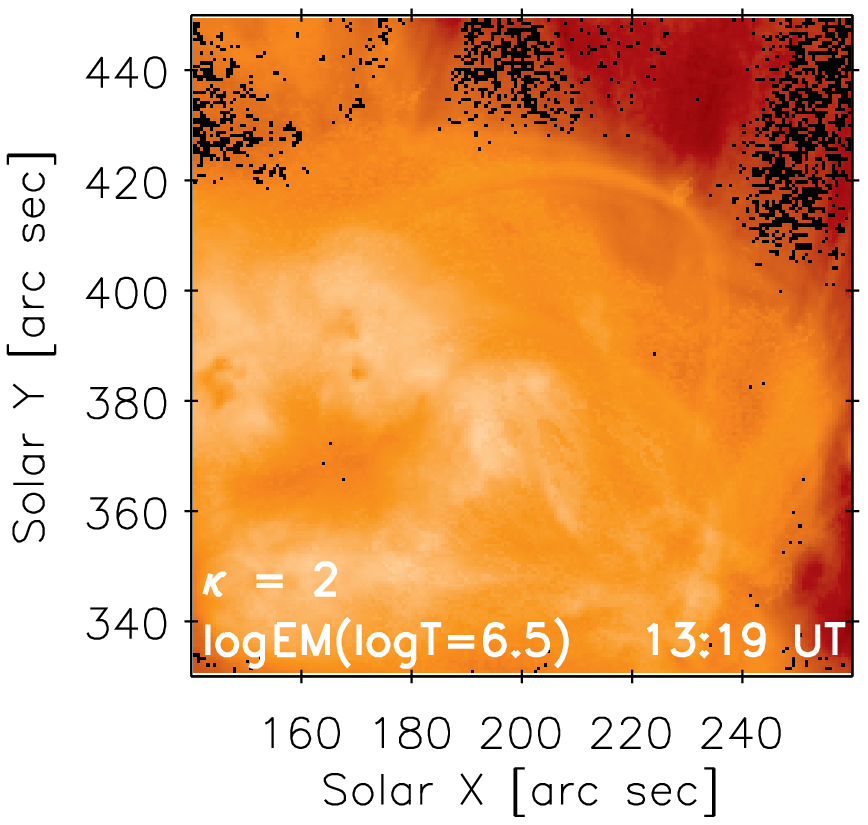}

       \includegraphics[width=4.20cm, bb=0  40 246 235, clip]{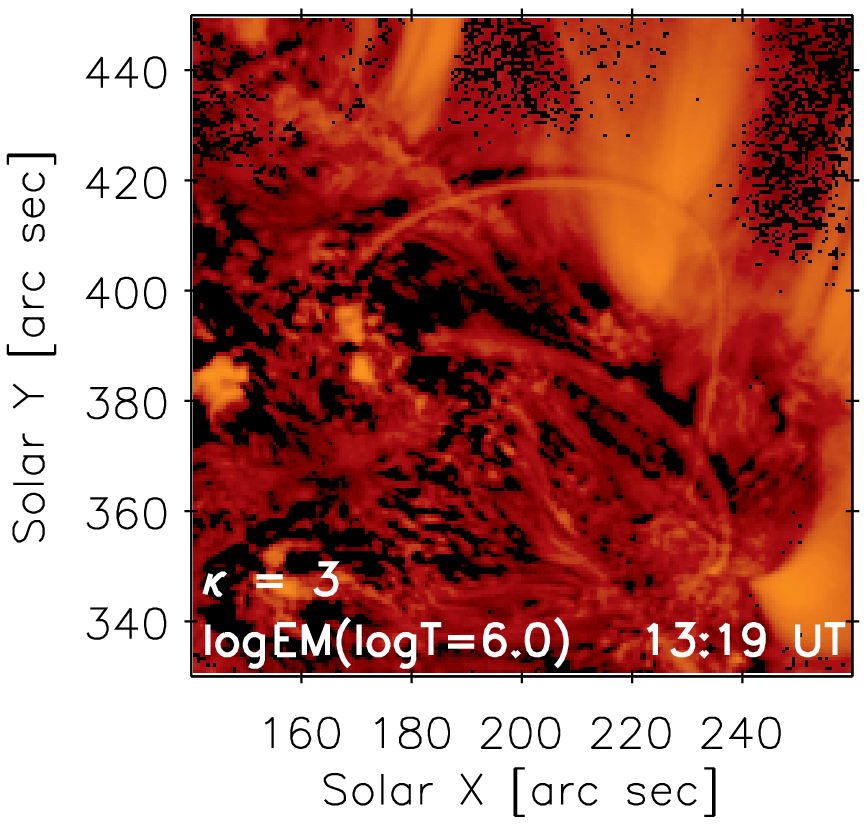}
       \includegraphics[width=3.30cm, bb=53 40 246 235, clip]{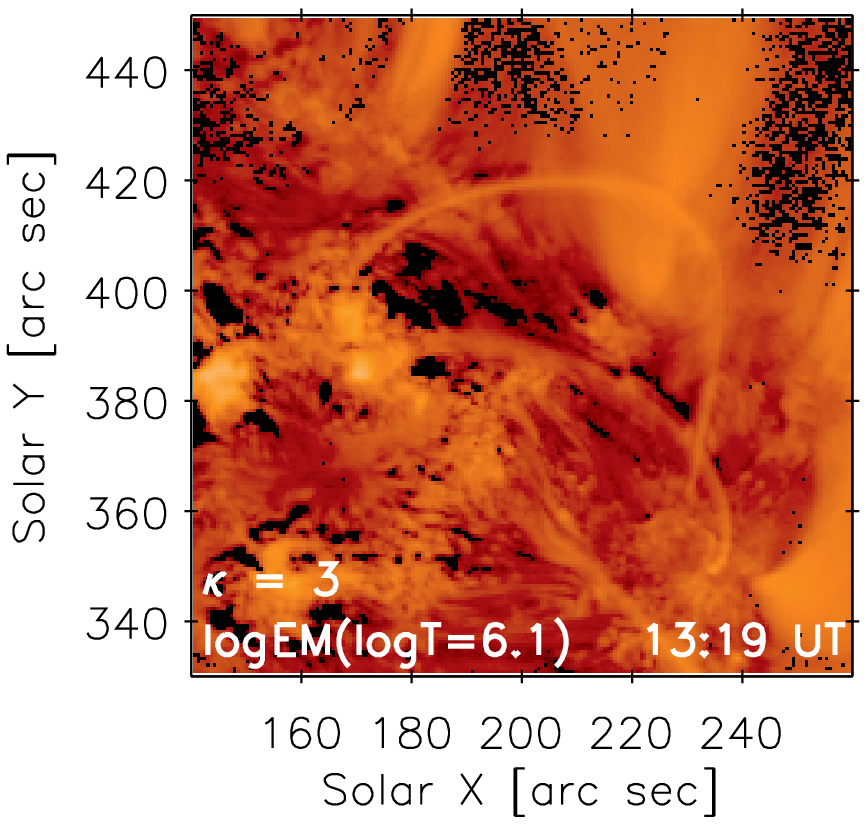}
       \includegraphics[width=3.30cm, bb=53 40 246 235, clip]{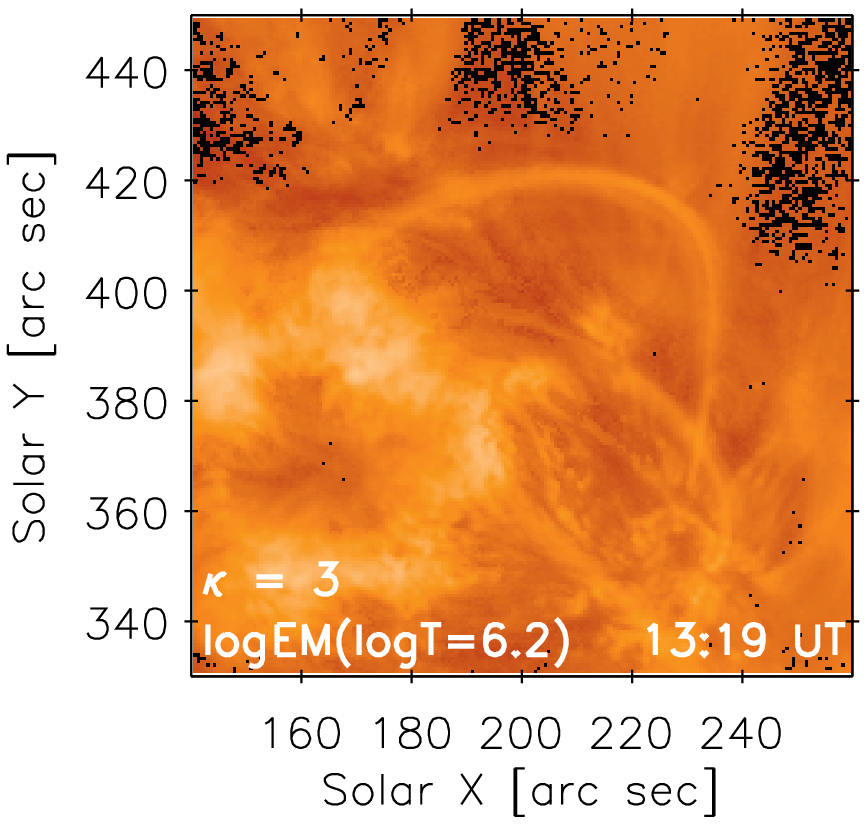}
       \includegraphics[width=3.30cm, bb=53 40 246 235, clip]{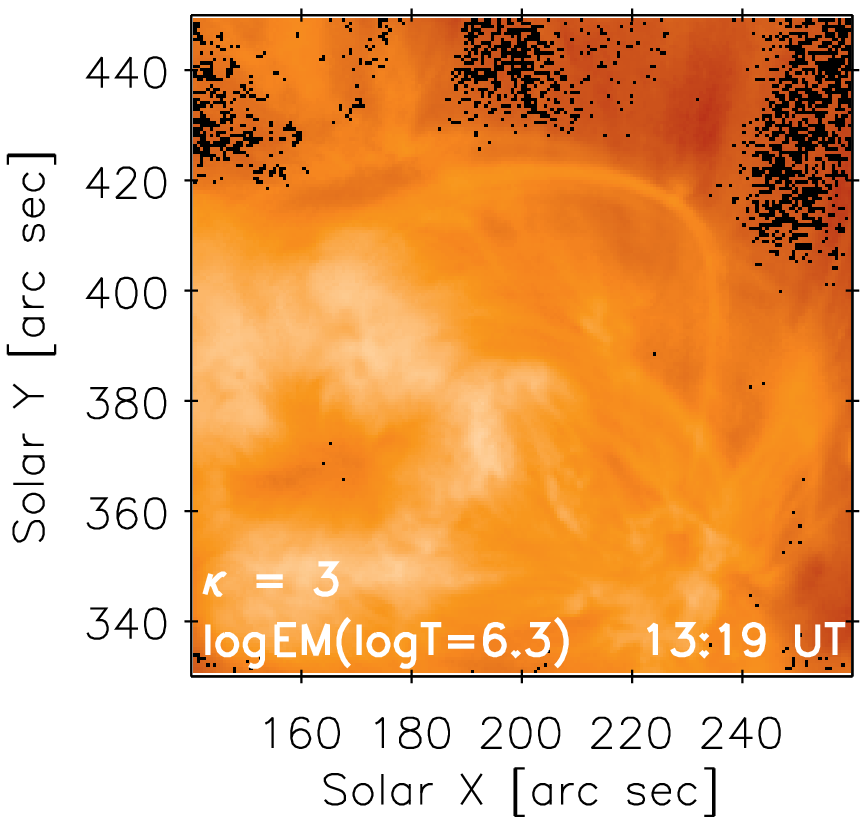}
       \includegraphics[width=3.30cm, bb=53 40 246 235, clip]{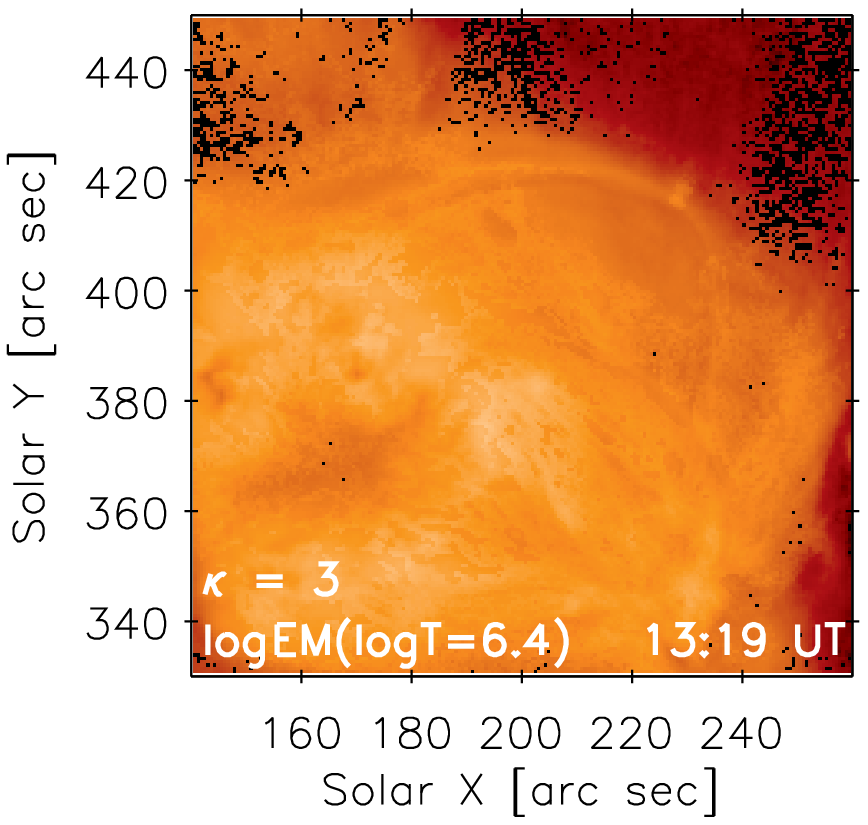}

       \includegraphics[width=4.20cm, bb=0  40 246 235, clip]{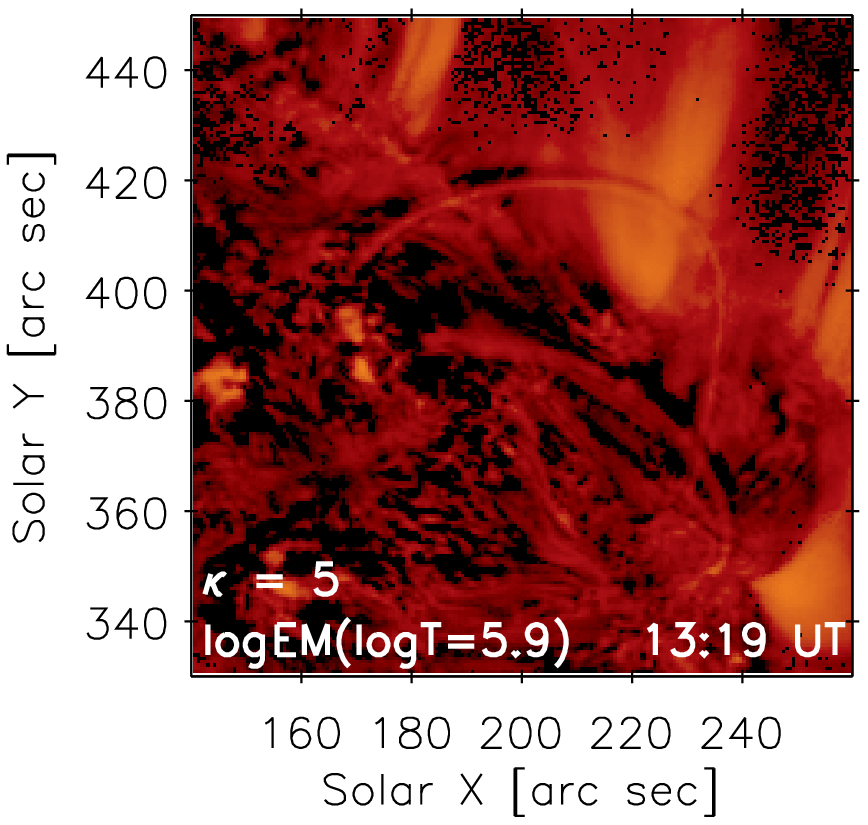}
       \includegraphics[width=3.30cm, bb=53 40 246 235, clip]{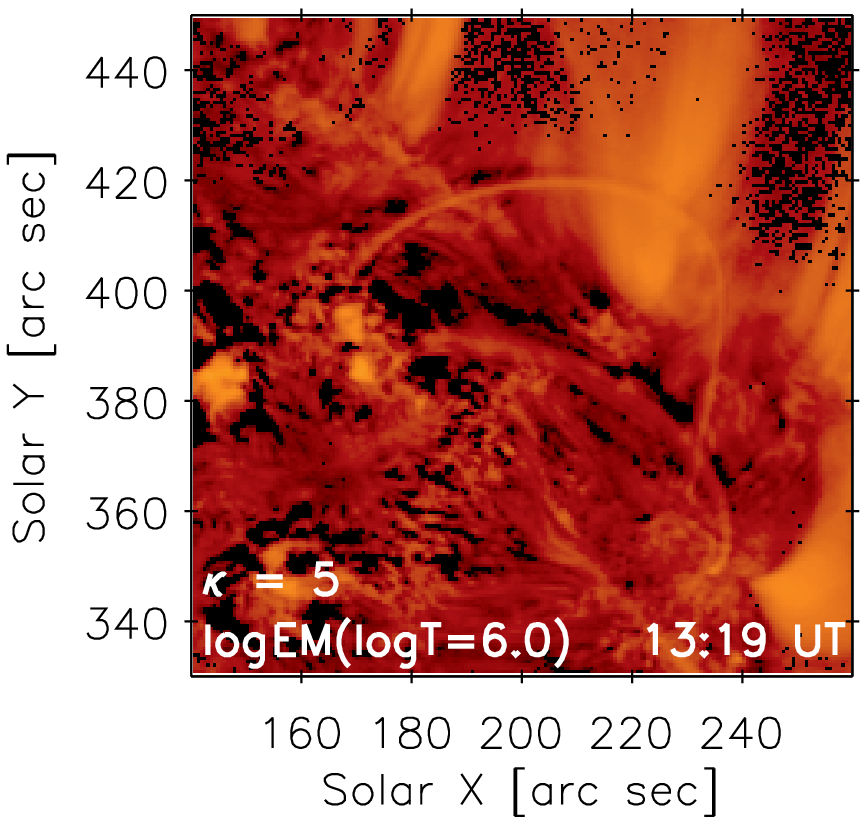}
       \includegraphics[width=3.30cm, bb=53 40 246 235, clip]{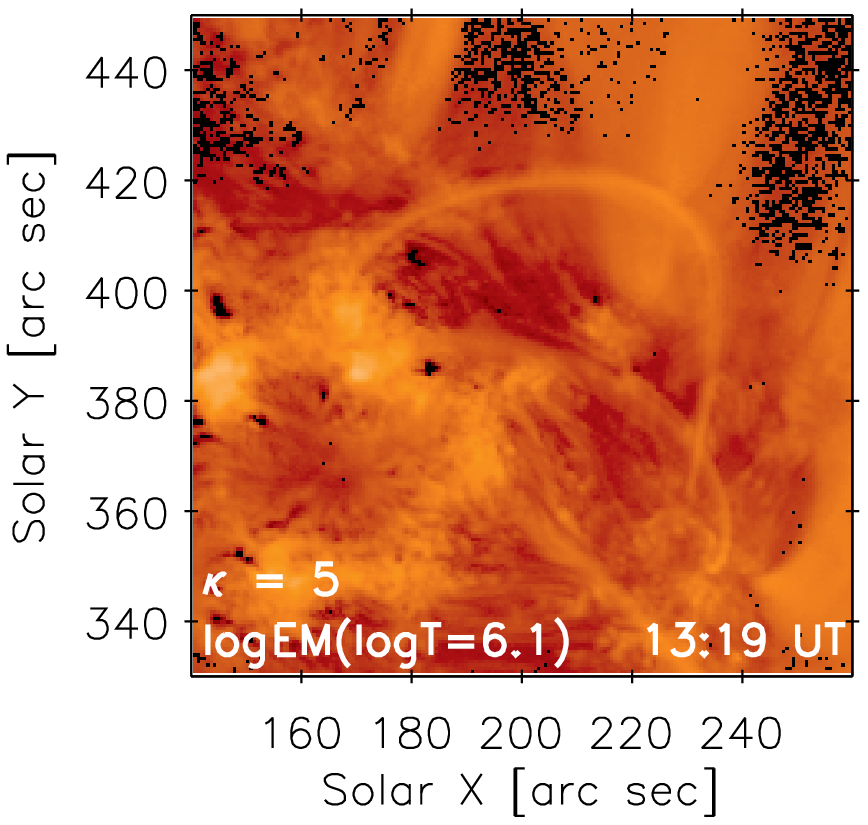}
       \includegraphics[width=3.30cm, bb=53 40 246 235, clip]{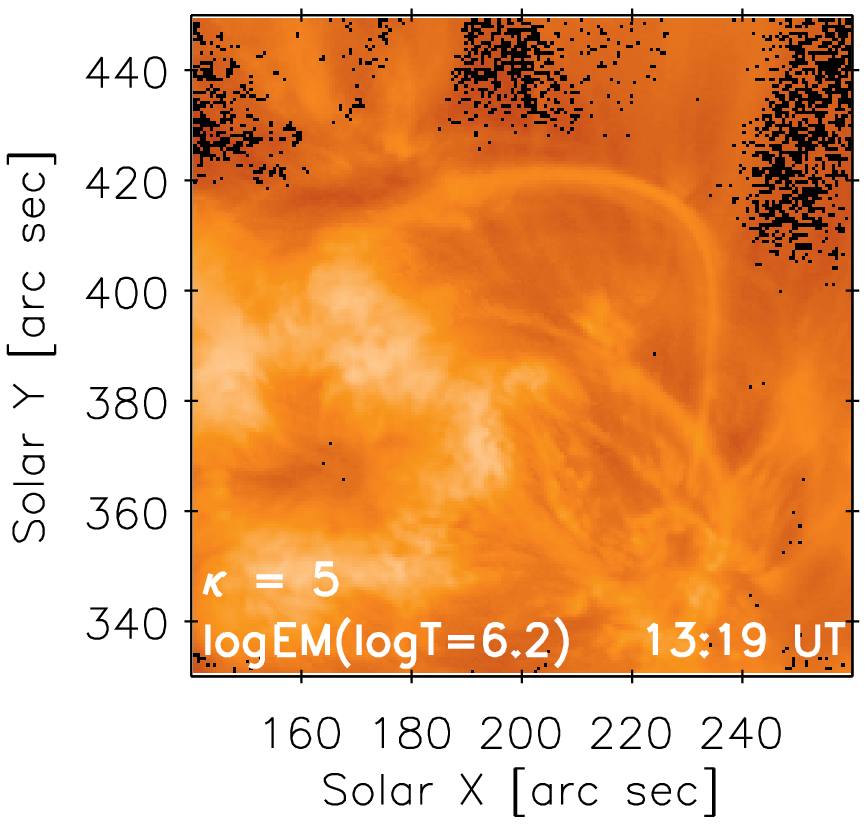}
       \includegraphics[width=3.30cm, bb=53 40 246 235, clip]{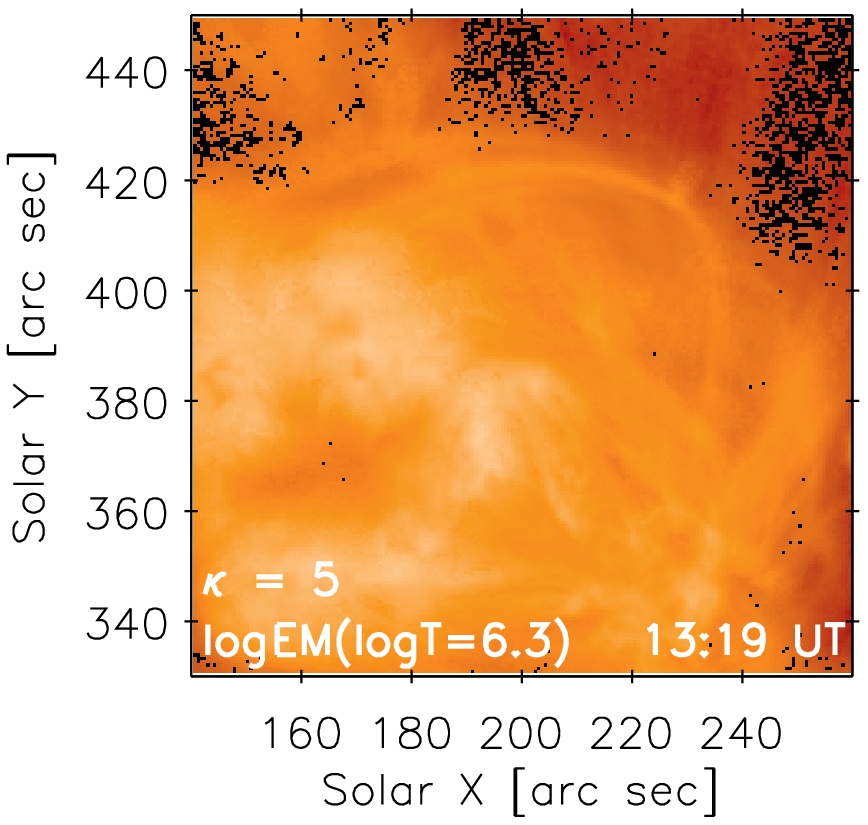}

       \includegraphics[width=4.20cm, bb=0   0 246 235, clip]{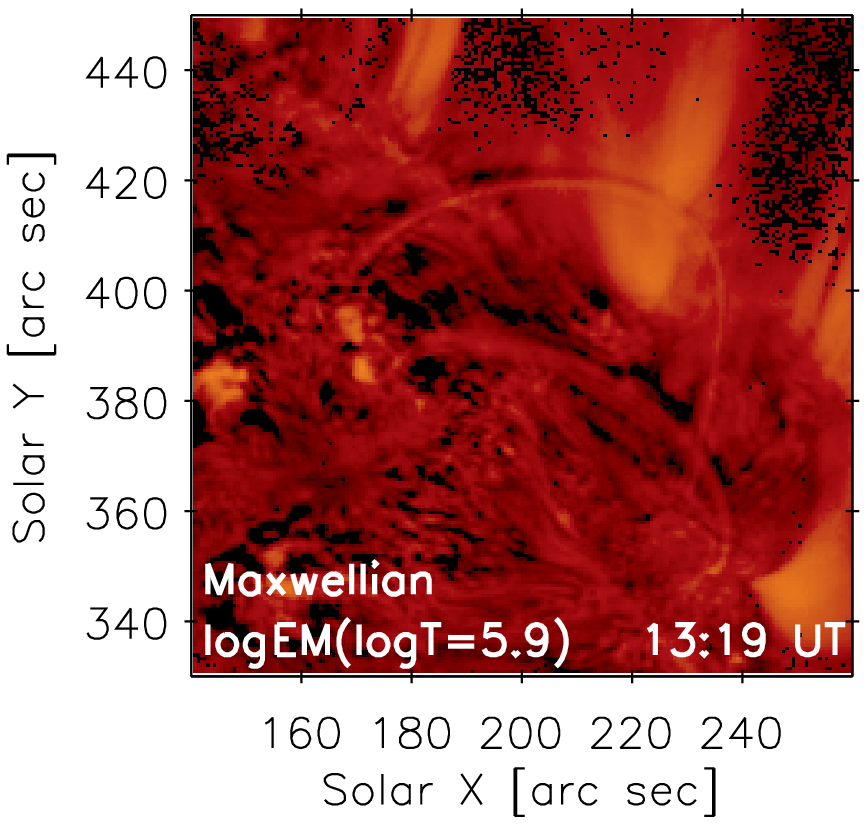}
       \includegraphics[width=3.30cm, bb=53  0 246 235, clip]{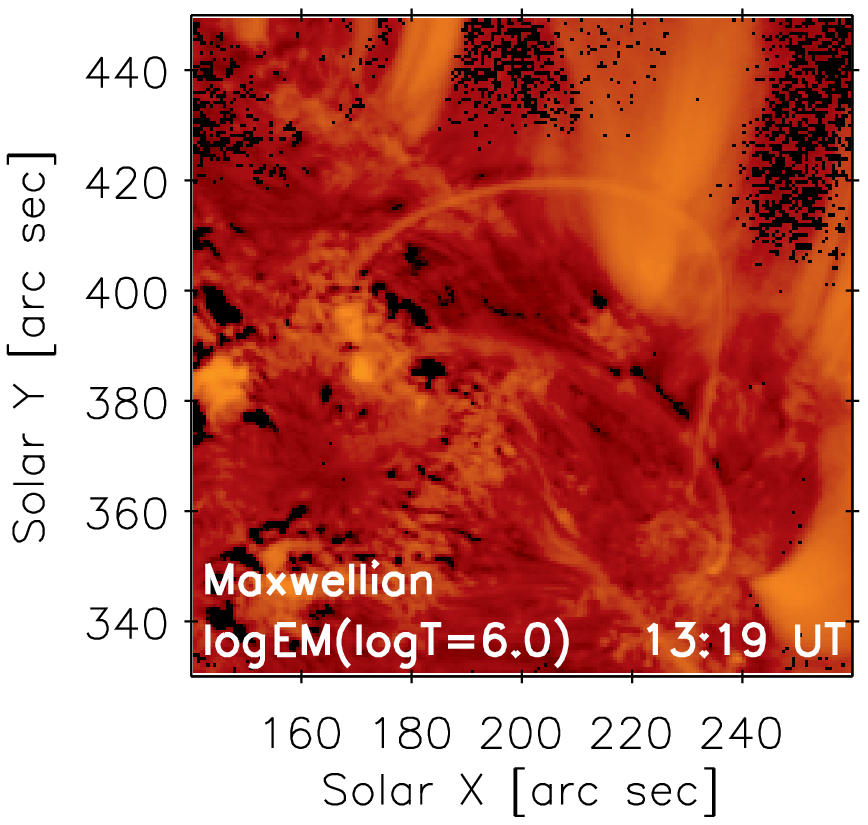}
       \includegraphics[width=3.30cm, bb=53  0 246 235, clip]{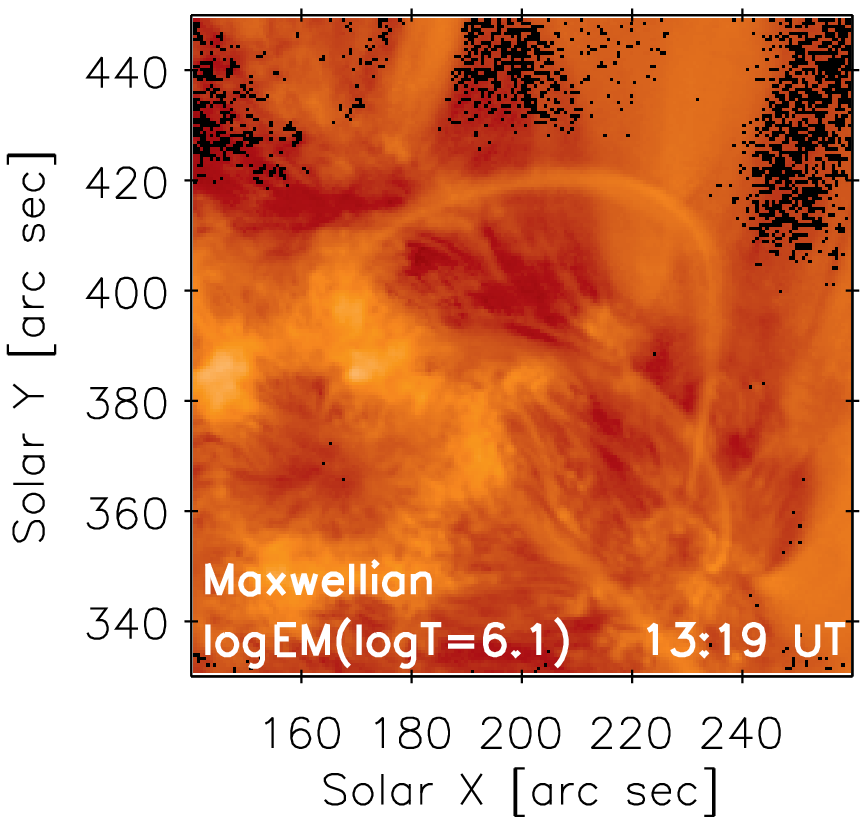}
       \includegraphics[width=3.30cm, bb=53  0 246 235, clip]{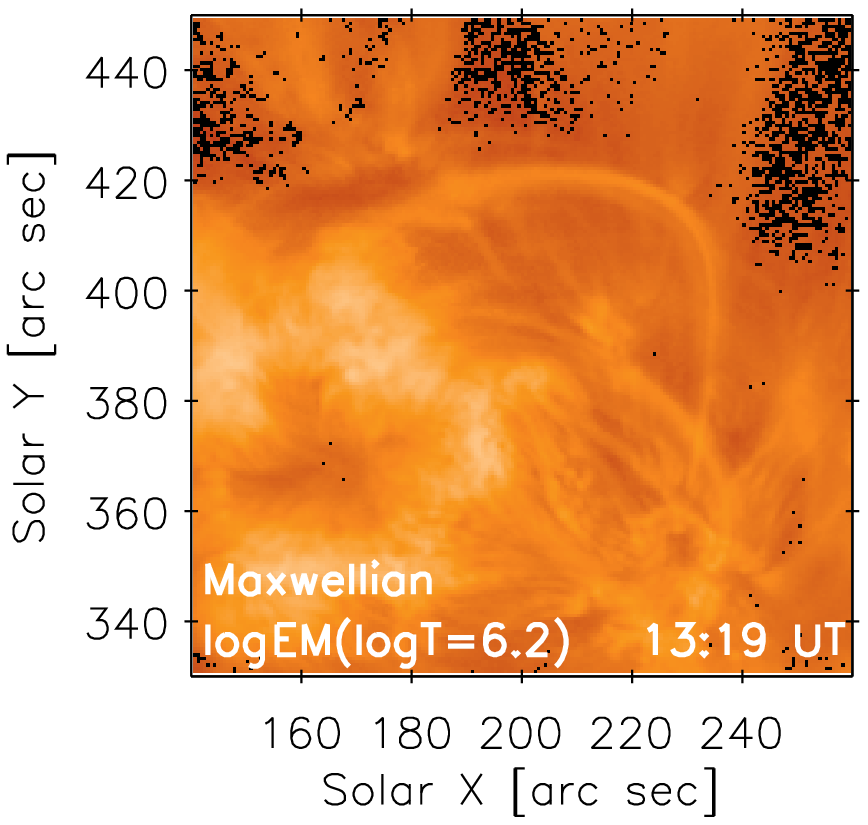}
       \includegraphics[width=3.30cm, bb=53  0 246 235, clip]{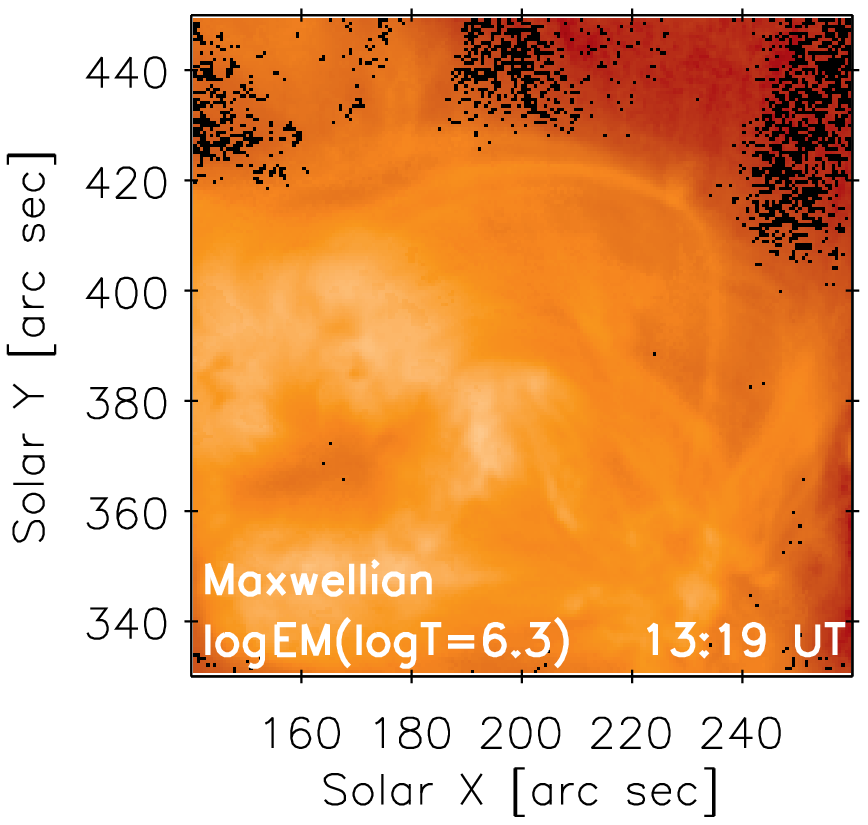}

       \caption{Emission measure EM$_\kappa(T)$ at different temperatures $T$ as a function of $\kappa$. The values of $\kappa$ and log$(T/$K) are indicated at each frame. \\ A color version of this image is available in the online journal. }
       \label{Fig:AIA_DEM_Loop}
   \end{figure*}
%
%
\section{\textit{SDO}/AIA Observations of a Transient Coronal Loop: Imaging}
\label{Sect:3}

The Atmospheric Imaging Assembly \citep[AIA,][]{Lemen12,Boerner12} on board NASA's \textit{Solar Dynamics Observatory} \citep[SDO,][]{Pesnell12} mission consists of 4 identical, normal-incidence two-channel telescopes with a diameter of 20\,cm. AIA provides multiple, near-simultaneous full-Sun images with high temporal (12\,s) and spatial resolution (1.5$\arcsec$, pixel size 0.6$\arcsec$). AIA images of the Sun are taken in 10 filters, 7 of which are centered on EUV wavelengths (94\AA,\,131\AA,\,171\AA,\,193\AA,\,211\AA,\,304\AA,\,and 335\AA), and 3 on UV or visible wavelengths (1600\AA,\,1700\AA, and 4500\AA). The EUV filters are centered on some of the strongest Fe lines in the solar EUV spectrum. However, other emission lines originating at different temperatures are present within each filter bandpass. This makes the temperature responses of the AIA EUV filters highly multi-thermal \citep[e.g.,][]{ODwyer10,DelZanna11c,Schmelz13}.

Compared to the Maxwellian distribution, the AIA responses for $\kappa$-distributions are even more multi-thermal \citep[][Fig.~8 therein]{Dzifcakova15}. This is mostly caused by the behaviour of the ionization equilibrium \citep{Dzifcakova13}, with the relative ion abundances of individual ions becoming wider and flatter with decreasing $\kappa$ (see also Fig.~\ref{Fig:Ioneq_AIA}). Peaks of the individual filter responses can also be shifted towards higher $T$, especially for low $\kappa$\,=\,2--3 \citep{Dzifcakova12,Dzifcakova15}.
Typically, in-depth understanding of the AIA observations requires differential emission measure (DEM) analysis, see, e.g., \citet{Hannah12} and \citet{Hannah13}.

In the following, we report on the AIA observations of a transient coronal loop and its spatial and temporal relation to the solar microflaring activity. DEM analysis of the AIA data is performed in Sects. \ref{Sect:3.3} and its influence on spectroscopic diagnostics is studied in Sect. \ref{Sect:4.3}.

\subsection{Relation of the Transient Loop to Microflaring Activity}
\label{Sect:3.1}

The transient loop studied here was observed in the active region (hereafter, AR) NOAA 11704. This active region was a small bipolar AR, of Hale class $\beta$/$\alpha$, characterized by two leading negative sunspots and dispersed plages of both polarities. On 2013 March 30, a compact B8.9-class microflare was observed within this AR, peaking at about 10:24\,UT (Fig.~\ref{Fig:Evolution}, \textit{top left}). A weaker B4.8 microflare was observed to peak at 13:21\,UT in a different AR NOAA 11708. Perhaps by coincidence, during the same time, we observe a transient loop within AR 11704 in the same place as the previous B8.9 microflare (Fig.~\ref{Fig:Evolution}). 

Figure \ref{Fig:Evolution} and Movie 1 (online) show the location of the B8.9 microflare and the transient loop within the AR 11704 in AIA 171\AA~channel. The 171\AA~channel is chosen since it provides the best representation of the morphology of all features. The B8.9 flare was related to a failed eruption of a long intermediate filament. One end of the filament lies within AR 11704, with the filament extending further away in the North-East direction (Fig.~\ref{Fig:Evolution}). The activation of the filament starts at about 10:00 as a series of brightenings within the filament and its immediate neighborhood, subsequently leading to apparently untwisting motions of bright threads. The filament does not erupt however, perhaps due to the overlying field seen e.g. in the AIA 193\AA~band.

The failed eruption is accompanied appearance of a flare arcade within AR 11704. The arcade first appears in the AIA 131\AA~band (Fig.~\ref{Fig:Evolution}, \textit{top right}) and subsequently in cooler AIA channels. The arcade is still visible in 171\AA~an hour later at 11:24\,UT and subsequently fades. The fading of the arcade is accompanied by the appearance of falling blobs along the arcade in 304\AA~(Movie 1) that are visible until about 12:30\,UT. After this time, evolving loops are still observed in the same location. The evolution of these loops (Movie 1) resembles slipping reconnection \citep[e.g.,][]{Aulanier06,Aulanier07,Aulanier12}, although this is ambiguous given the presence of many individual loops and other emitting structures. The loop emission at the loci of the flare arcade completely fades out of the AIA 171\AA~channel at about 12:53\,UT. However, at approximately 13:01\,UT, a transient coronal loop starts brightening up once more (Sect. \ref{Sect:3.2}), preceded again by activity in the filament (Fig.~\ref{Fig:Evolution}, \textit{bottom left}). No restructuring of the large-scale magnetic configuration of the AR 11704 is noticeable during the period studied here (09:00--15:00\,UT, Fig.~\ref{Fig:Evolution}).

%
\subsection{Multi-wavelength Evolution of the Loop}
\label{Sect:3.2}

We now focus on the evolution of the transient loop. The multi-wavelength evolution of the EUV emission is shown in Figure \ref{Fig:AIA_Loop} with a cadence of 20 minutes. The emission of the transient loop is observed in AIA 171\AA~for nearly 2 hours (see also online Movie 1). The images in Fig.~\ref{Fig:AIA_Loop} are chosen to include the time of 13:19\,UT, during which the EIS spectrometer observes a part of the transient loop (Sect. \ref{Sect:4}).

Upon careful examination, we find spatial misalignments of unknown origin between various AIA bands, especially in Solar $Y$ direction. These misalignments are corrected for manually by matching the positions of the low-lying moss emission. In this correction, the AIA 171\AA~and 193\AA~bands are taken as the reference ones, since no detecable misalignment between these two bands is found. The AIA 211\AA~is corrected by shifting it by $\Delta Y$\,=\,$-1.6$\,px (1\,px\,$\equiv$\,0.6$\arcsec$) to match the 193\AA~band, which has similar emission morphology. Subsequently, the 335\AA~band is corrected by shifting it by $\Delta Y$\,=\,$2.13$\,px to match the corrected 211\AA~band, again due to similar emission morphology. The 131\AA~band is shifted $\Delta Y$\,=\,1.5\,px to match the 171\AA~band. No correction is necessary for the 94\AA~band. After these corrections, the location of the moss emission is the same throughout the AIA bands, and the spatial correspondence of loops observed in different bands is substantially improved as well. We also note that the AIA data shown in Fig.\,\ref{Fig:AIA_Loop} are corrected for solar differential rotation using the \textit{drot\_map} routine available within Solar Soft under IDL.

Before the loop appears in AIA 171\AA, it is visible in AIA 335\AA, 211\AA, and 193\AA~(Fig.~\ref{Fig:AIA_Loop}, \textit{top row}). The transient loop belongs to the series of evolving loops already present at the same location (Movie 1). At about 13:01\,UT (Fig.~\ref{Fig:AIA_Lightcurves}), the loop is discernible as a brightening structure in all AIA EUV channels. In the vicinity of its right footpoint at X=235", Y=370", we observe heightened activity (see Movie 1), with the appearance of many bright, short, short-lived closed loops and jetting activity, likely indicating ongoing magnetic reconnection. The transient loop continues to brighten, and at 13:19\,UT, corresponding to the time of the EIS observation, the loop is already a prominent emitting structure (Fig.~\ref{Fig:AIA_Loop}, \textit{second row}). The temporal evolution of the loop emission is shown in Fig.~\ref{Fig:AIA_Lightcurves}. The \textit{top} panel of this figure shows background-subtracted lightcurves in a small black box of 4$\times$5 AIA pixels at the position of the loop. The background is represented by the green box of 4$\times$5 AIA pixels located to the right of the black box  (Fig.~\ref{Fig:AIA_Loop}, \textit{second row}). Its location is chosen so that its intensity is not contaminated by the jet emission (Fig. \ref{Fig:AIA_Lightcurves}).

The loop emission first peaks in the 335\AA~filter at about 13:20\,UT, followed by 211\AA~and 193\AA~at 13:26\,UT, and subsequently by 94\AA~at 13:27\,UT, 171\AA~at 13:29\,UT, and 131\AA~at 13:32\,UT that order. I.e., the emission first peaks in \ion{Fe}{16} and subsequently cools down to \ion{Fe}{8} observed in 131\AA. This is important, since these results show that no hot flare-like emission is present. If it were, it would be detected as \ion{Fe}{21} and \ion{Fe}{18} emission in the AIA 131\AA~and 94\AA~bands \citep[e.g.,][]{DelZanna11d,Petkaki12} prior to the maximum of the 335\AA~lightcurve. The 94\AA~emission is observed to be very weak, of the order of several DN\,s$^{-1}$\,px$^{-1}$, suggesting absence of strong, dense plasma emitting in \ion{Fe}{18}. The 131\AA~emission closely follows the 171\AA, showing that it is dominated by \ion{Fe}{8}. This conclusion is confirmed using the time-distance plots constructed as a function of Solar $X$ using a slit placed at Solar $Y$\,=\,392.7$\arcsec$, i.e., through the centres of the black and green boxes (Fig.~\ref{Fig:AIA_Loop}, \textit{second row}).

We note that the lightcurves show multiple peaks. This is most likely due to the multi-thermal nature of the AIA bands. In principle, DEM analysis can help identify the contributions to individual peaks, if the distribution function (i.e., value of $\kappa$) is known. Since $\kappa$ cannot be determined from AIA data alone (Sect. \ref{Sect:3.3}), we do not perform this analysis.


We also note that the loop evolution is accompanied by presence of multiple jets originating near the loop's right (western) footpoint (Movie 1). With increasing height, these jets diverge from the loop and thus do not contaminate the loop emission. In Fig.~\ref{Fig:AIA_Lightcurves}, \textit{bottom}, the jets appear as a series of fast-moving, short-lived brightenings near the position of the loop (black horizontal lines). The jetting activity starts at about 13:14\,UT and extends beyond the lifetime of the transient loop.

With increasing time, the loop evolves into a series of individual threads and subsequently fades away (Figs. \ref{Fig:AIA_Loop}, \ref{Fig:AIA_Lightcurves}, and Movie 1).

%
\subsection{Multithermality of the Loop as a function of $\kappa$}
\label{Sect:3.3}

Since the loop is observed in all six AIA coronal EUV bands, it is expected to be multi-thermal. We used the regularized inversion method of \citet{Hannah12,Hannah13} to obtain the DEMs as a function of $\kappa$. The results are shown in Figs. \ref{Fig:AIA_DEM} and \ref{Fig:AIA_DEM_Loop}.

The DEM$_\kappa$ reconstruction is performed for each pixel of the 1-minute averaged AIA data observed at 13:19 UT (Fig. \ref{Fig:AIA_Loop}, \textit{second row}). Before the DEM reconstruction, we remove the stray light from the AIA data using the method of \citet{Poduval13}. In the DEM reconstruction, a temperature interval of log$(T/$K)\,=\,5.5--7.4 ($\Delta$log$(T/$K)\,=\,0.1) is used. The value of $\kappa$ is assumed to be constant throughout the field of view. We note that such assumption may not be justified, as individual emitting and indeed overlapping structures could in principle have different values of $\kappa$. However, in the absence of simultaneous stereoscopic \textit{and} spectroscopic observations throughout the field of view, it is not possible to obtain the value of $\kappa$ for each emitting structure in each pixel at a given time; an assumption is therefore necessary. By adopting a single value of $\kappa$ for the entire field of view, we demonstrate the feasibility of the regularized inversion method to recover the non-Maxwellian DEMs from AIA observations using the corresponding responses for the $\kappa$-distributions \citep{Dzifcakova15}, which (to our knowledge) has not been done before.

The background-subtracted DEMs averaged over the small box along the loop (Fig.~\ref{Fig:AIA_Loop}) are shown in red in Fig.~\ref{Fig:AIA_DEM}. DEMs corresponding to the black box are shown in black, while the background DEMs corresponding to the green box in Fig.~\ref{Fig:AIA_Loop} are shown in green. The background-subtracted DEMs peak at log$(T/$K)\,=\,6.2 for the Maxwellian distribution and $\kappa$\,=\,5. The peak of the DEM is shifted to higher $T$ for $\kappa$\,=\,2--3, being at log$(T/$K)\,=\,6.3 and 6.4 for $\kappa$\,=\,3 and 2, respectively. This behaviour of the DEMs is mainly the result of the shifts in the ionization equilibrium to higher $T$, see Fig.~\ref{Fig:Ioneq_AIA} and \citet{Mackovjak14}. The DEMs are multi-thermal, with significant amount of emission originating at temperatures lower than the peak of the DEMs, down to about log$(T/$K)\,=\,5.9--6.1.

The pixel-by-pixel reconstructed EM$_\kappa(T)$ are shown in Fig.~\ref{Fig:AIA_DEM_Loop} for the five temperature bins where the background-subtracted DEM$_{\kappa}(T)$ is the highest (Fig. \ref{Fig:AIA_DEM})  Note that the width of the temperature bin is $\Delta$log$(T/$K)\,=\,0.1. We see that the loop is present in all of these temperature bins. At lower temperatures, it is relatively well-defined, with the EM$_\kappa(T)$ becoming more fuzzy with increasing $T$ for all $\kappa$. This behaviour comes from the emission morphology in the progressively hotter AIA bands.

The DEMs for the Maxwellian distribution and $\kappa$\,=\,5 contain a spurious high-temperature peak at log$(T/$K)\,$\geqq$\,7.0. This peak is about a factor of $\approx$30 weaker than the main one and was reported for on-disk AIA observations also by \citet[][Sect. 3 therein]{Dudik14c}. This peak is present also in the background DEMs, which may suggest that it is an artifact of the method. The peak gets progressively suppressed with decreasing $\kappa$, until it is absent from the background-subtracted DEMs for $\kappa$\,=\,2.

We note that in principle, a combination of AIA filter ratios permits diagnostics of $\kappa$, but only for isothermal or near-isothermal structures \citep{Dudik09,Dzifcakova12}. This is because such color-color diagram are not monotonic, but contain complicated curves. Since the DEMs obtained are significantly multi-thermal for all $\kappa$ studied, it is not possible to use the combinations of AIA filter ratios to diagnose $\kappa$. Instead, the diagnostic of $\kappa$ has to be performed using ratios of individual spectral lines that produce monotonic ratio-ratio diagrams (see Sect. 4.3).

%
\begin{table*}[!ht]
\begin{center}
\caption{Background-subtracted EIS line intensities averaged over the pixels listed. See text for details of the background subtraction. The standard deviations $\sigma_{10\%}$ and $\sigma_{20\%}$ are obtained by considering a 10\% and 20\% calibration uncertainty, respectively. Selfblends are indicated, with very weak transitions in parentheses.
\label{Table:1}}
\begin{tabular}{cllccccccc}
\tableline
\tableline
		&		  &				& \multicolumn{3}{c}{Loop (288:317)}	&  \multicolumn{3}{c}{Loop (300:309)}  \\
Ion		& $\lambda$ [\AA] & selfblending transitions [\AA]& \hspace{0.5cm} $I$ & $\sigma_{10\%}(I)$	& $\sigma_{20\%}(I)$ &
								  \hspace{0.5cm} $I$ & $\sigma_{10\%}(I)$	& $\sigma_{20\%}(I)$ \\
\tableline
\ion{Fe}{11}	& 182.167& --					& \hspace{0.5cm}  795&	 99& 169& \hspace{0.5cm}  934& 117& 199 \\
\ion{Fe}{11}	& 188.216& --					& \hspace{0.5cm} 1638&	172& 332& \hspace{0.5cm} 1947& 204& 394 \\
\ion{Fe}{11}	& 257.554& 257.538, 257.547, 257.558		& \hspace{0.5cm}  398&	 45&  82& \hspace{0.5cm}  414&  47&  86 \\
\ion{Fe}{11}	& 257.772& 257.725				& \hspace{0.5cm}  178&	 23&  38& \hspace{0.5cm}  234&  30&  50 \\
\ion{Fe}{12}	& 186.887& 186.854, (186.931)			& \hspace{0.5cm} 1406&	145& 283& \hspace{0.5cm} 1498& 154& 302 \\
\ion{Fe}{12}	& 195.119& 195.179, (195.078), (195.221)	& \hspace{0.5cm} 2256&	228& 453& \hspace{0.5cm} 2506& 254& 503 \\
\ion{Fe}{13}	& 196.525& --			 		& \hspace{0.5cm}  261&	 27&  53& \hspace{0.5cm}  223&  23&  45 \\
\ion{Fe}{13}	& 202.044& --					& \hspace{0.5cm} 1346&	153& 279& \hspace{0.5cm} 1779& 202& 368 \\
\ion{Fe}{13}	& 203.826& 203.772, 203.795, 203.835		& \hspace{0.5cm} 2591&	270& 524& \hspace{0.5cm} 2532& 264& 512 \\
\tableline
\tableline
\end{tabular}
\end{center}
\end{table*}
%
   \begin{figure}[!t]
	\centering
	\includegraphics[width=4.00cm, clip]{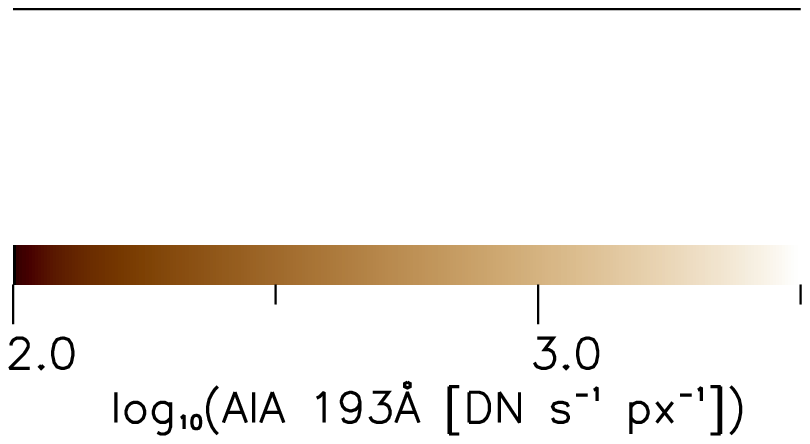}
	\includegraphics[width=4.00cm, clip]{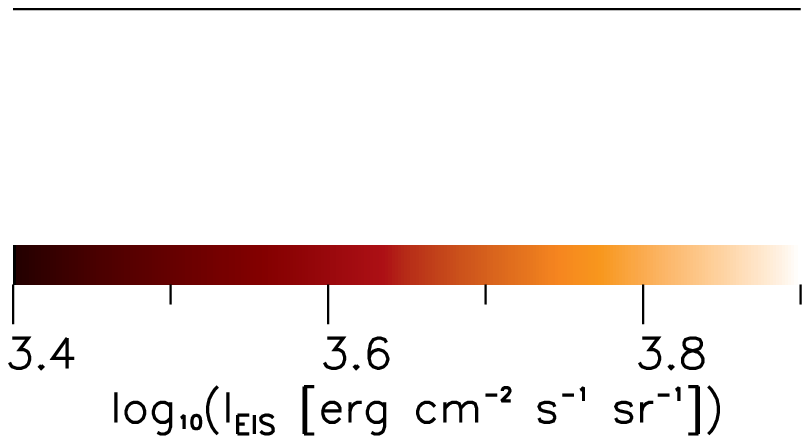}

	\includegraphics[width=4.00cm, clip]{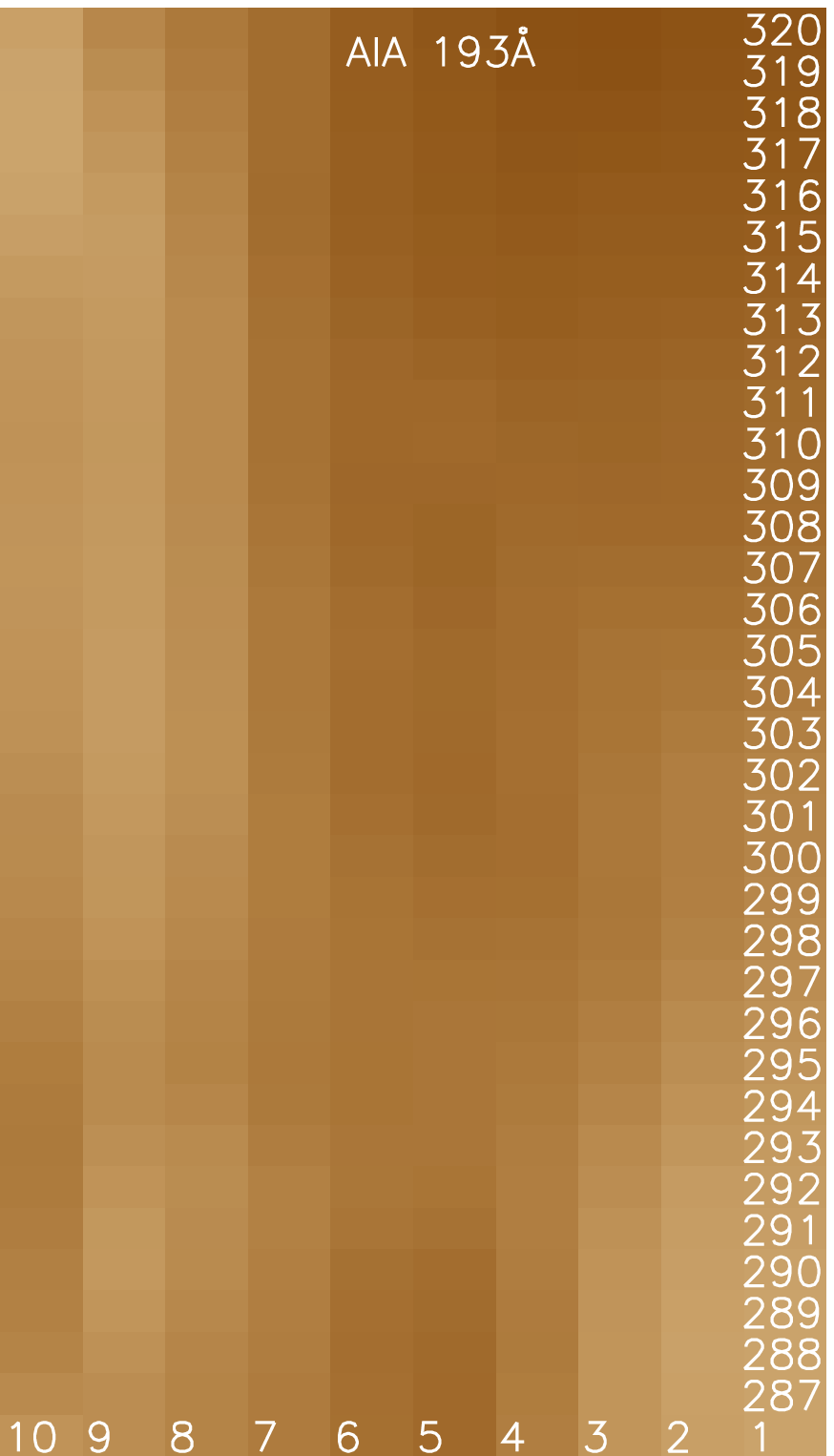}
	\includegraphics[width=4.00cm, clip]{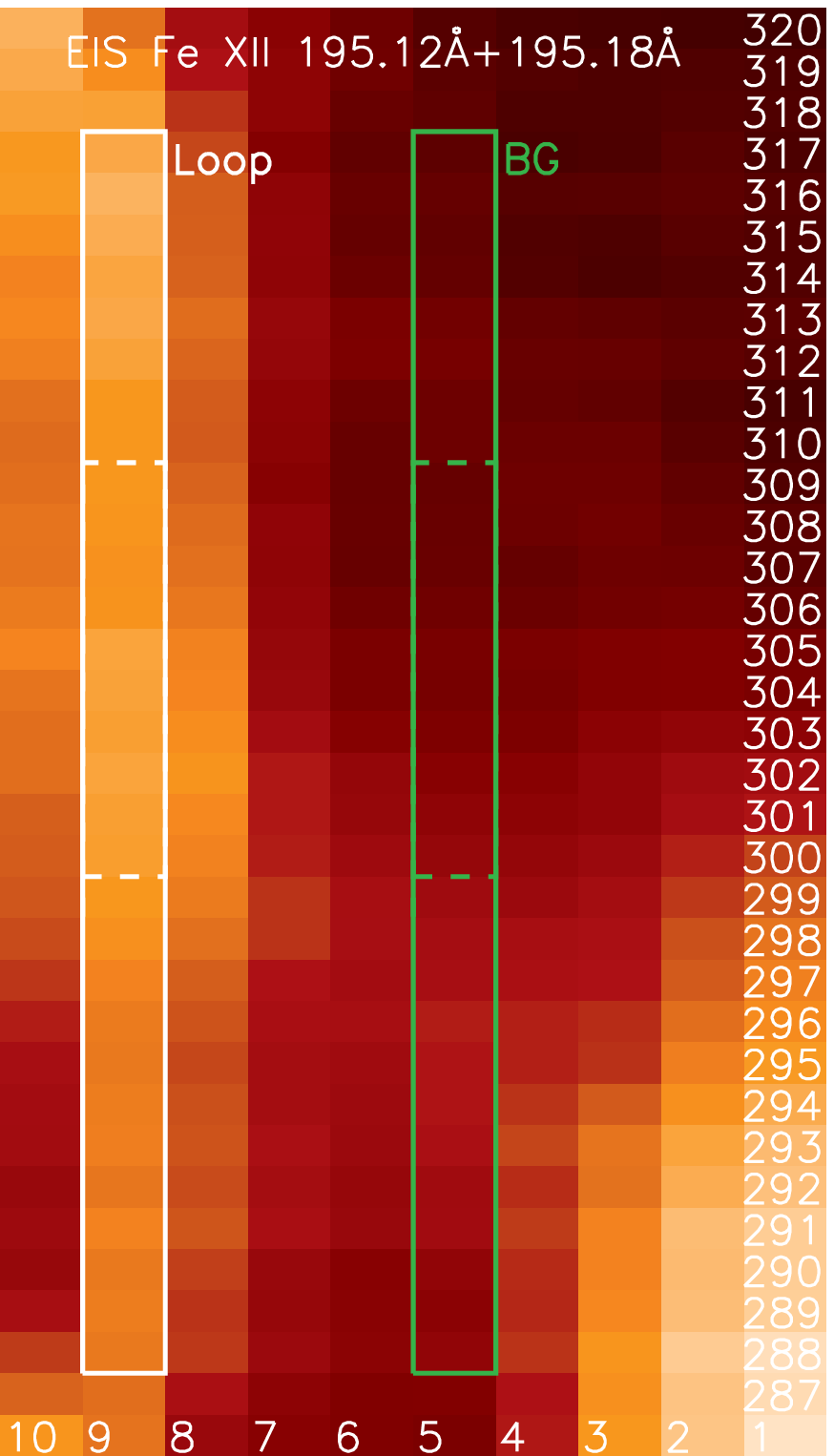}
	\caption{A portion of the EIS raster selected for analysis, corresponding to the field of view shown by white box in Fig.~\ref{Fig:AIA_Loop}. Numbers give the pixel coordinates in the raster. \textit{Left}: AIA 193\AA~pseudo-raster corresponding to the EIS FOV. \textit{Right}: EIS intensities of the \ion{Fe}{12} 195.12\AA\,+195.18\AA~selfblend. Pixel size is 2$\arcsec$\,$\times$\,1$\arcsec$. 
	      \\ A color version of this image is available in the online journal.}
       \label{Fig:EIS_raster}
   \end{figure}
%
%
%
%
   \begin{figure*}[!t]
	\centering
	\includegraphics[width=8.4cm,clip]{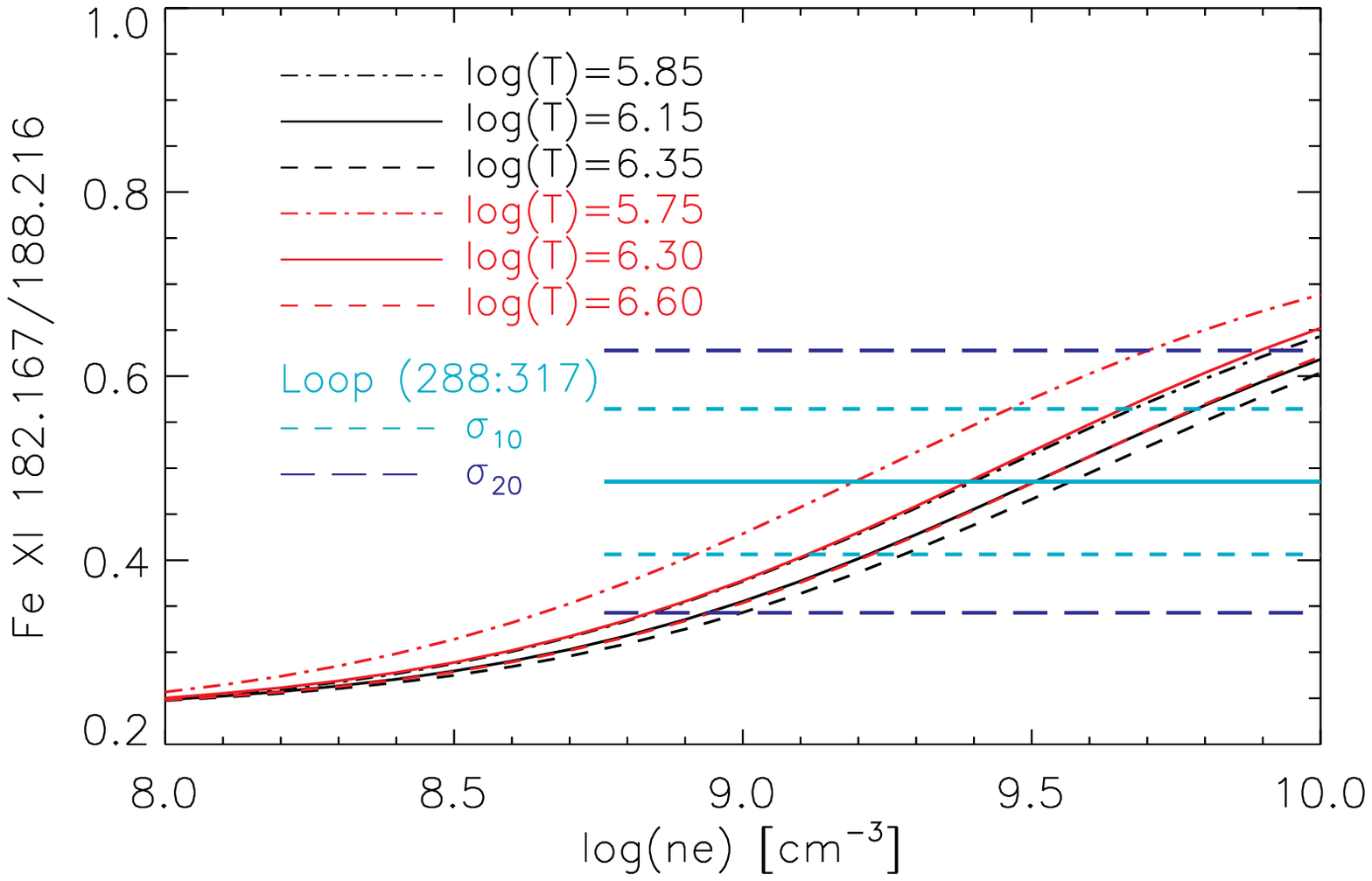}
	\includegraphics[width=8.4cm,clip]{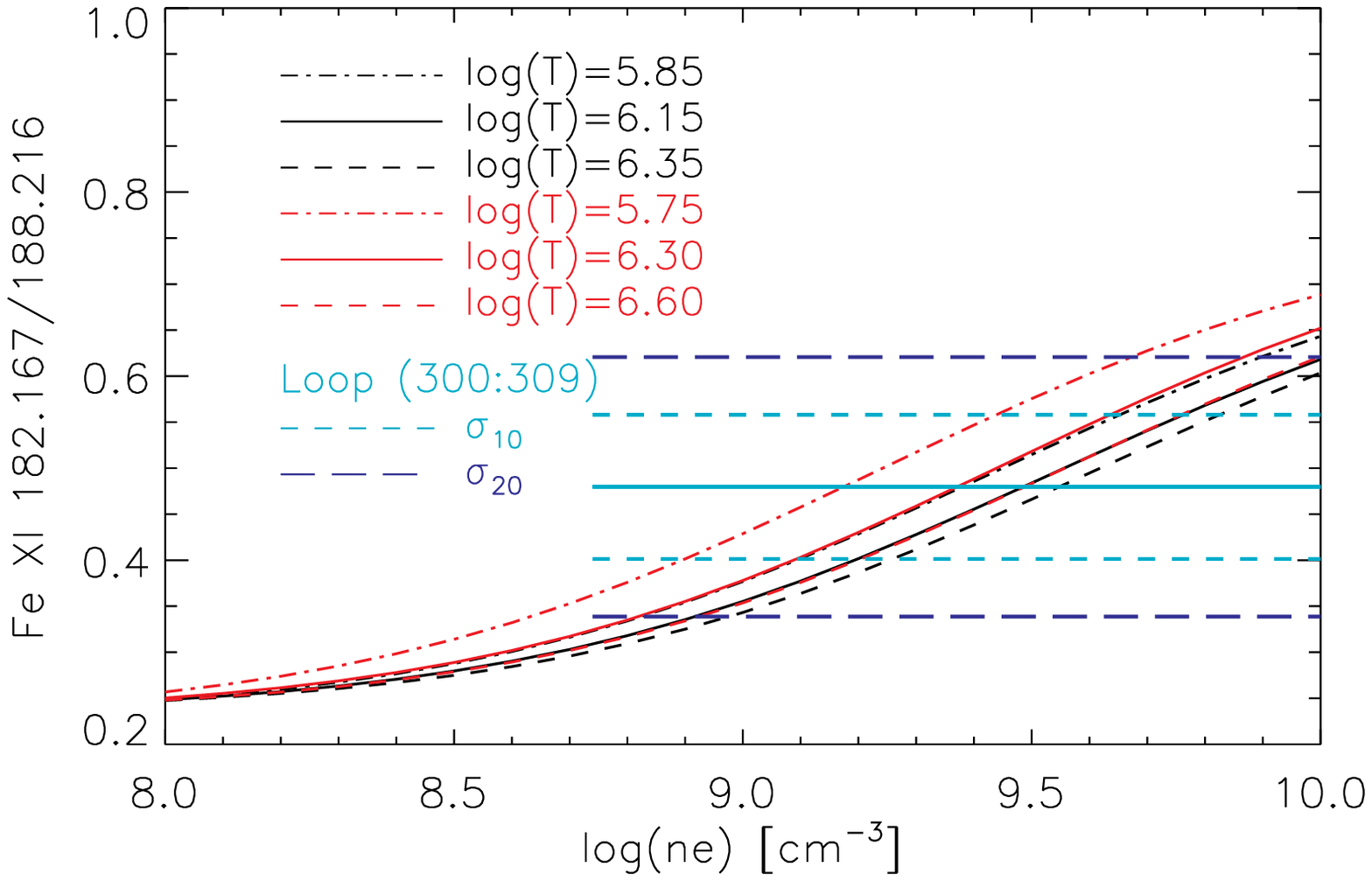}
	\includegraphics[width=8.4cm,clip]{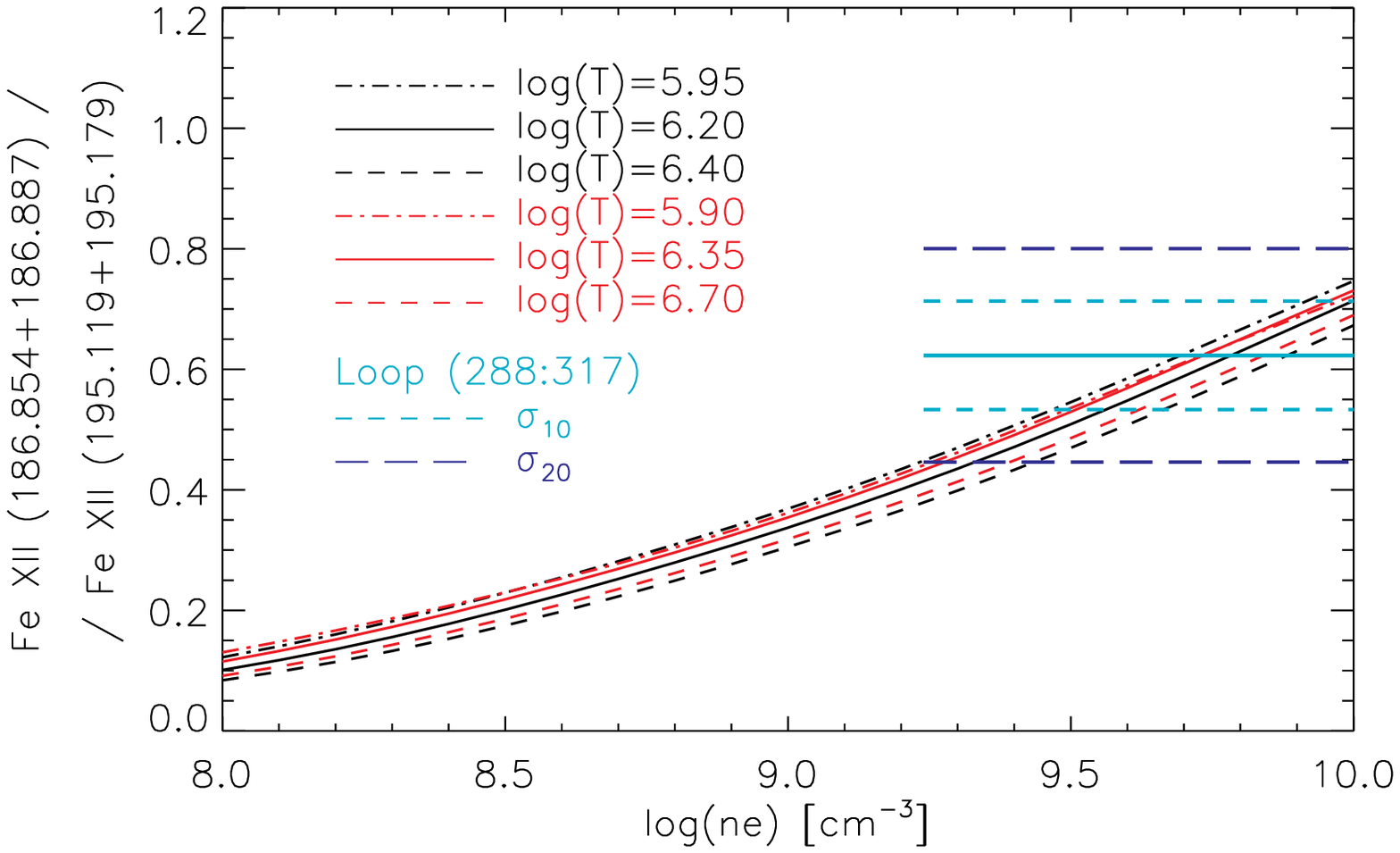}
	\includegraphics[width=8.4cm,clip]{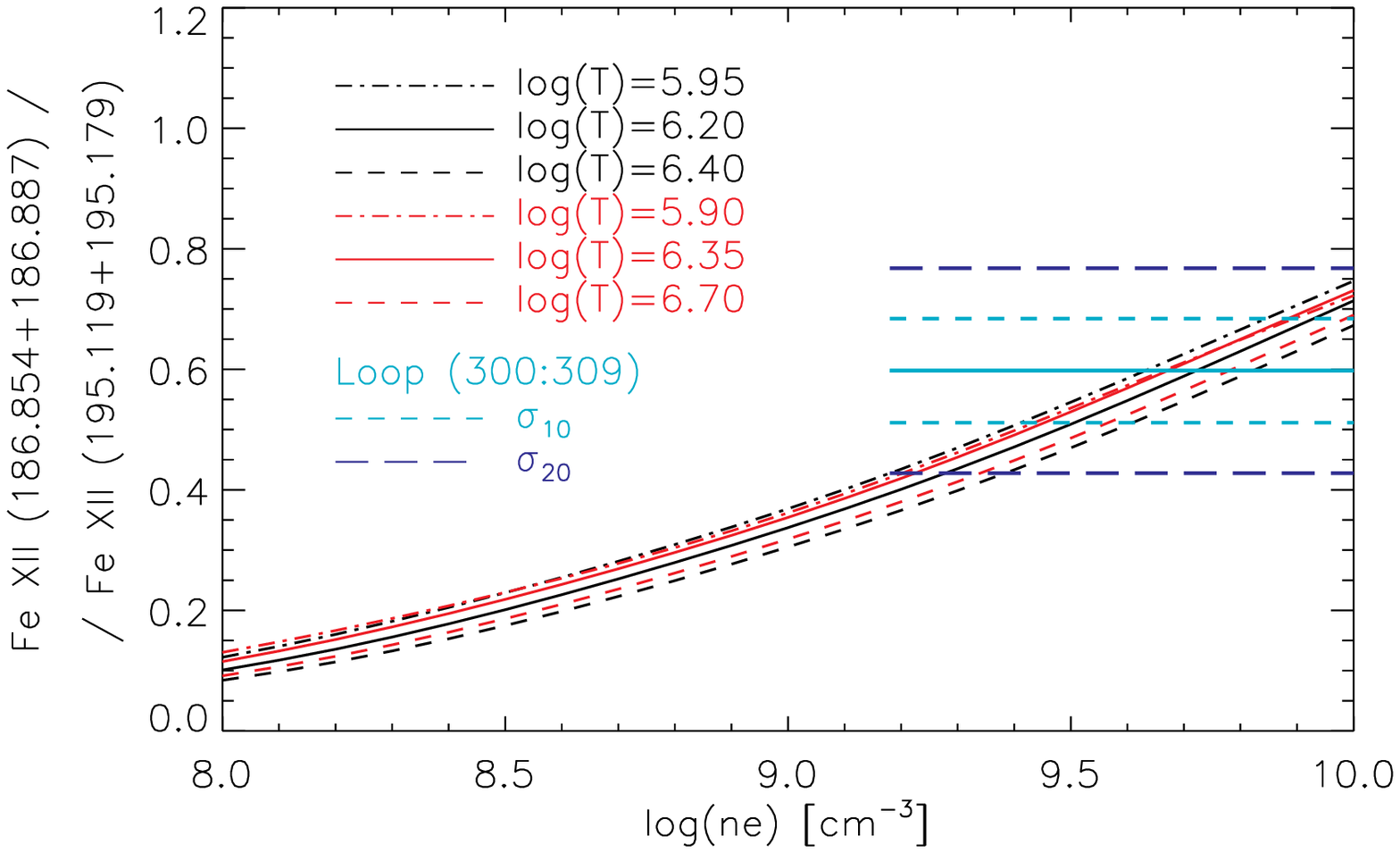}
	\includegraphics[width=8.4cm,clip]{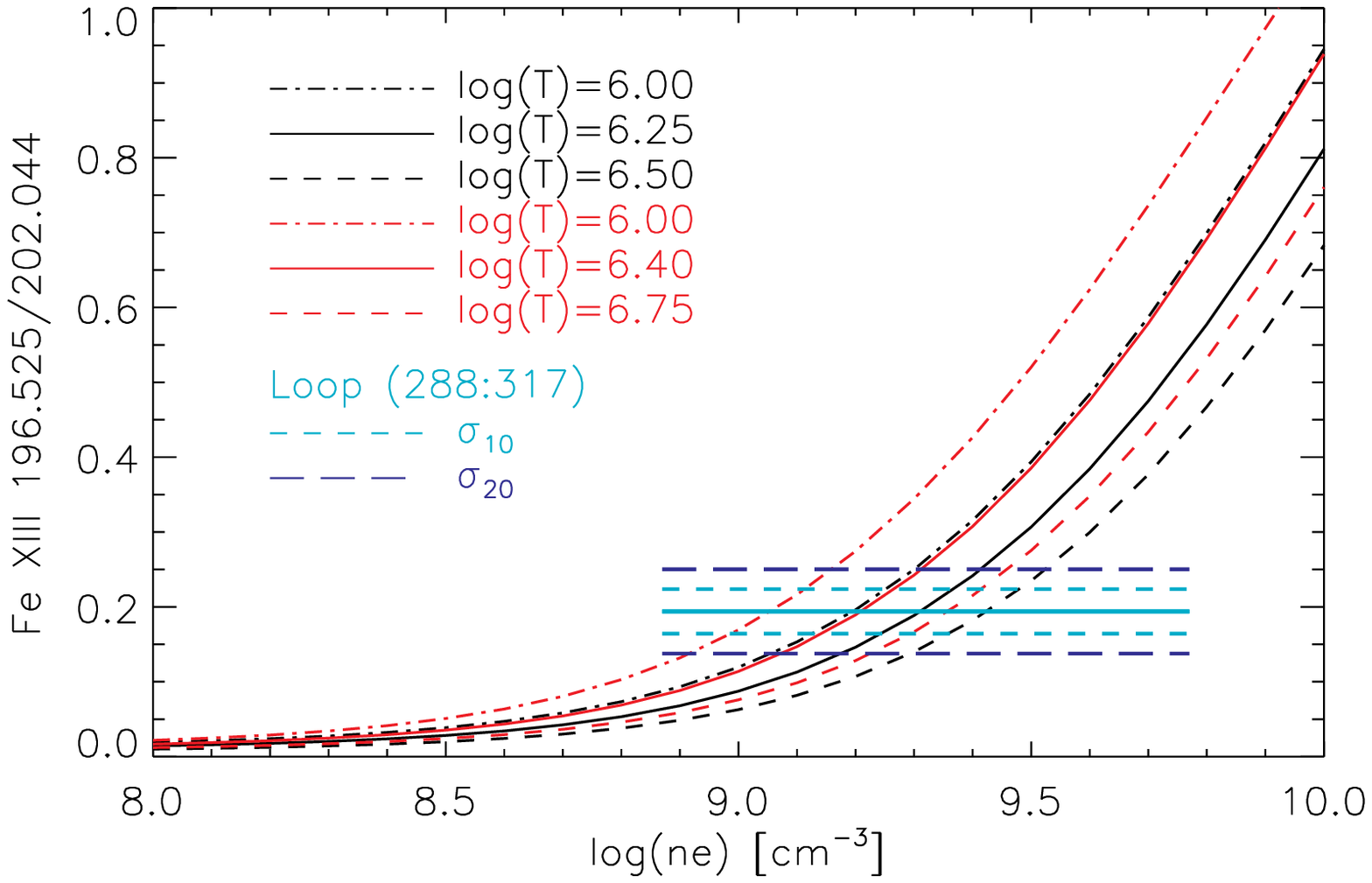}
	\includegraphics[width=8.4cm,clip]{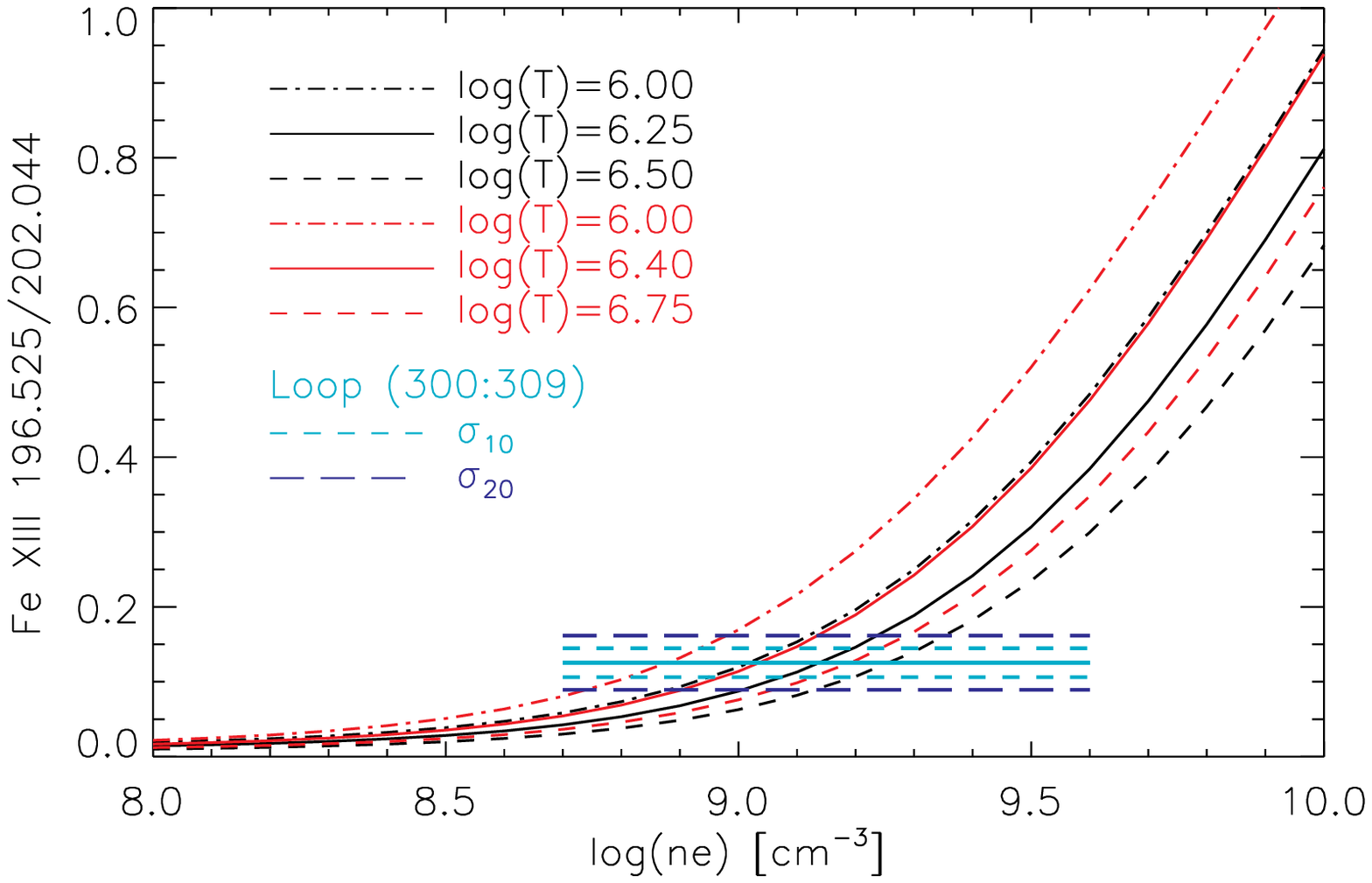}
	\includegraphics[width=8.4cm,clip]{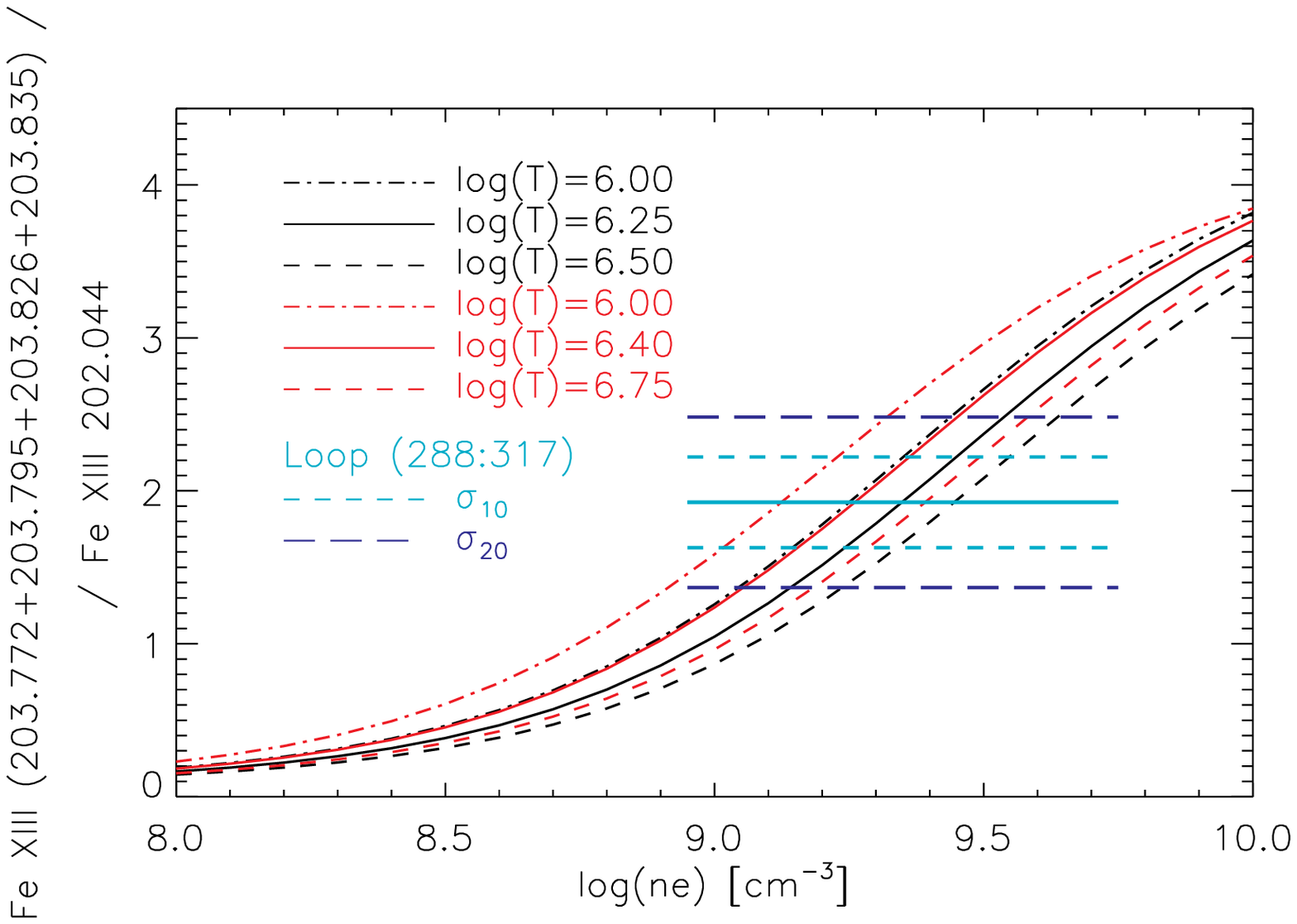}
	\includegraphics[width=8.4cm,clip]{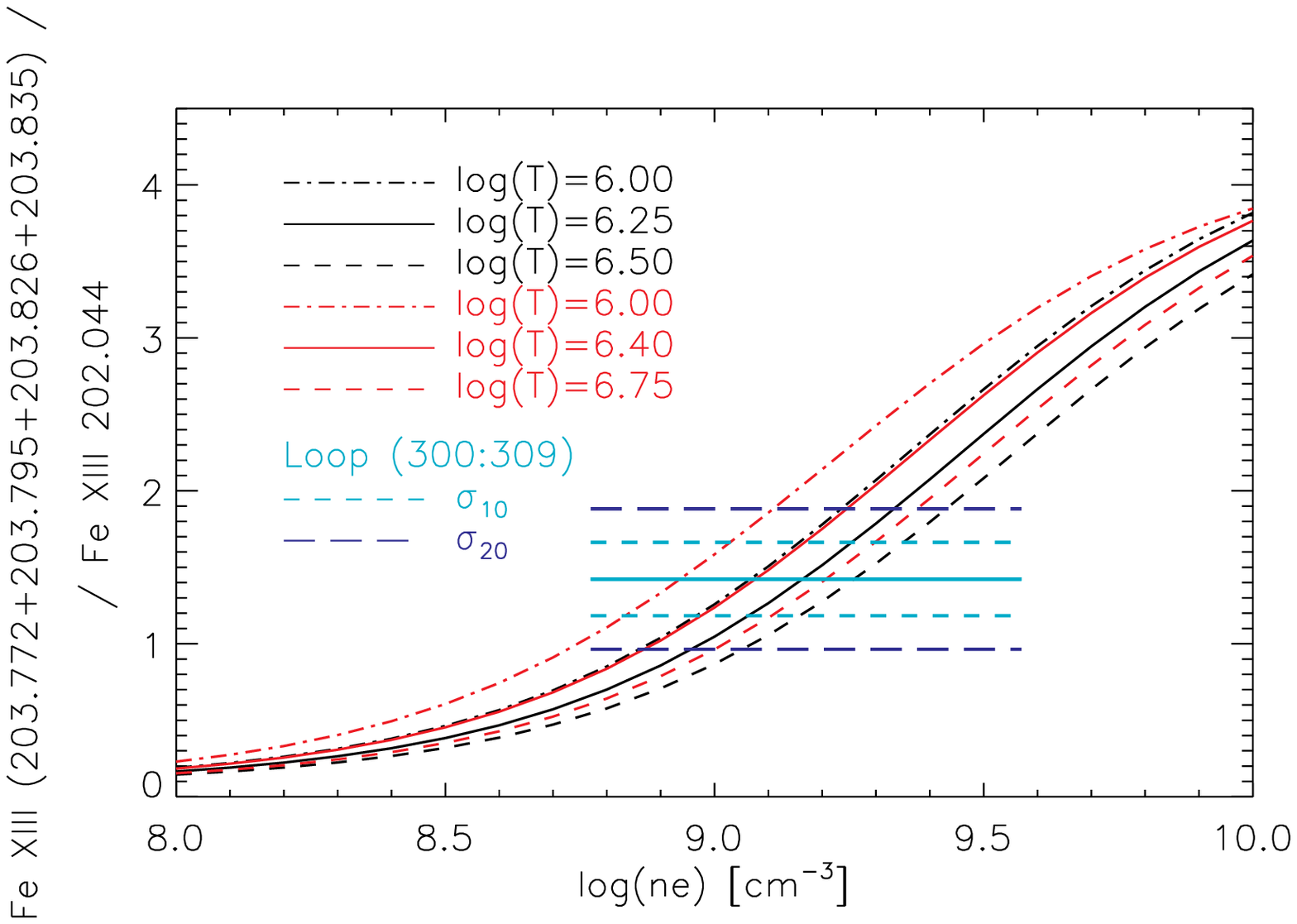}
	\caption{Diagnostics of density using \ion{Fe}{11}, \ion{Fe}{12}, and \ion{Fe}{13} ratios. Black color corresponds to the Maxwellian distribution; red stands for $\kappa$\,=\,2. Different linestyles correspond to different $T$ for each distribution (see \citet{Dudik14b}, Figs. 5--7 therein): Full lines stand for the temperature corresponding to the peak of the ion abundance, dashed and dot-dashed lines for the temperatures at which the relative ion abundance is 1\% of its maximum. Observed value of the ratio is plotted as the full cyan line. Uncertainties including calibration uncertainties are plotted for illustration. \\ A color version of this image is available in the online journal.}
        \label{Fig:Diag_ne}
   \end{figure*}
%
%
   \begin{figure*}[!ht]
	\centering
	\includegraphics[width=7.4cm,clip]{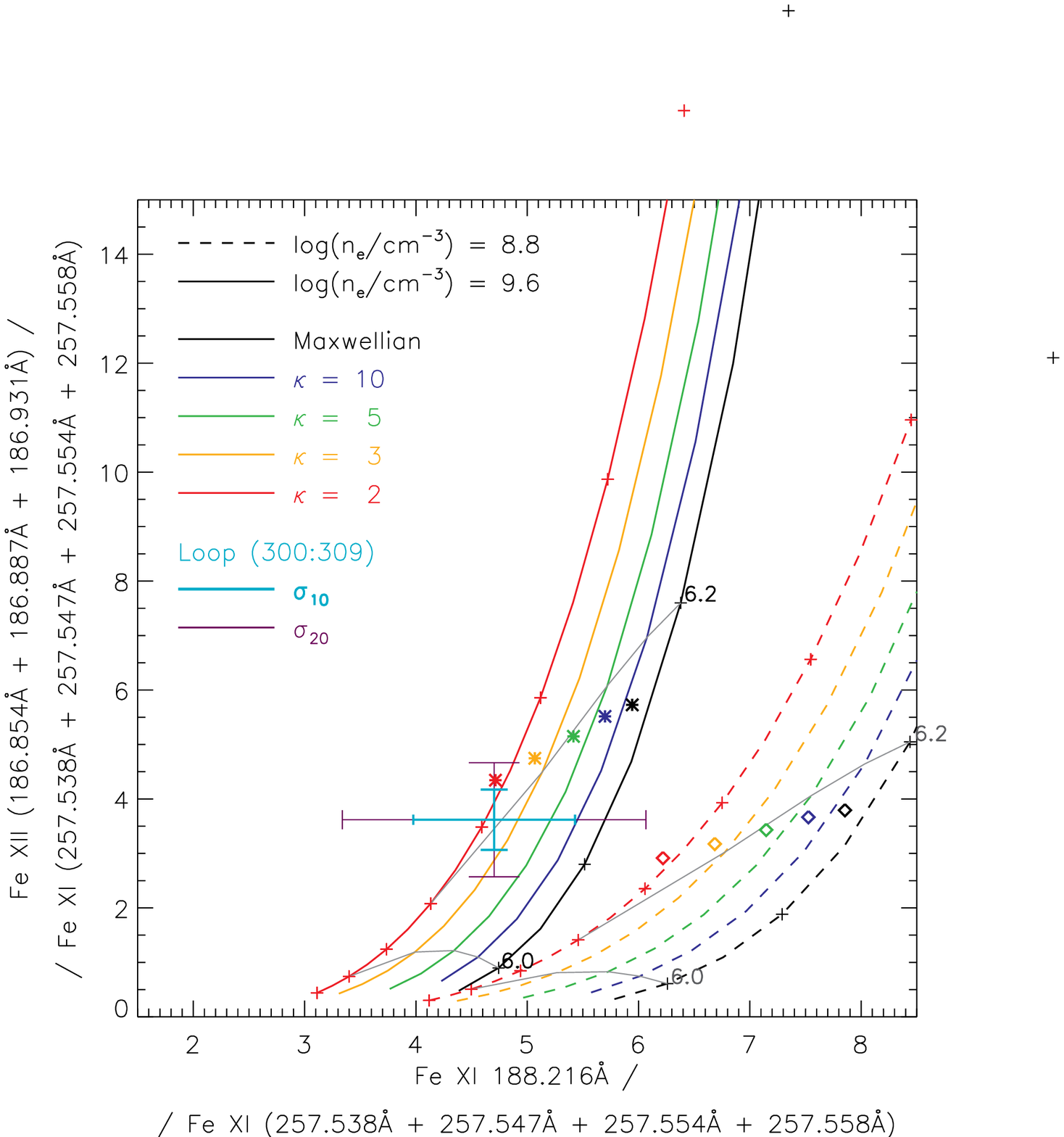}
	\includegraphics[width=7.4cm,clip]{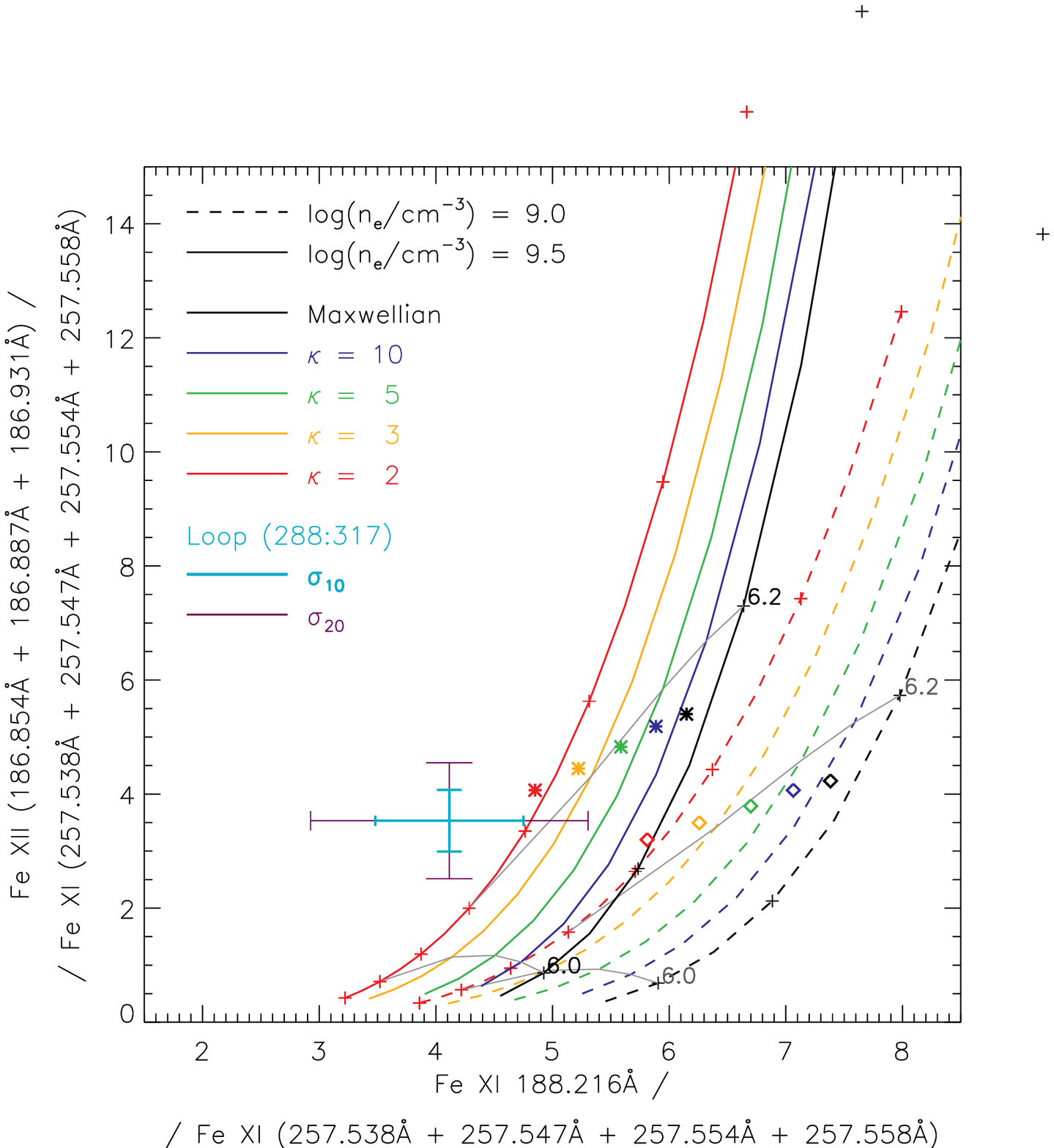}
	\includegraphics[width=7.4cm,clip]{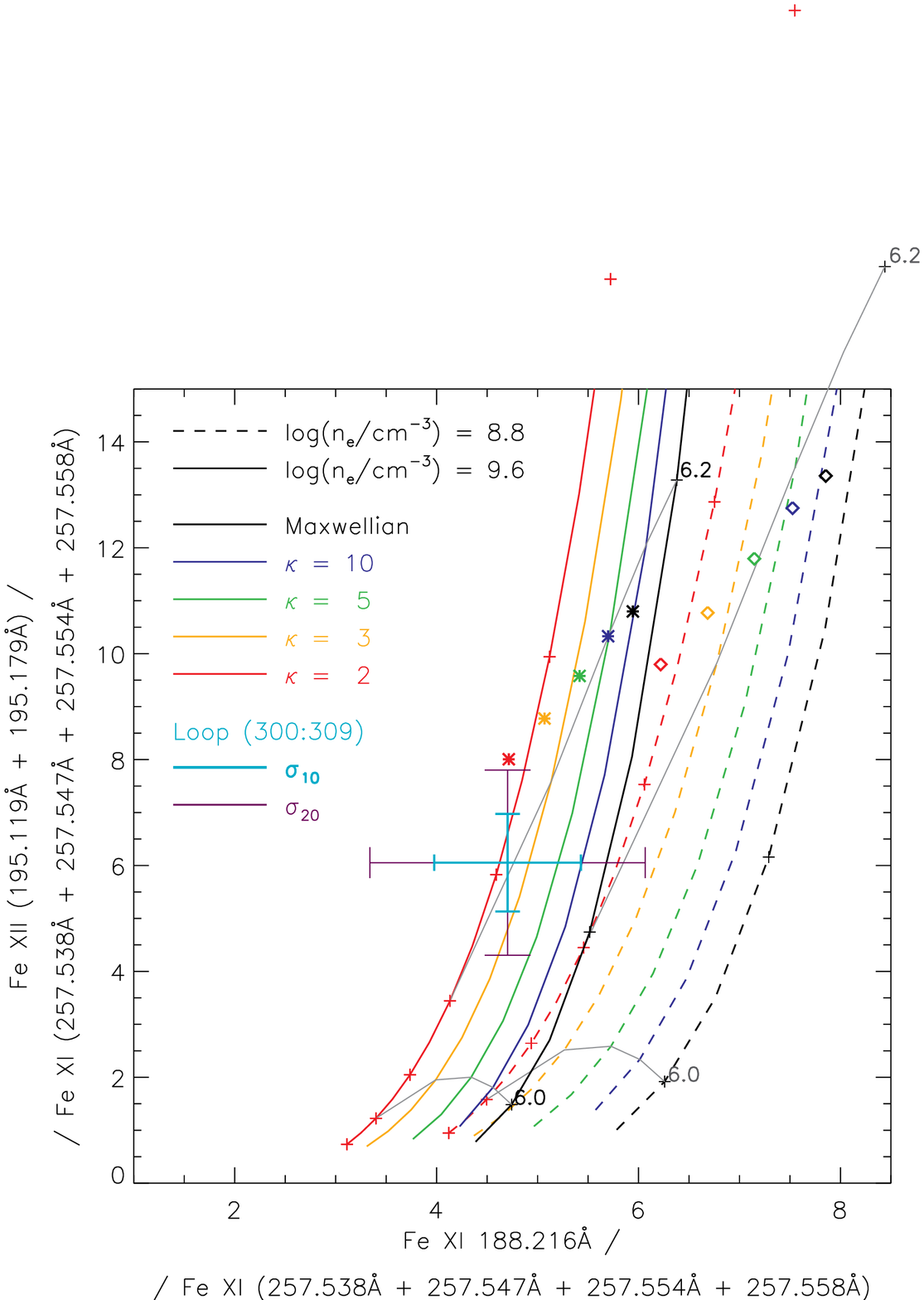}
	\includegraphics[width=7.4cm,clip]{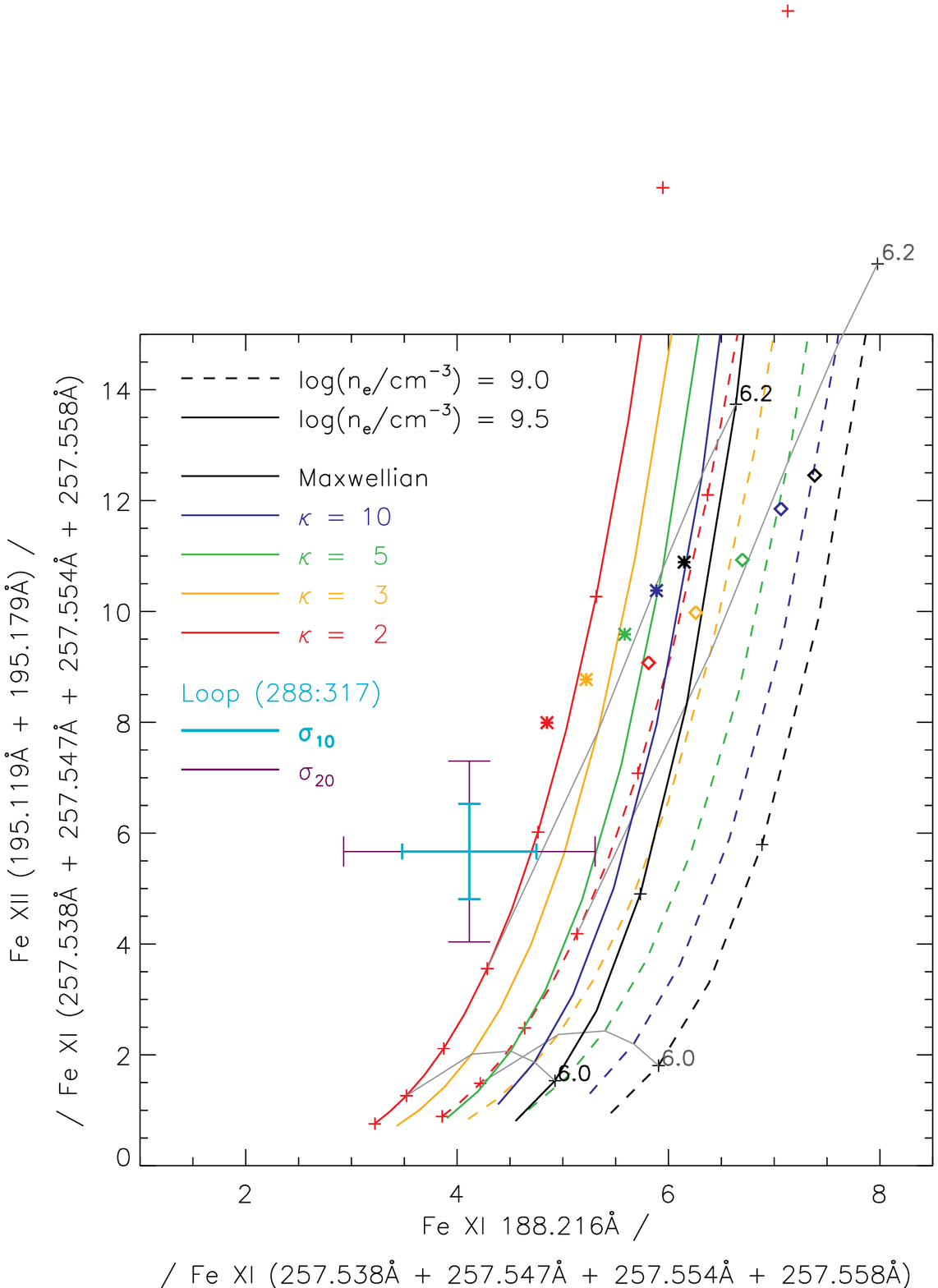}
	\includegraphics[width=7.4cm,clip]{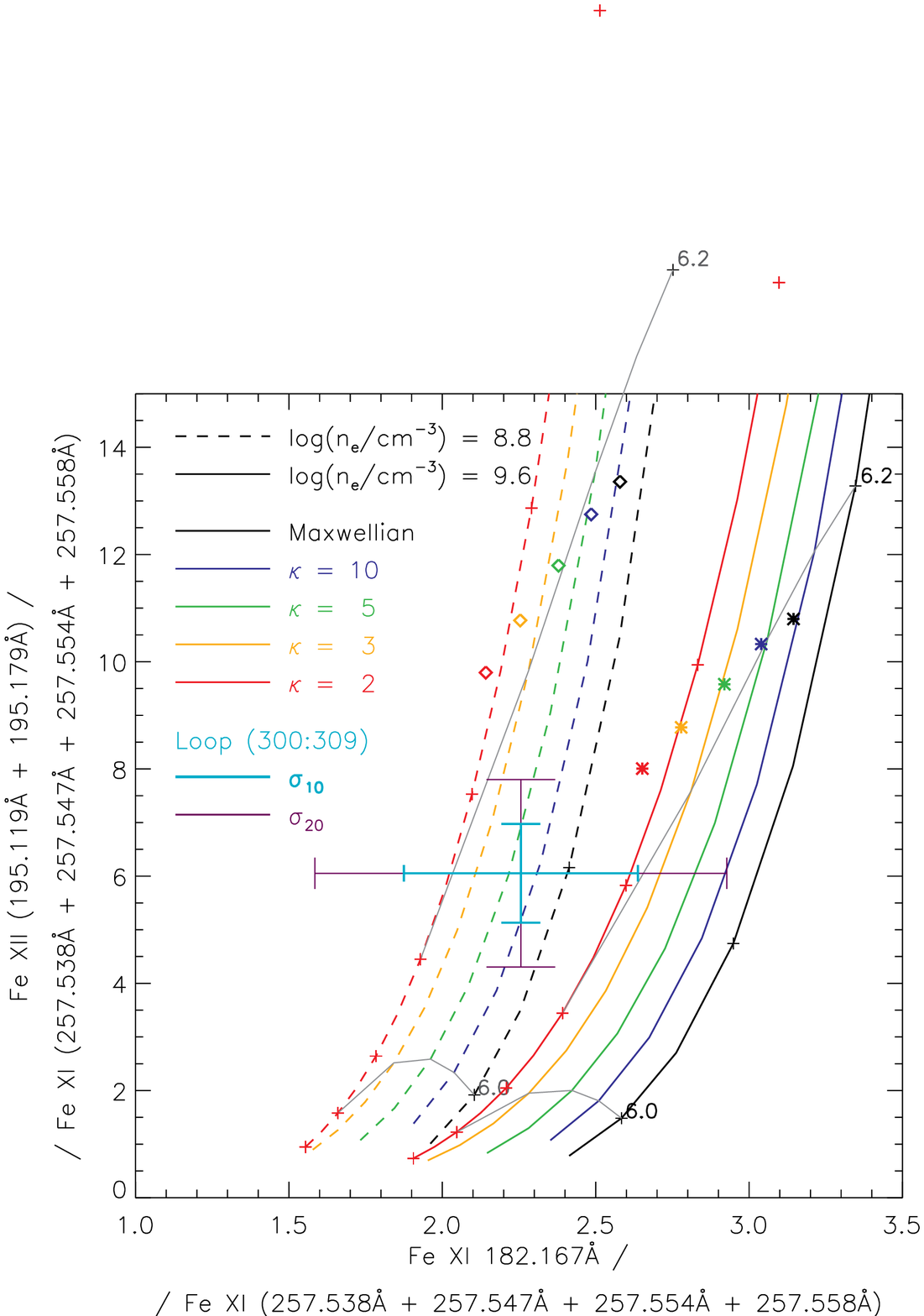}
	\includegraphics[width=7.4cm,clip]{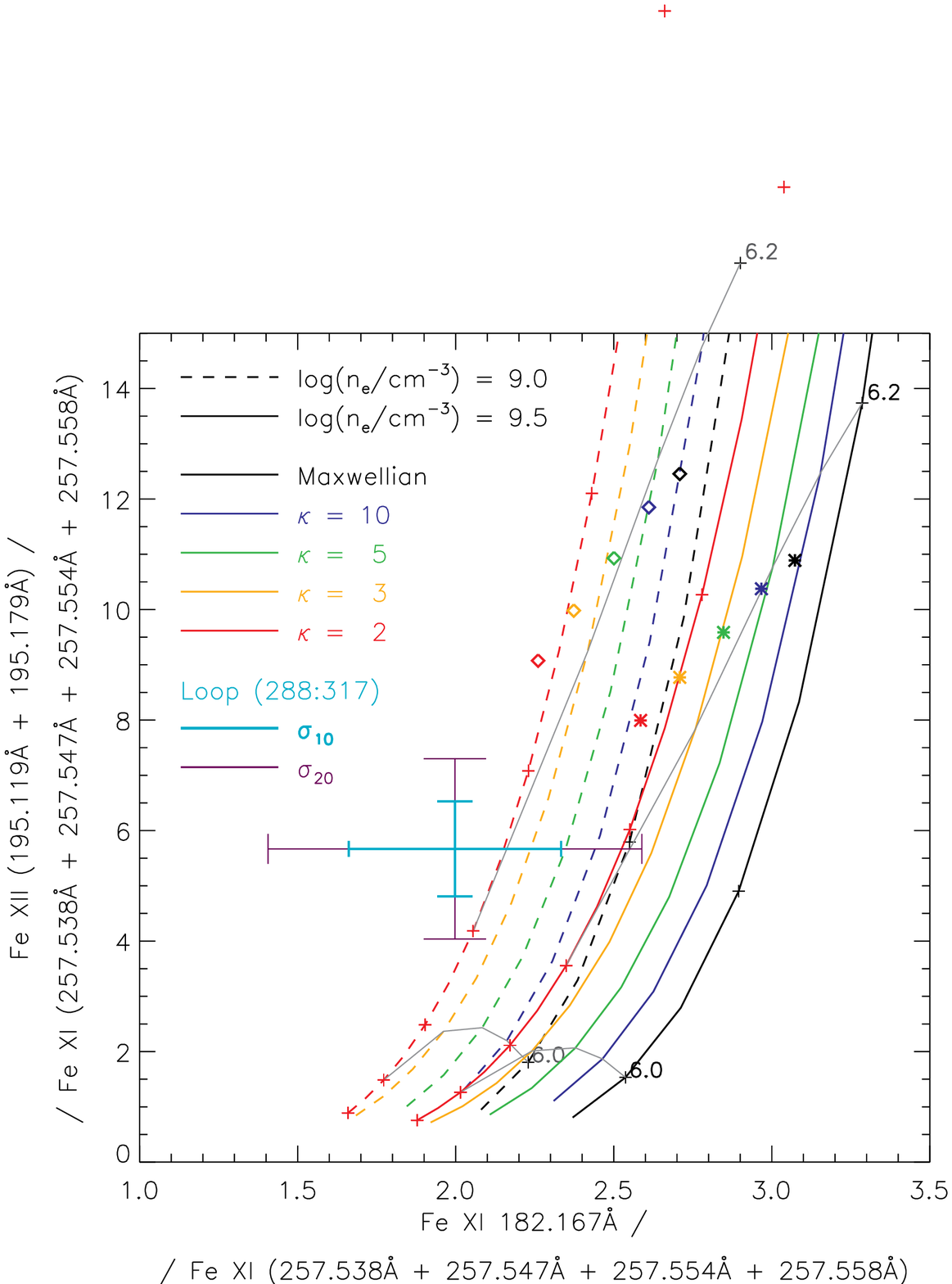}
	\caption{Diagnostics of $\kappa$ involving the \ion{Fe}{11} 257.554\AA~selfblend. Different colors correspond to different $\kappa$, linestyles denote electron density. Diamonds and asterisks denote the predicted ratios based on DEM analysis. \\ A color version of this image is available in the online journal.}
        \label{Fig:Diag_TK1}
   \end{figure*}
%
   \begin{figure*}[!h]
	\centering
	\includegraphics[width=7.4cm,clip]{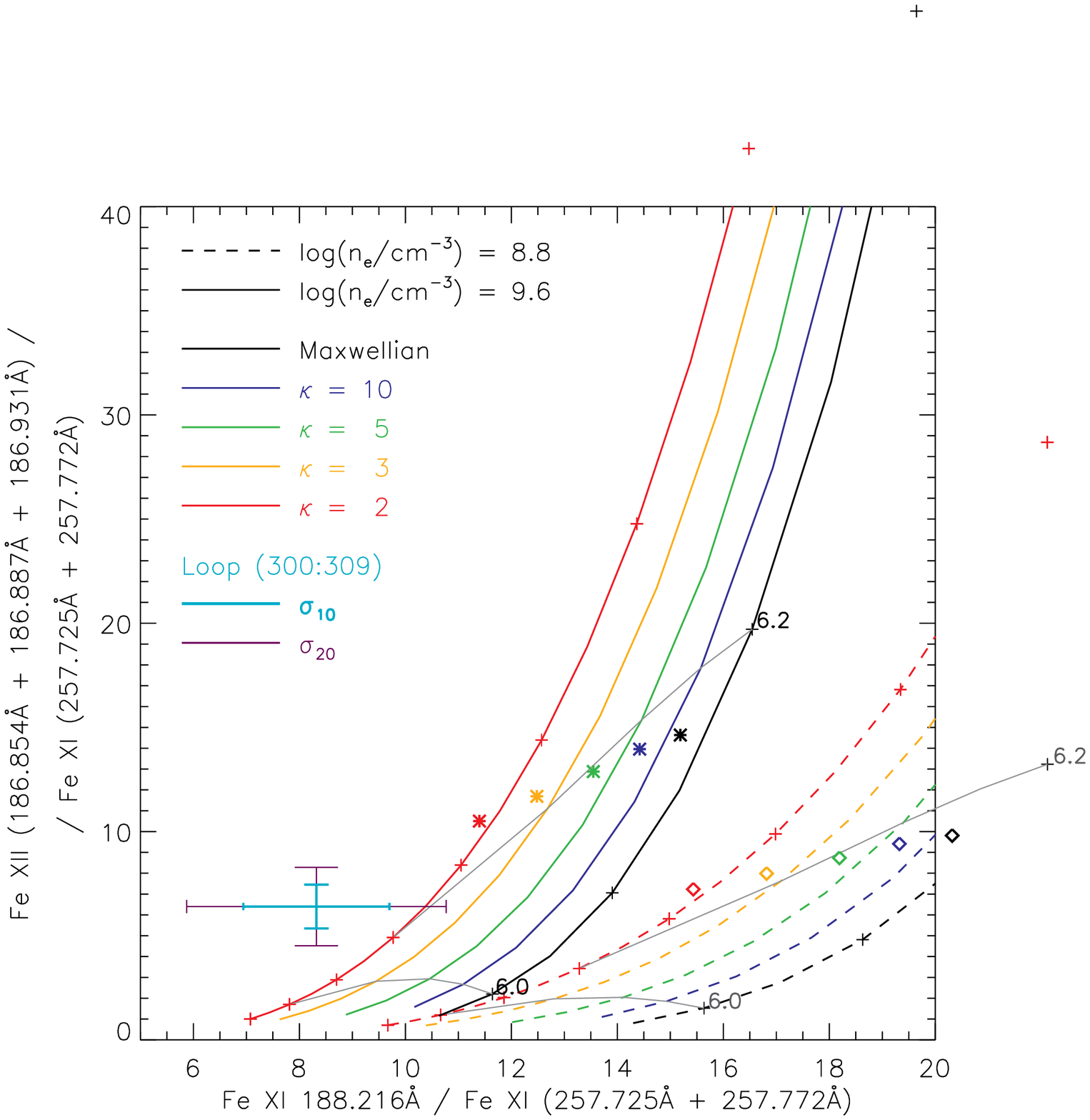}
	\includegraphics[width=7.4cm,clip]{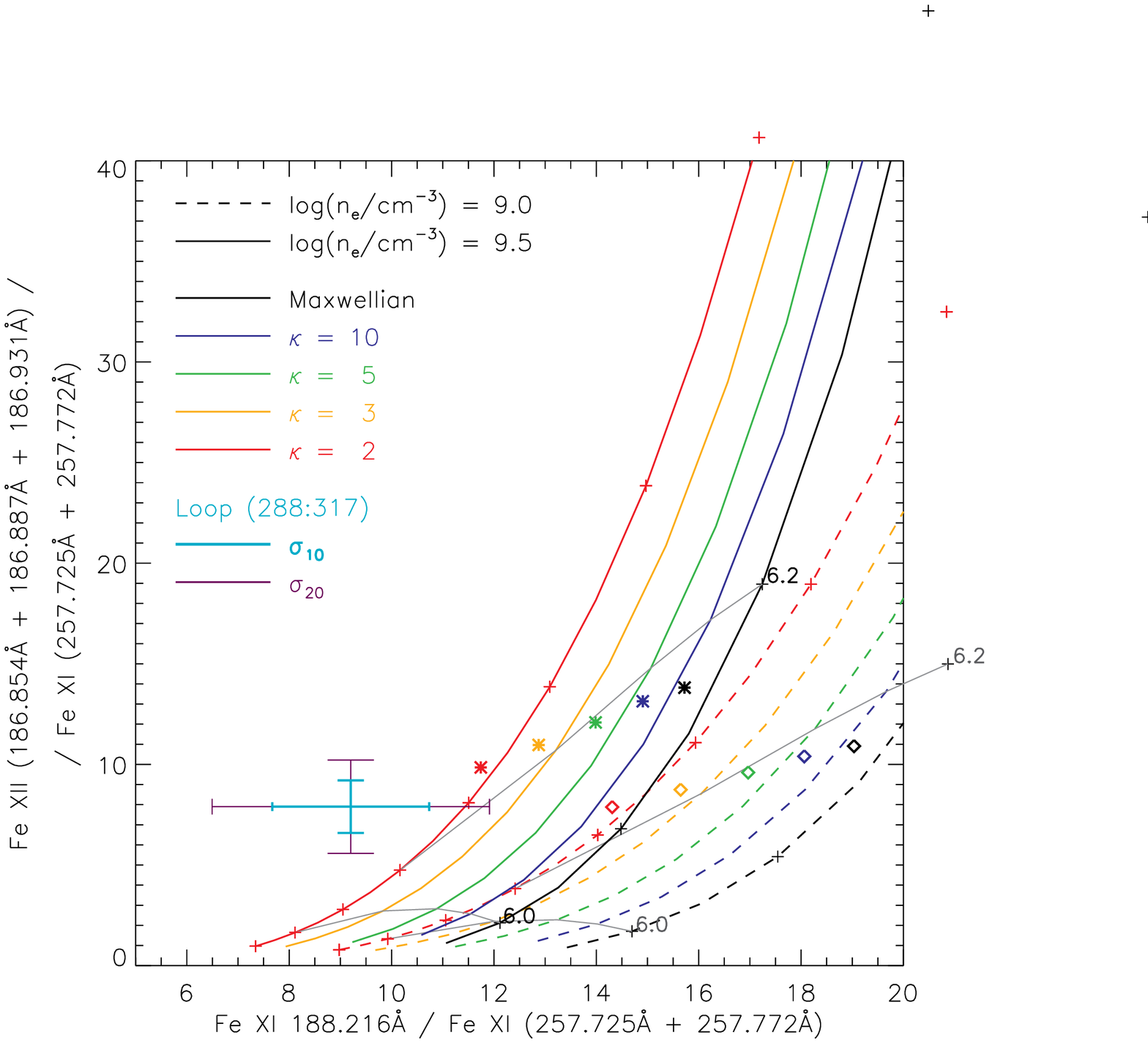}
	\includegraphics[width=7.4cm,clip]{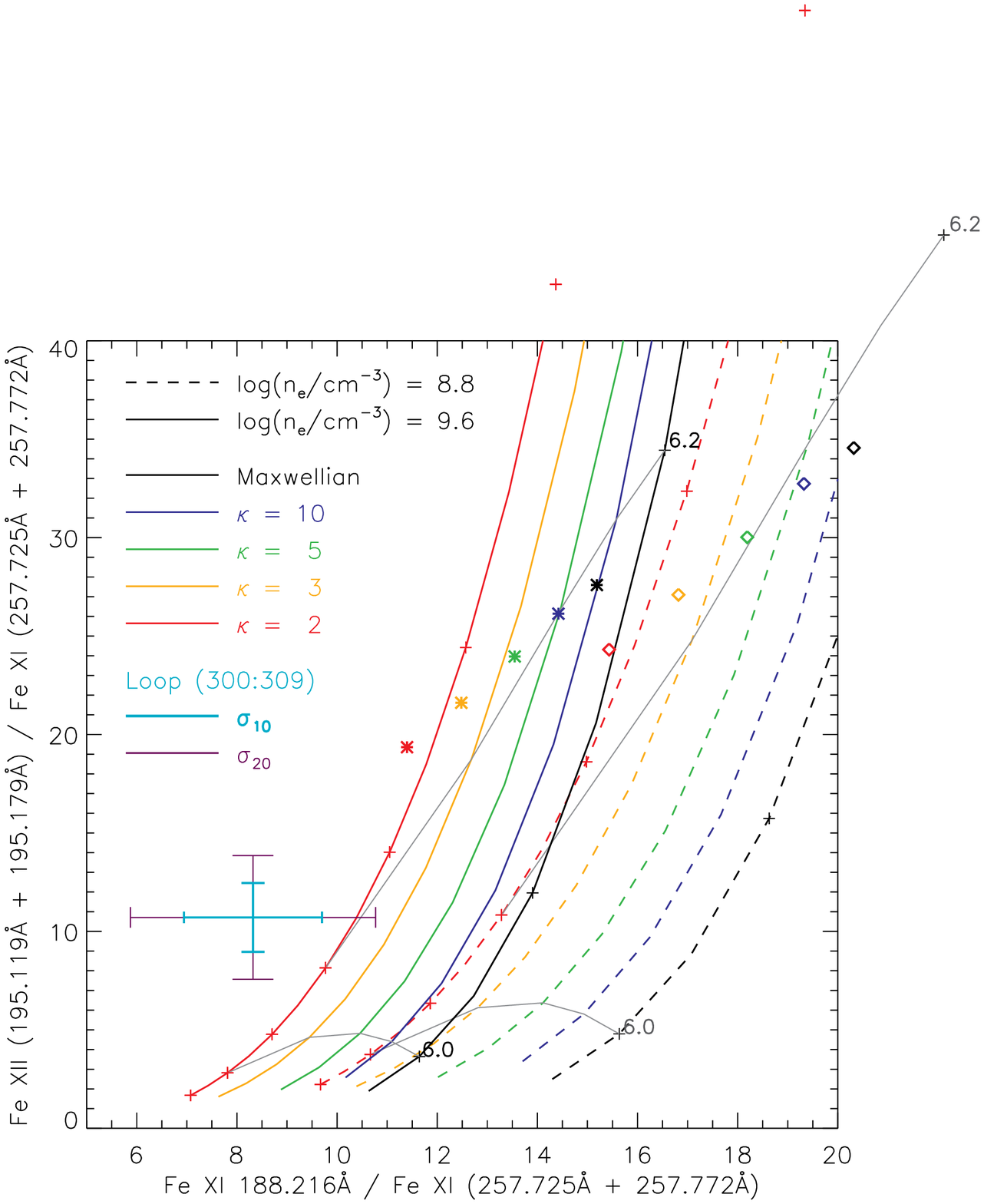}
	\includegraphics[width=7.4cm,clip]{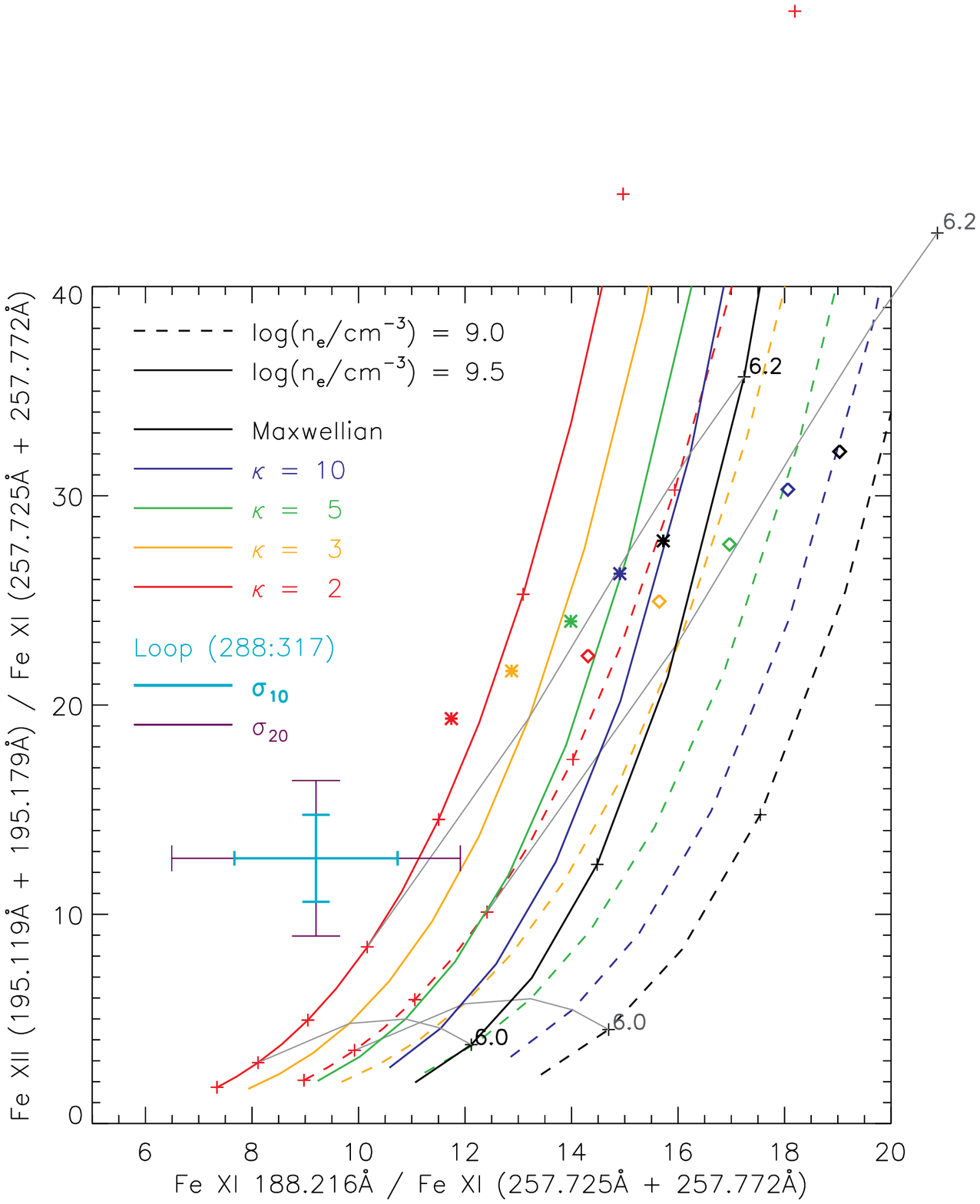}
	\includegraphics[width=7.4cm,clip]{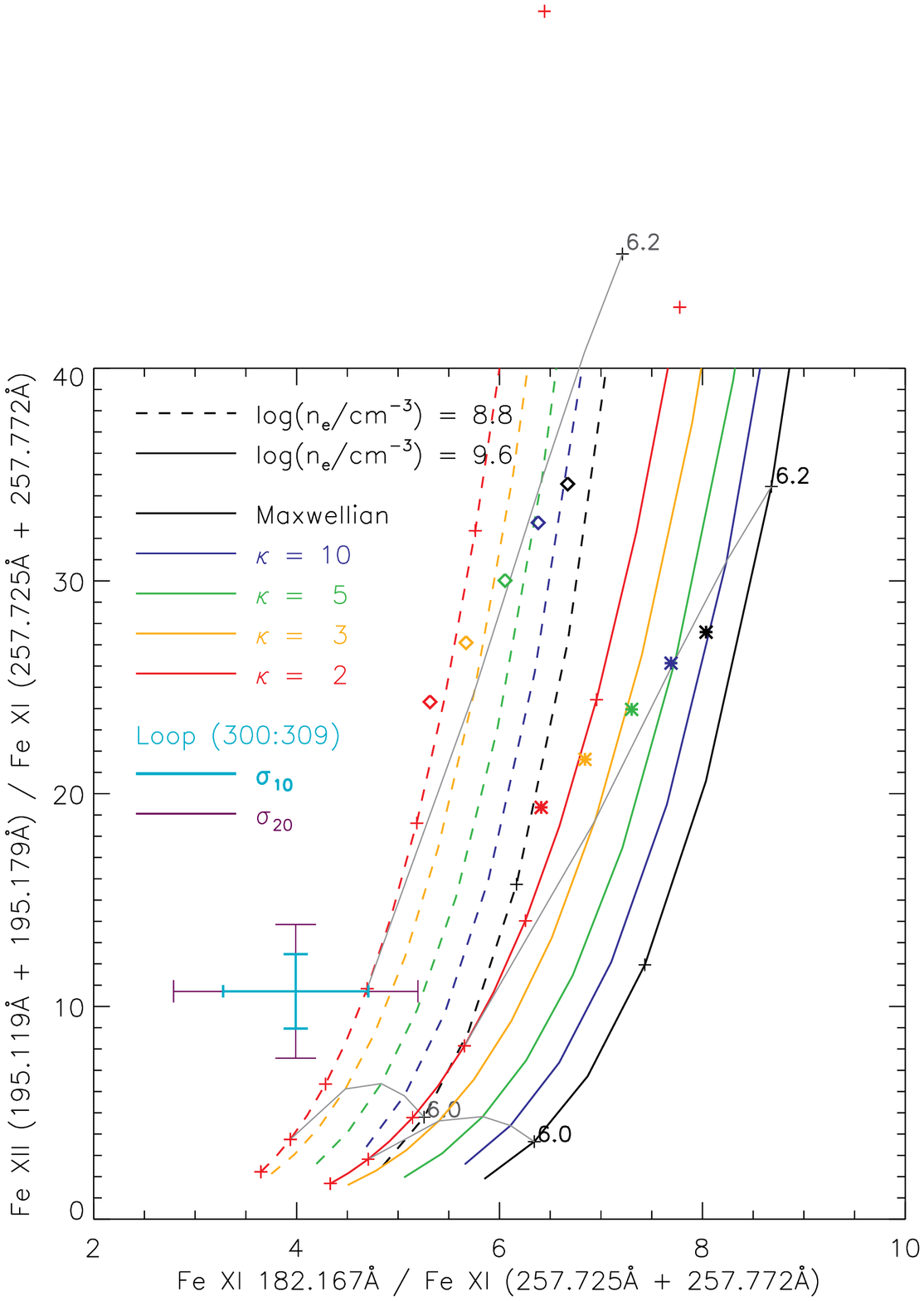}
	\includegraphics[width=7.4cm,clip]{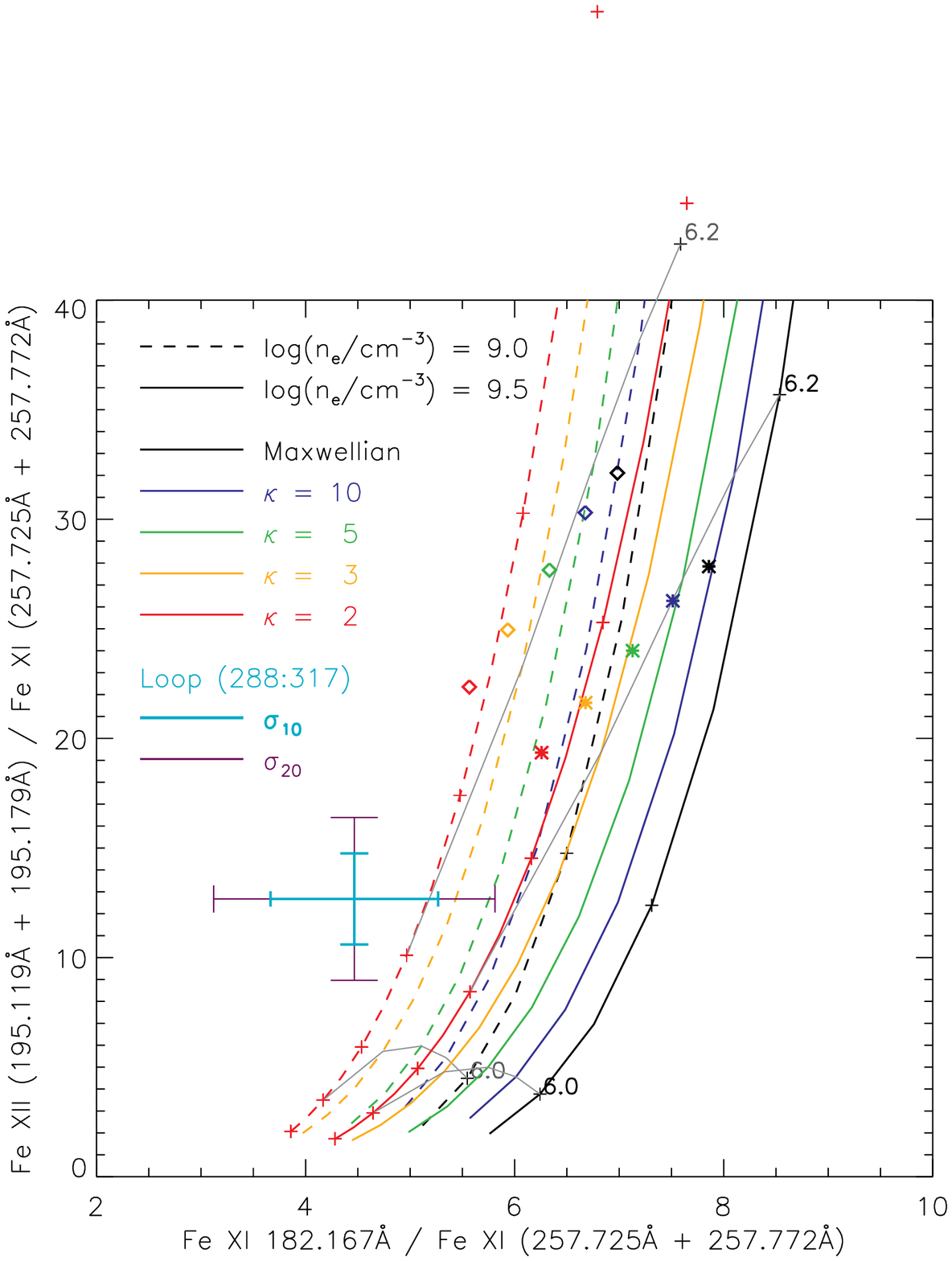}
	\caption{Diagnostics of $\kappa$ involving the \ion{Fe}{11} 257.772\AA~selfblend. Different colors correspond to different $\kappa$, linestyles denote electron density. Diamonds and asterisks denote the predicted ratios based on DEM analysis. \\ A color version of this image is available in the online journal.}
        \label{Fig:Diag_TK2}
   \end{figure*}
%
   \begin{figure*}[!h]
	\centering
	\includegraphics[width=7.4cm,clip]{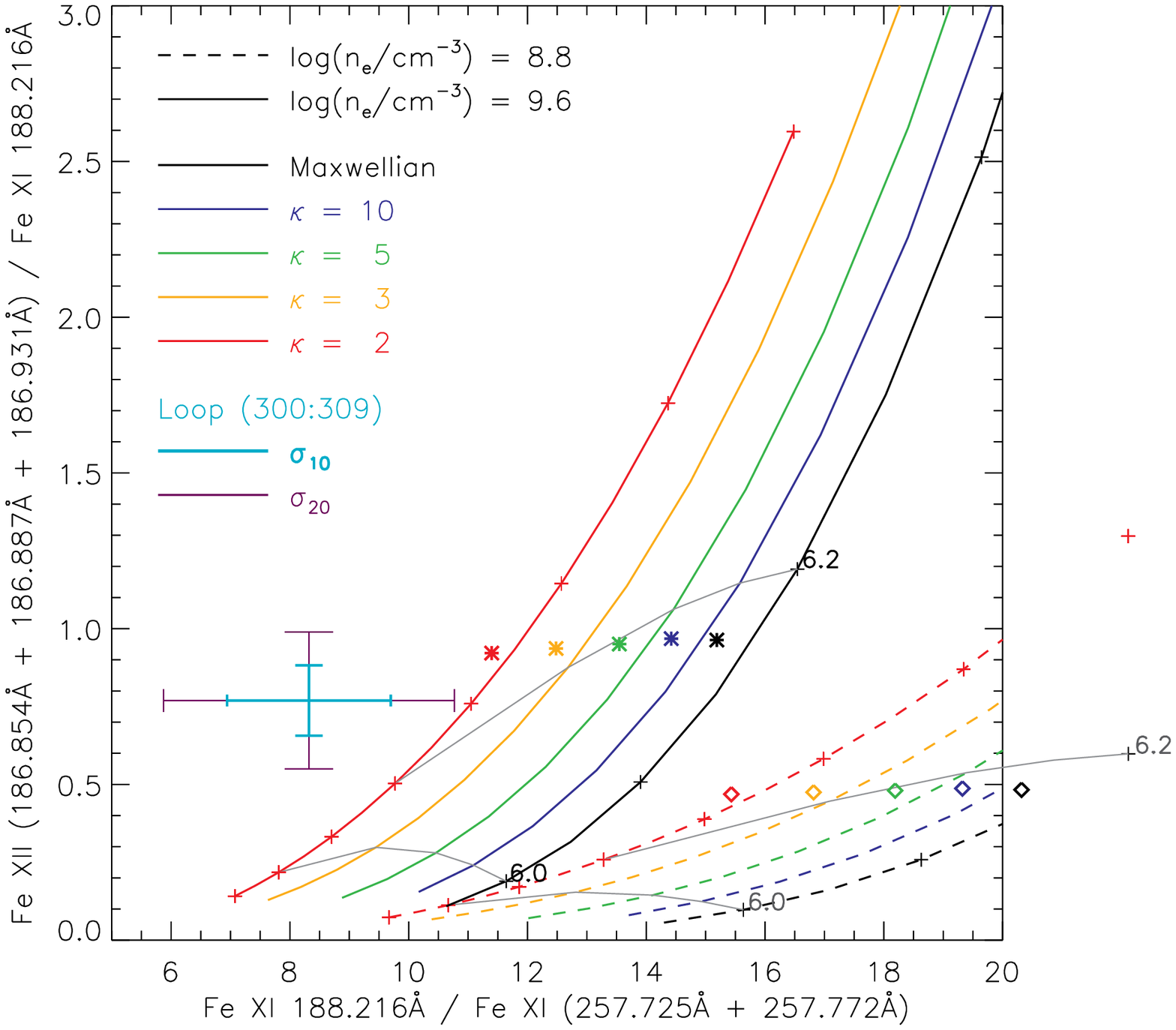}
	\includegraphics[width=7.4cm,clip]{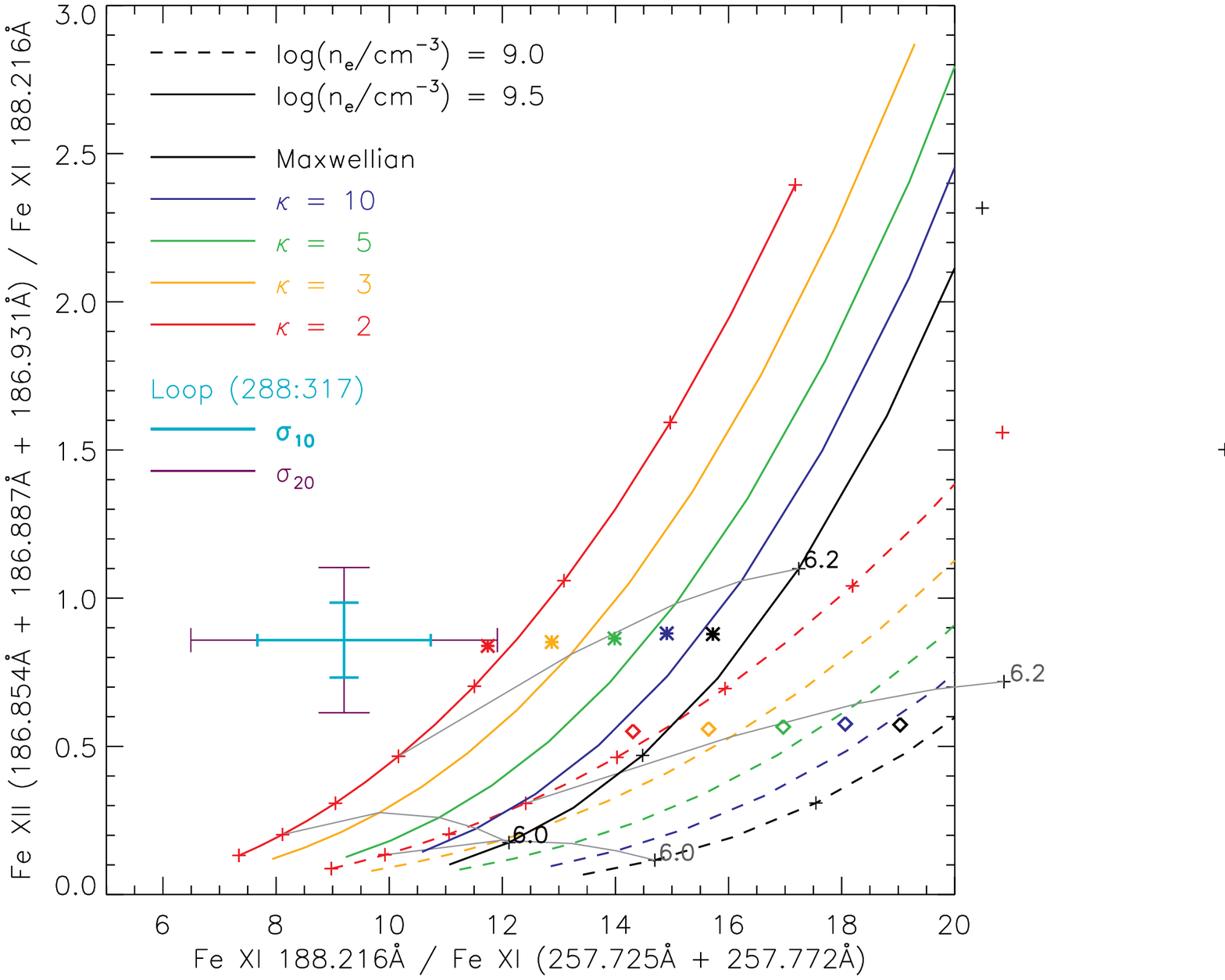}
	\includegraphics[width=7.4cm,clip]{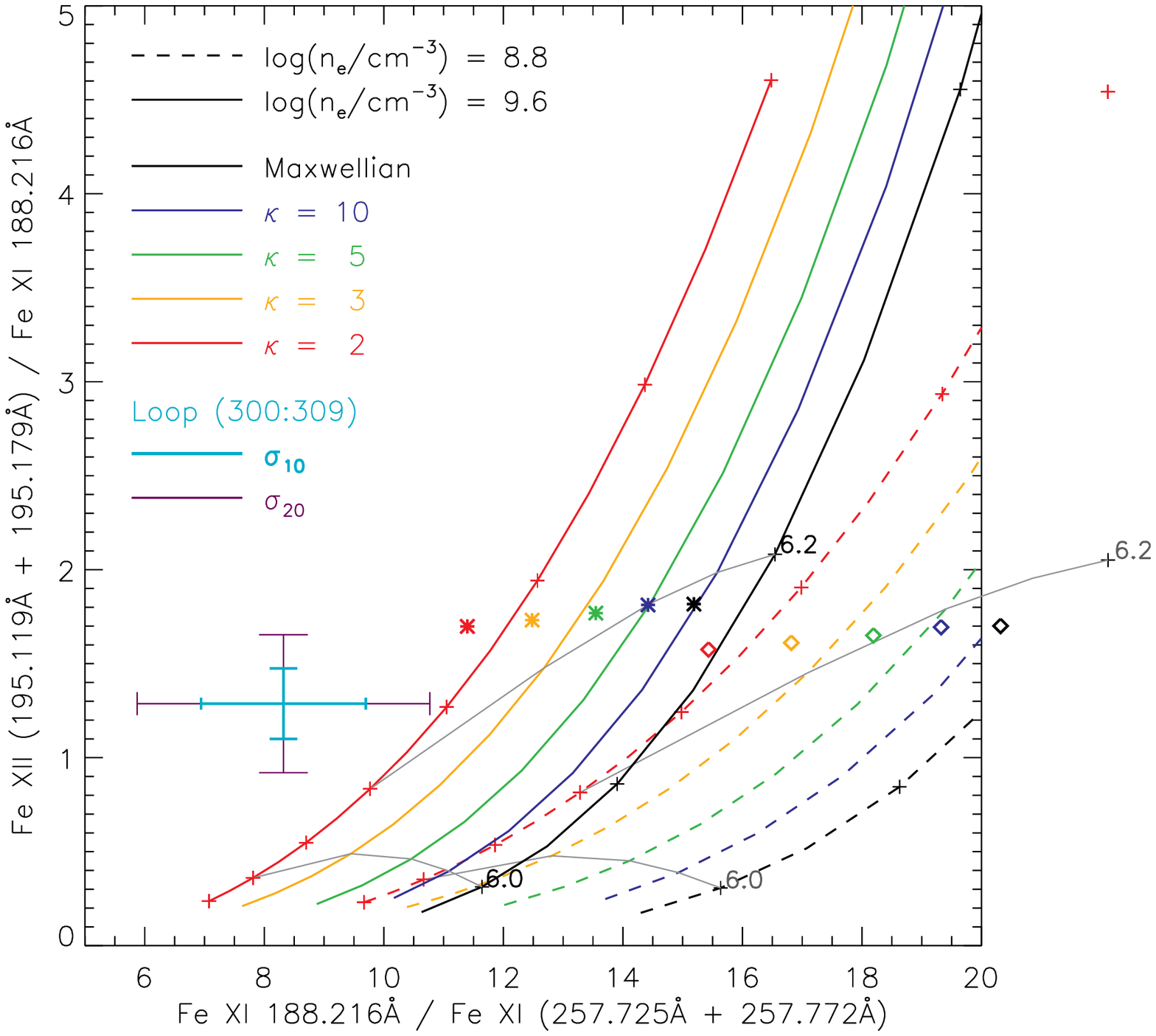}
	\includegraphics[width=7.4cm,clip]{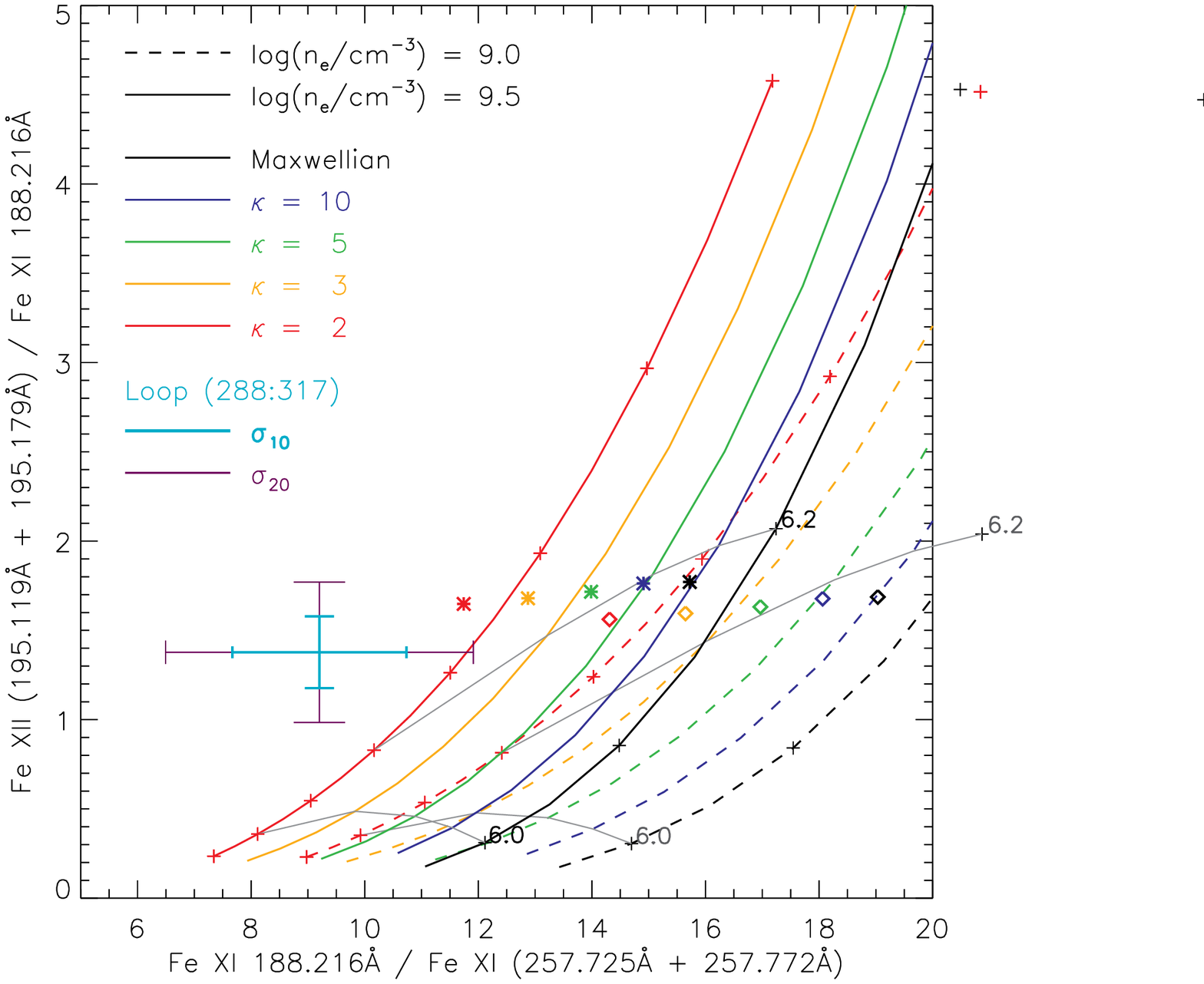}
	\includegraphics[width=7.4cm,clip]{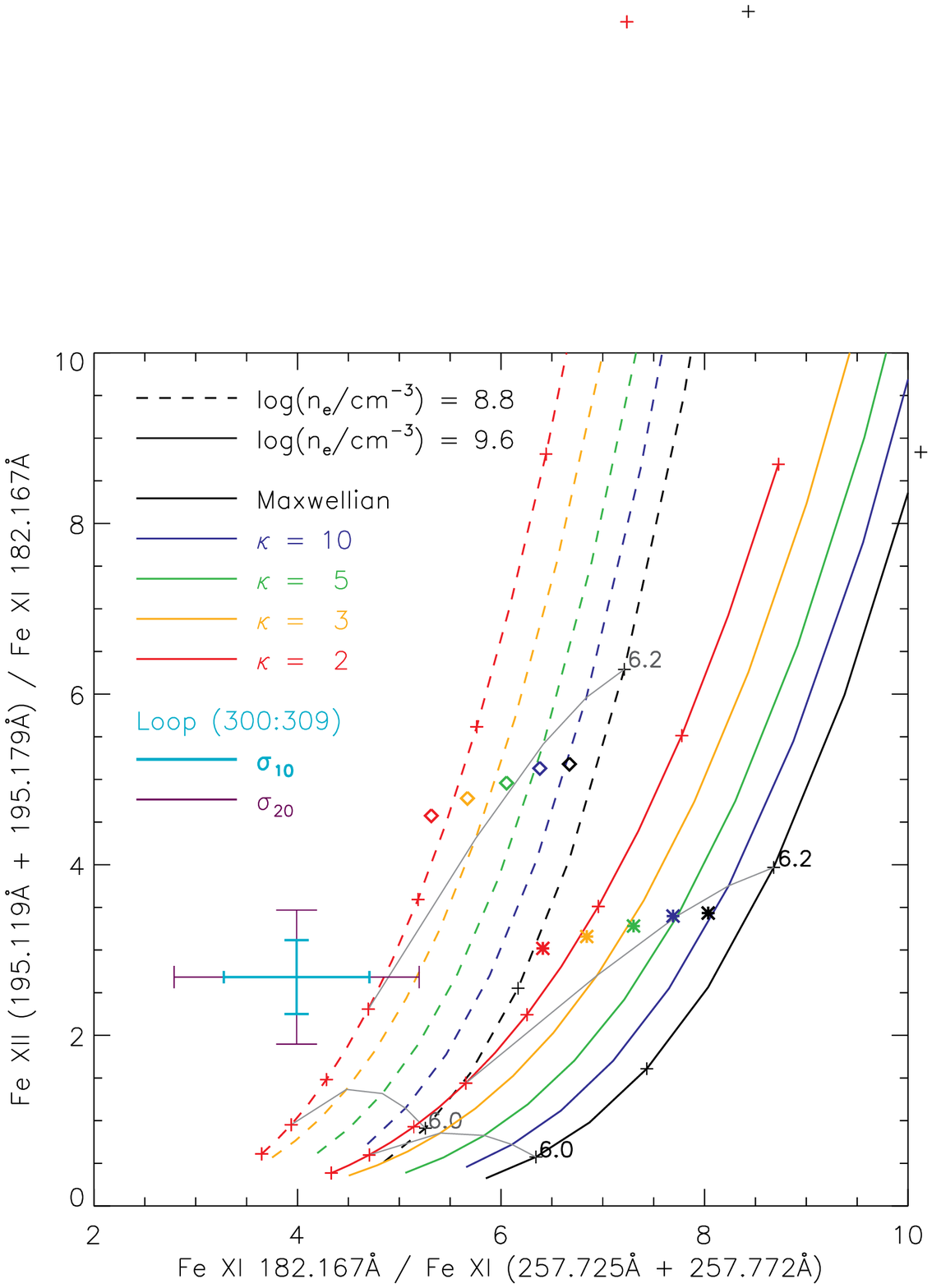}
	\includegraphics[width=7.4cm,clip]{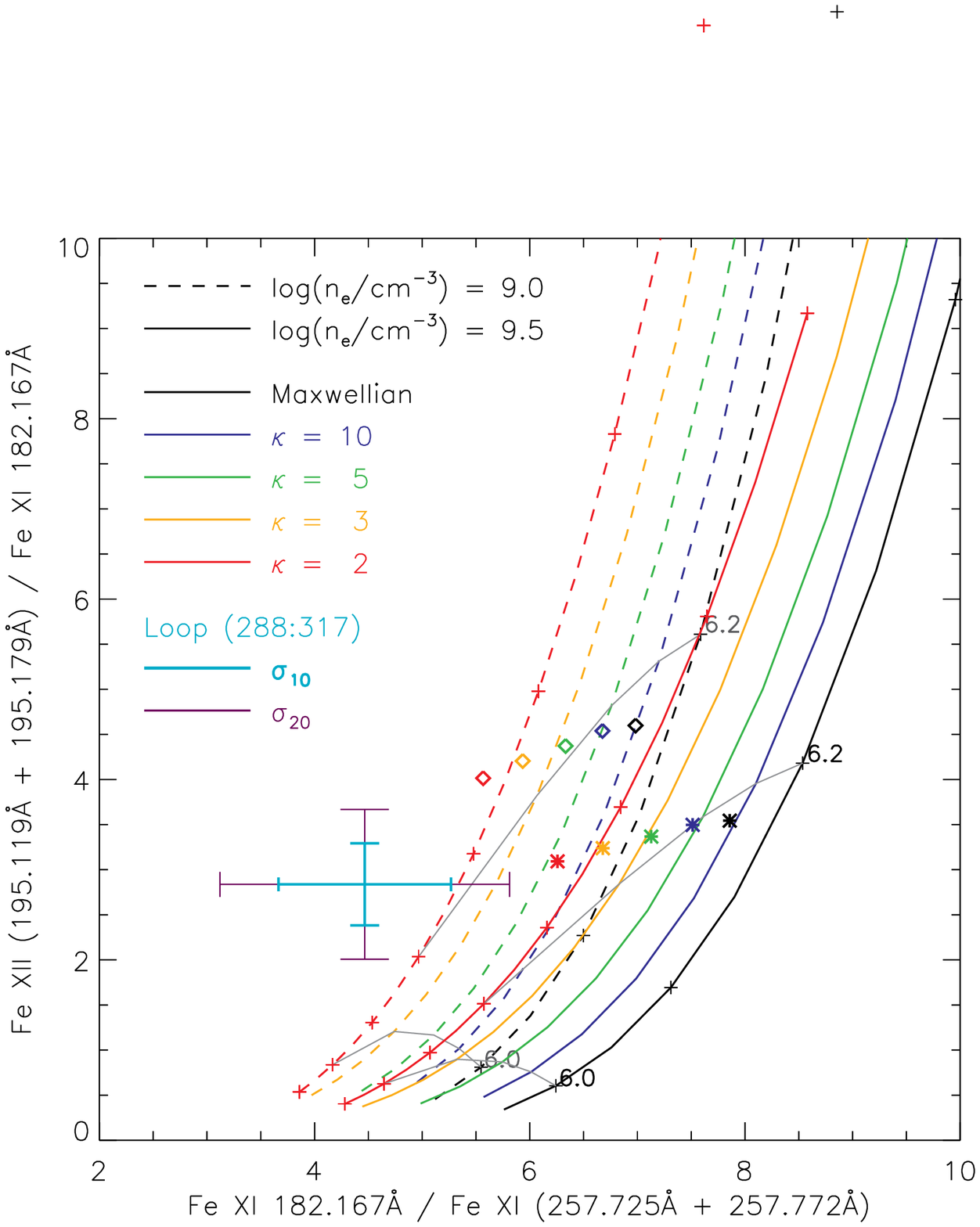}
	\caption{Diagnostics of $\kappa$ involving the \ion{Fe}{11} 257.772\AA~selfblend and \ion{Fe}{12}\,/\,\ion{Fe}{11} ratios using lines only from the EIS short-wavelength channel. Different colors correspond to different $\kappa$, linestyles denote electron density. Diamonds and asterisks denote the predicted ratios based on DEM analysis. \\ A color version of this image is available in the online journal.}
        \label{Fig:Diag_TK3}
   \end{figure*}
%
%
   \begin{figure*}[!h]
	\centering
	\includegraphics[width=7.4cm,clip]{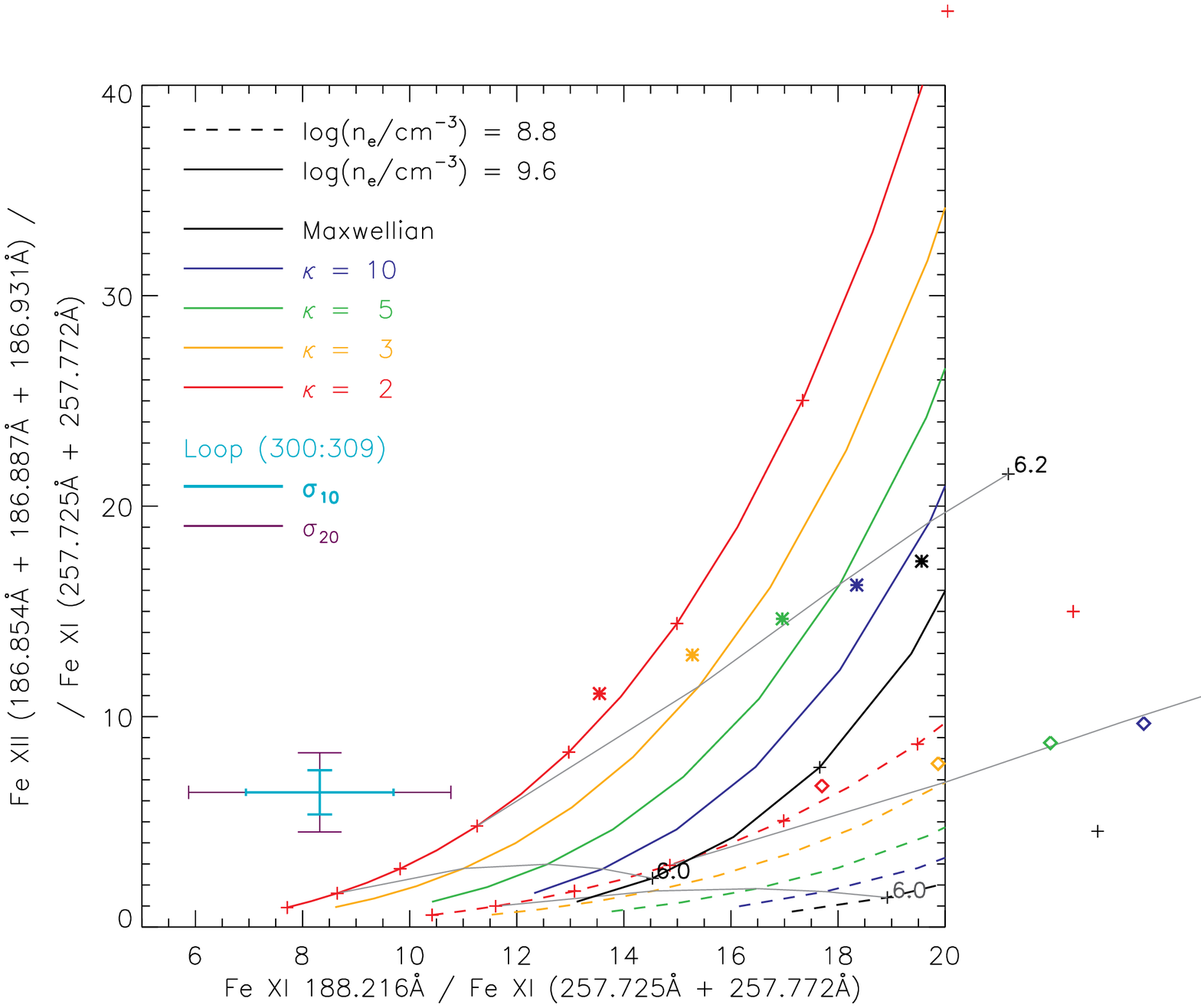}
	\includegraphics[width=7.4cm,clip]{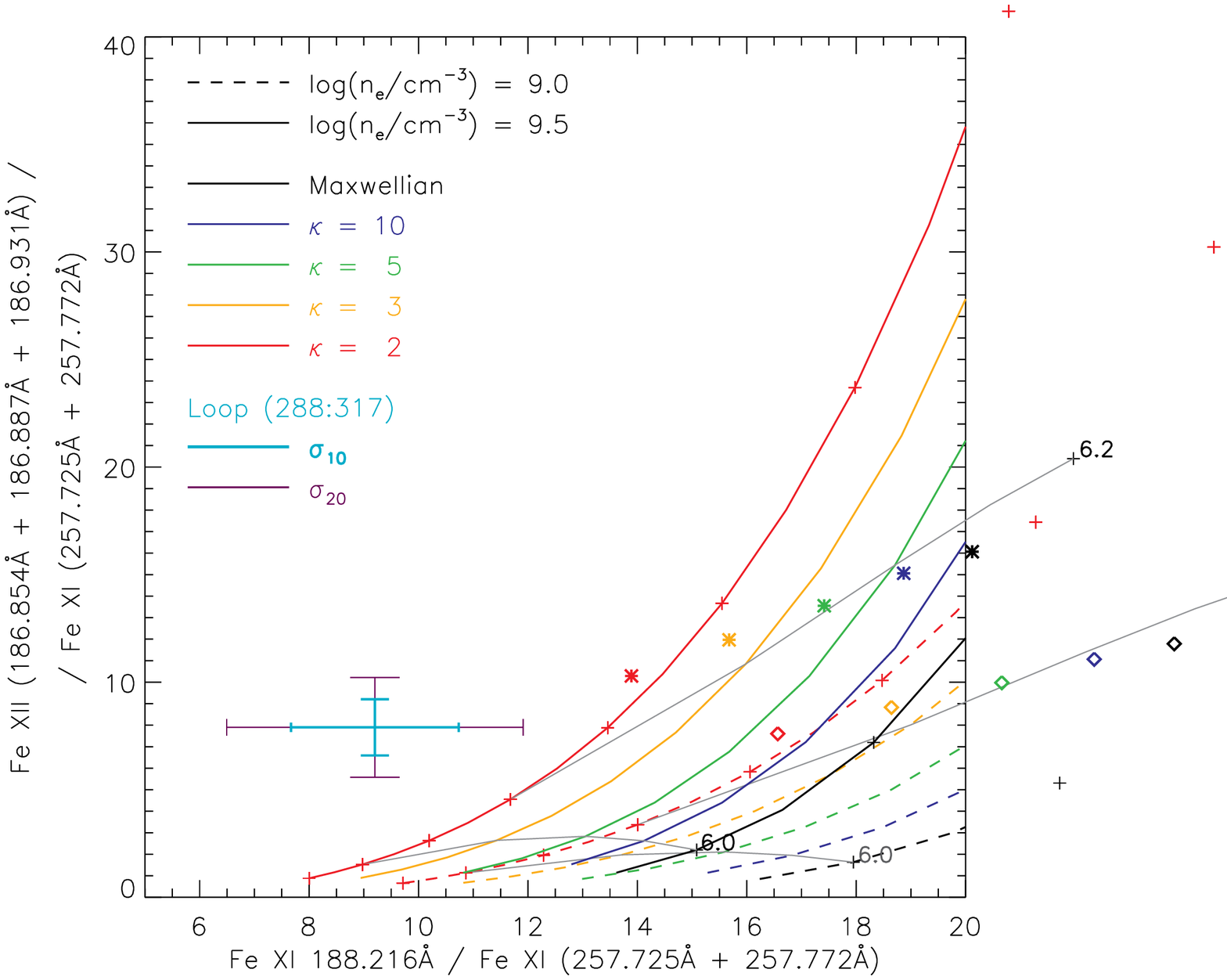}
	\includegraphics[width=7.4cm,clip]{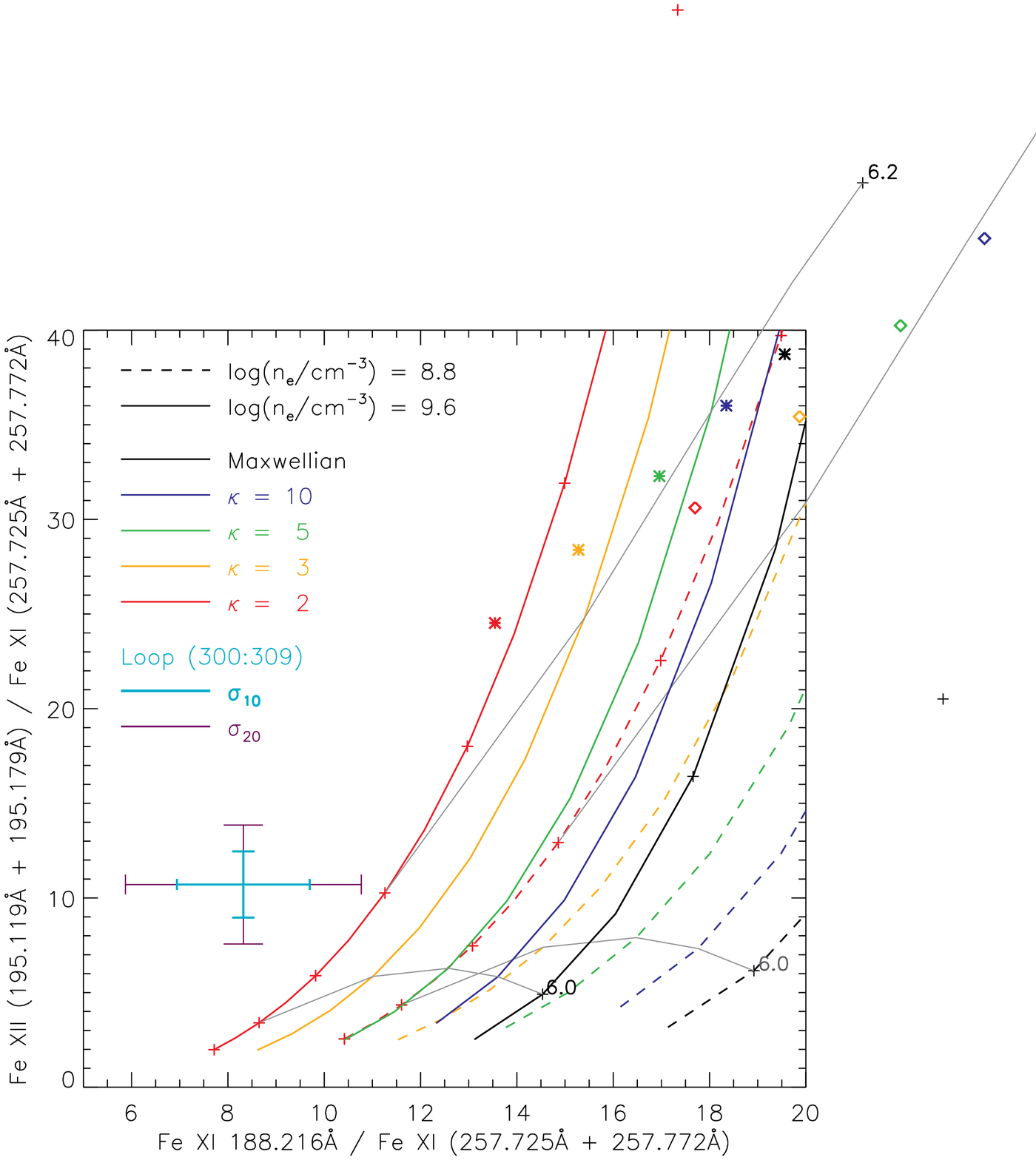}
	\includegraphics[width=7.4cm,clip]{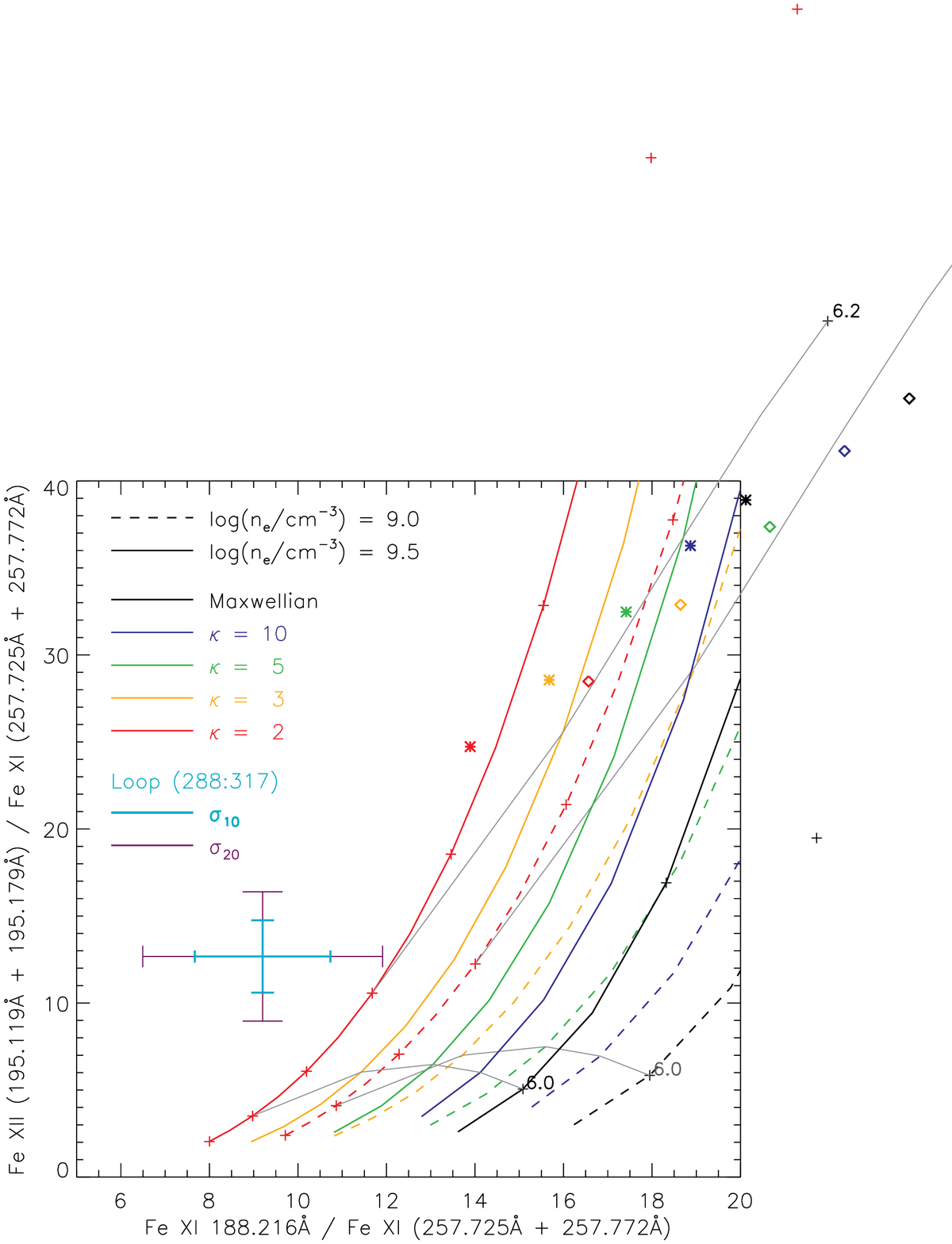}
	\includegraphics[width=7.4cm,clip]{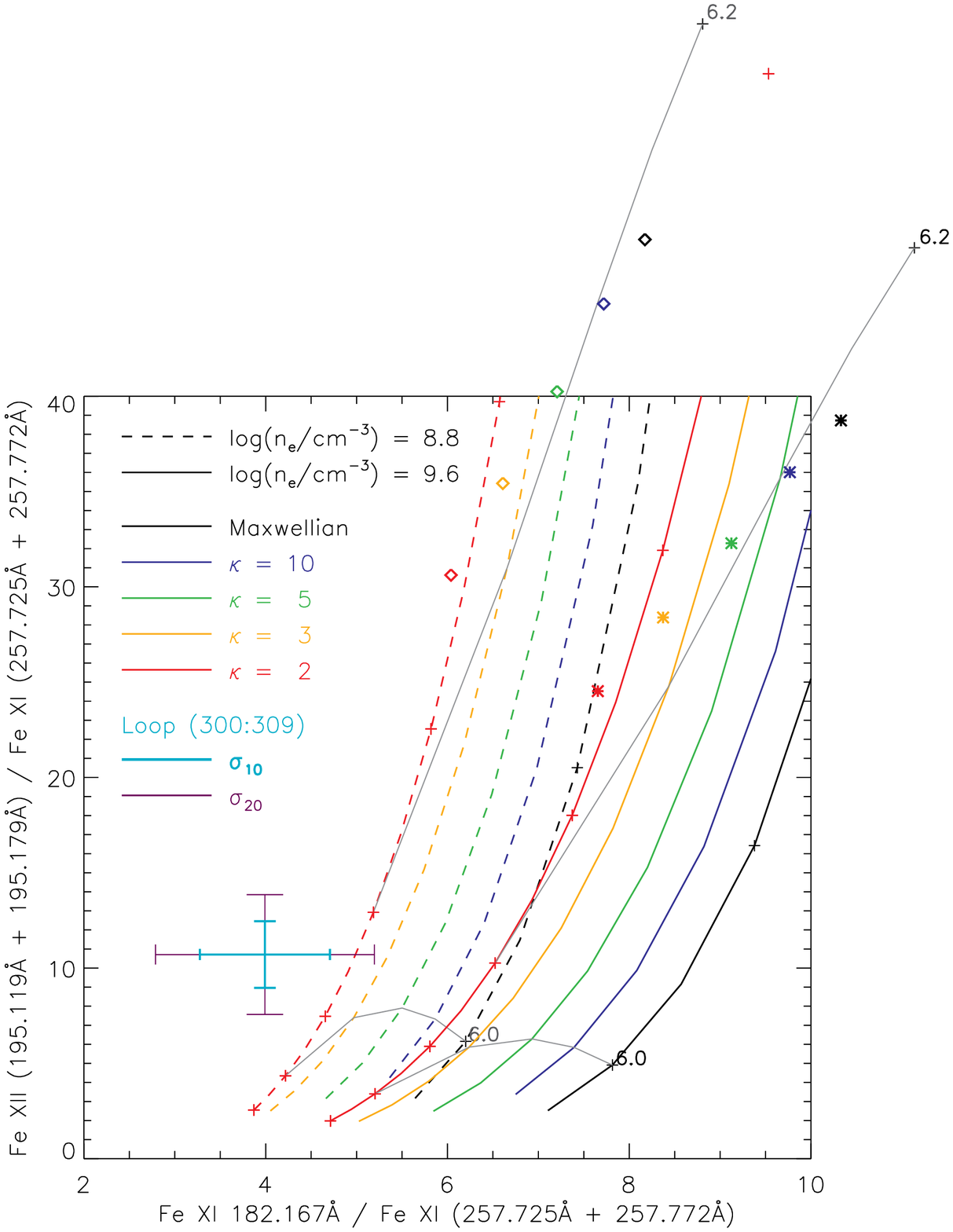}
	\includegraphics[width=7.4cm,clip]{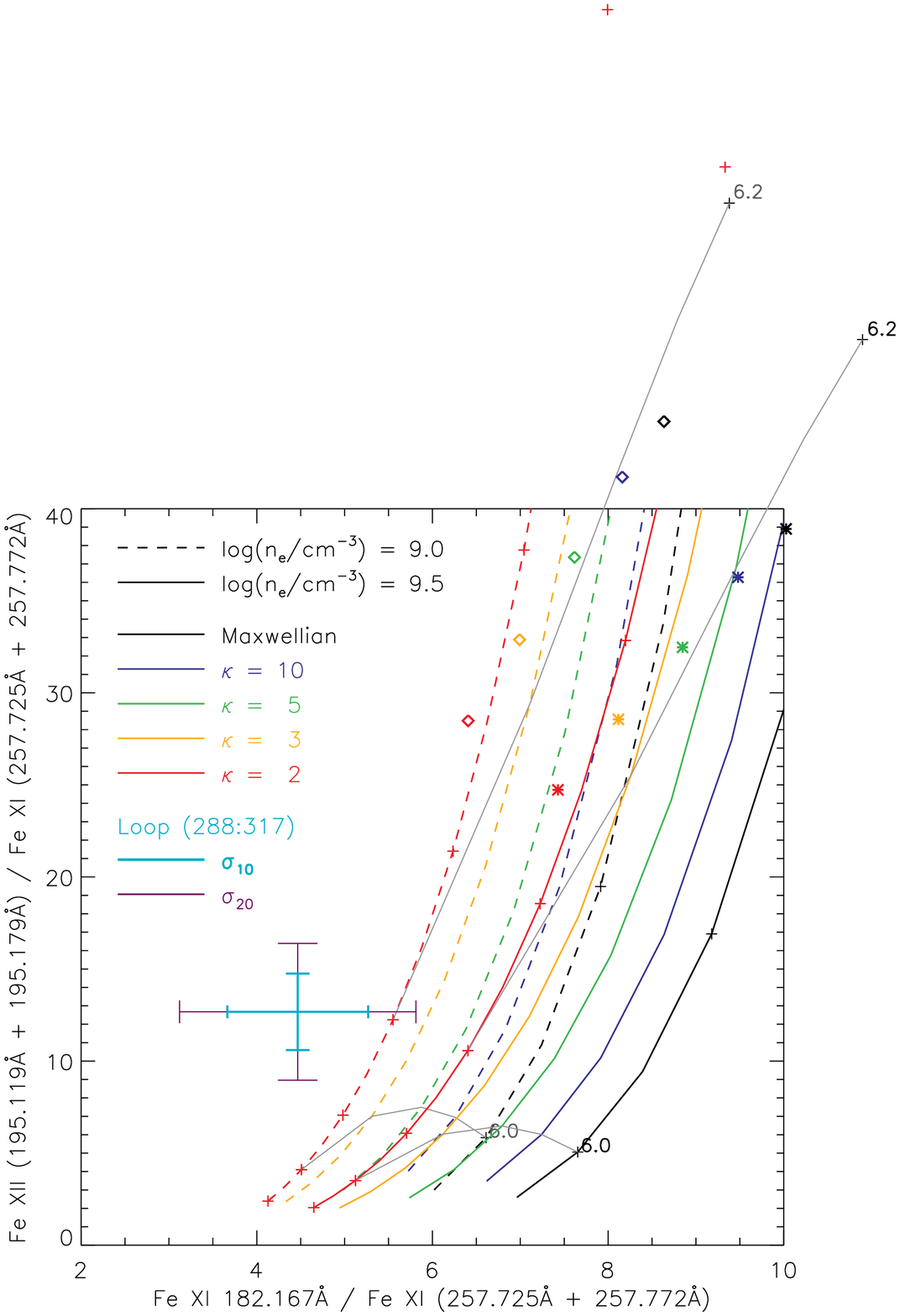}
	\caption{Same as Fig.~\ref{Fig:Diag_TK2}, but with atomic data corresponding to CHIANTI 7.1 and the KAPPA package. \\ A color version of this image is available in the online journal.}
        \label{Fig:Diag_TK2_old}
   \end{figure*}
%
%
%
\begin{table*}[!ht]
\begin{center}
\caption{Electron densities diagnosed using individual ratios.}
\label{Table:2}
\begin{tabular}{lcccccc}
\tableline
\tableline
							& \multicolumn{3}{c}{Loop (288:317)}	&  \multicolumn{3}{c}{Loop (300:309)}  \\
      Line ratio                            		&$I_1/I_2$ & log($n_\mathrm{e})_\mathrm{Maxw.}$ 	 &  log($n_\mathrm{e})_{\kappa=2}$  
							&$I_1/I_2$ & log($n_\mathrm{e})_\mathrm{Maxw.}$ 	 &  log($n_\mathrm{e})_{\kappa=2}$ \\  
\tableline
\ion{Fe}{11} 182.167\,/\,188.215              		& 0.49    & 9.45$^{+0.06}_{-0.10}$ & 9.34$^{+0.12}_{-0.20}$ 
							& 0.48 	  & 9.43$^{+0.06}_{-0.10}$ & 9.32$^{+0.12}_{-0.19}$ \\ 
\ion{Fe}{12} (186.854+186.887)\,/\,(195.119+195.179)	& 0.62    & 9.86$^{+0.09}_{-0.08}$ & 9.83$^{+0.03}_{-0.09}$ 
							& 0.60 	  & 9.79$^{+0.10}_{-0.08}$ & 9.76$^{+0.09}_{-0.02}$ \\  
\ion{Fe}{13} 196.525\,/\,202.044           		& 0.19    & 9.30$^{+0.11}_{-0.11}$ & 9.20$^{+0.15}_{-0.15}$ 
							& 0.13 	  & 9.12$^{+0.12}_{-0.12}$ & 9.02$^{+0.15}_{-0.16}$ \\  
\ion{Fe}{13} (203.826 sbl)\,/\,202.044 		  	& 1.92    & 9.35$^{+0.10}_{-0.10}$ & 9.27$^{+0.13}_{-0.14}$ 
							& 1.42 	  & 9.16$^{+0.10}_{-0.10}$ & 9.08$^{+0.13}_{-0.14}$ \\
\tableline
adopted value						&	  \multicolumn{3}{c}{9.0 -- 9.5}	
							&   	  \multicolumn{3}{c}{8.8 -- 9.6}		    \\
\tableline
\tableline
\end{tabular}
\end{center}
\end{table*}
%
%
\section{\textit{Hinode}/EIS Observations During HOP 226}
\label{Sect:4}

The Extreme-Ultraviolet Imaging Spectrometer \citep[EIS,][]{Culhane07} onboard the \textit{Hinode} mission \citep{Kosugi07} provides EUV spectra of the Sun in the wavelength ranges 171-212\AA~and 245-291\AA~with a spectral resolution of about 22\,m\AA~and spatial resolution down to 1--2$\arcsec$, corresponding to the width of the slit chosen.

On 2013 March 30, EIS was observing portion of AR 11704 as a part of the Hinode Operation Plan (HOP) 226. HOP 226 was originally designed for diagnostics of non-Maxwellian $\kappa$-distributions using weak \ion{O}{4}--\ion{O}{5} and \ion{S}{10}--\ion{S}{11} lines \citep[see][]{Mackovjak13}. Since these lines are weak, EIS was rastering an area within the AR core using 2$\arcsec$ slit and long exposures (60\,s and 600\,s). Each 60\,s and 600\,s raster contained 10 exposures of the 2$\arcsec$ slit at contiguous positions in Solar X, covering 512$\arcsec$ in Solar Y (heliocentric co-ordinates). For context, as well as density diagnostics, the raster contained additional lines, notably several strong \ion{Fe}{11}--\ion{Fe}{13} lines. 

Upon examination of the data, we find that the 600\,s rasters suffer badly from an accumulation of cosmic rays, and that the weak O and S lines do not have sufficient intensities for diagnostics. However, the last 60\,s raster, starting at 13:11\,UT and progressively rastering in the West--East direction, captures a portion of the coronal loop in its ninth exposure at 13:19\,UT (position 9 in Fig.~\ref{Fig:EIS_raster}), with the portion of the loop lying directly along the EIS slit. Some of the O and S lines proposed for diagnostics are again too weak in this 60s exposure. Nevertheless, the strong \ion{Fe}{11}--\ion{Fe}{13} lines  (Table \ref{Table:1}) are sufficient to perform diagnostics of the $\kappa$-distributions. It is these data that are analyzed here.

\subsection{EIS Data Processing and Calibration}
\label{Sect:4.1}

We first performed coalignment of the raster with the AIA data. To do this, we use the \ion{Fe}{12} 195.119\AA~selfblend (hereafter, ``195.119\AA~sbl'') and the AIA 193\AA~data uncorrected for solar differential rotation, but with removed stray light \citep{Poduval13}. These AIA data are used to build an AIA 193\AA~pseudo-raster image by extracting the EIS field of view during each exposure within the raster, averaging individial AIA 193\AA~frames within the duration of each raster exposure, and convolving with the EIS point-spread function, which is assumed to be a Gaussian with FWHM of 2$\arcsec$ in both $X$ and $Y$ directions \citep[][Appendix A therein]{DelZanna11c}. By comparing this AIA 193\AA~pseudo-raster image with the EIS 195.119\AA~sbl, we found shifts in the EIS positioning of $\Delta X$\,=\,13$\arcsec$ and $\Delta Y$\,=\,$-9.5\arcsec$ with respect to AIA. Furthermore, the EIS slit is found to be rotated by $+1^\circ$ with respect to the Solar $Y$. Note that since the EIS raster has only 10 exposures in Solar $X$, the EIS raster image cannot be rotated. Rather, the AIA data have to be rotated by $-1^\circ$ before constructing the AIA 193\AA~pseudo-raster image. Because of the relative rotation between the two instruments, as well as for simplicity, the positions of the individual EIS pixels will be given in pixel units $x$ and $y$ rather than Solar $X$ and $Y$.

The EIS data were calibrated using the standard \textit{eis\_prep.pro} routine available within the SolarSoft platform running under the Interactive Data Language (IDL). During the calibration, the latest in-flight radiometric calibration of \citet{DelZanna13a} was used. The calibrated data contain intensities in physical units [erg\,cm$^{-2}$\,s$^{-1}$\,sr$^{-1}$\,\AA$^{-1}$] and the 1-$\sigma$ errors on the observed intensities at each spatial and wavelength pixel. The intensities and uncertainties are stored in the separate files, with the error file also containing flags for missing pixels (dusty, hot, warm pixels and pixels affected by cosmic rays). We note that usually, these missing pixels are interpolated during the calibration procedure. However, the number of missing pixels in the raster is relatively high, up to 30\%. This brings up concerns about uncertainties in the fitting of the observed intensities at each single spatial pixel \citep{Young10}.

In this work, the pixels flagged as missing were interpolated only for purposes of visualising the raster field of view containing the loop (Fig.~\ref{Fig:EIS_raster}, \textit{right}). The missing pixels were excluded from any further analysis, since they cannot always be interpolated with accuracy due to their high number. Instead, we rely on intensities obtained by averaging along a selected loop segment observed in position 9 of the EIS slit. The intensity averaging is done over $y$ at each wavelength pixel using exclusively pixels not flagged as missing. The same is done at position 5 of the EIS slit, which we chose to represent the background. We selected two loop segments over which the averaging is performed. The first one consists of pixels $y$\,=\,288 to 317, denoted simply as ``loop (288:317)'', and representing the average loop spectrum. The second one is much shorter, consisting of pixels $y$\,=\,300 to 309, denoted as ``loop (300:309)'', and representing a much shorter segment of the loop.
We chose to average over longer rather than shorter loop segments in order to minimize errors from photon noise. We also note that the effective EIS resolution in $y$ is only about 3$\arcsec$--4$\arcsec$ because of the instrument point-spread function.

Subsequently, the average spectrum of the background at position 5 is subtracted from the average loop spectrum. The resulting background-subtracted loop spectrum is fitted with Gaussian line profiles and locally linear continua, taking into account known blends and selfblends \citep[e.g.,][]{DelZanna10c,DelZanna11,DelZanna12d,Dudik14b}. The intensities obtained are listed in Table \ref{Table:1}. The uncertainties on these intensities are obtained by error propagation of the uncertainties of the photon noise and dark-current subtraction that are the output of the \textit{eis\_prep.pro}. These uncertainties are then added in quadrature with the uncertainty of the EIS radiometric calibration. We consider two values of the radiometric calibration uncertainty, 10\% and 20\%. The 20\% uncertainty is the uncertainty of the ground calibration \citep{Culhane07}, while \citet{Wang11} argues that the in-flight uncertainty is smaller, about 10\%. The final uncertainties on the line intensities are denoted $\sigma_{10\%}$ and $\sigma_{20\%}$, respectively, and are also listed in Table \ref{Table:1}.

We note that we employed the revised radiometric calibration of \citet{DelZanna13a} rather than the ground calibration \citep{Culhane07} or the calibration of \citet{Warren14}. \citet{DelZanna13a} showed that significant departures from the EIS ground calibration occurred over time, especially with the lines in the long-wavelength channel of EIS being underestimated by about a factor of two for observations after 2010. Line ratios were used to obtain a calibration corrected for decrease of sensitivity. \citet{Warren14} obtained similar results, comparing the EIS radiances from the whole Sun with the \textit{SDO}/EVE irradiances together with the quiet-Sun DEM modeling. However, the EVE calibration can be an additional source of uncertainty. Furthermore, the quiet-Sun DEM modelling used by \citet{Warren14} can in itself be sensitive to $\kappa$ \citep{Mackovjak14}, which could entangle the calibration to diagnostics and overcomplicate the problem. The \citet{DelZanna13a} calibration is sufficient for our purposes, as the diagnostics of electron density (Sect. \ref{Sect:4.2}) as well as the diagnostics of $\kappa$ (Sect. \ref{Sect:4.3}) rely only on line ratios \citep{Dzifcakova10,Mackovjak13,Dudik14b}, and thus do not require absolute calibration.

We also note that the decrease of sensitivity of the long-wavelength channel of EIS by about a factor of 2 is clearly significant for our diagnostic purposes. This is because the diagnostics of $\kappa$ in Sect. \ref{Sect:4.3} involve lines from both the short-wavelength and long-wavelength EIS channels. We note that the \citet{DelZanna13a} calibration applies only to observations up to September 2012, therefore we have assumed that no further degradation occurred since. If the long-wavelength channel sensitivity further degraded significantly, the radiances of the long-wavelength channel would be underestimated, and the ratios shown in Figs. \ref{Fig:Diag_TK1}--\ref{Fig:Diag_TK3} would decrease.

%
\subsection{Density Diagnostics}
\label{Sect:4.2}

Diagnostics of electron density are a neccessary prerequisite to diagnostics of $\kappa$ (Sect. \ref{Sect:4.3}). The Fe lines observed by EIS (Table \ref{Table:1}) contain several combinations of lines sensitive to electron density \citep[e.g.,][]{Watanabe09,Young09,DelZanna10c,DelZanna11,DelZanna12a,Dudik14b}. The density-sensitive line ratios are listed in Table \ref{Table:2}, where the diagnosed densities are also listed.

The diagnostics of density are performed using the method developed first by \citet{Dzifcakova10}. The theoretical line ratios were calculated by \citet{Dudik14b} using the latest available atomic data (see also Sect. \ref{Sect:2.2}). This density diagnostics method employs ratios of lines arising in the same ion. The sensitivity of these ratios to $n_\mathrm{e}$ cannot be significantly influenced by their sensitivity to $T$ and $\kappa$, otherwise the density diagnostics would be precluded. Typically, it was found that density-sensitive ratios commonly used for the Maxwellian distribution can be used under the assumption of $\kappa$-distributions in the same manner, albeit the resulting densities can be up to 0.1 dex lower for $\kappa$\,=\,2 (red color in Fig.~\ref{Fig:Diag_ne}. This uncertainty due to the unknown value of $\kappa$ is however comparable or smaller than the uncertainty due to the dependence of individual ratios on $T$ \citep{Dzifcakova10,Mackovjak13,Dudik14a,Dudik14b}. The overall uncertainty due to the a-priori unknown values of $T$ and $\kappa$ can be in some instances as large as 0.4 dex in log$(n_\mathrm{e}/\mathrm{cm}^{-3})$ (e.g., Fig.~\ref{Fig:Diag_ne}).

In principle, calibration uncertainties compound the uncertainty of the diagnosed density. However, the EIS density-sensitive lines belonging to the same ion have typically similar wavelengths, and are all observed within the same EIS channel. It is unlikely for the corresponding line ratios to be severely affected by calibration uncertainties. Therefore, we chose to ignore the calibration uncertainties when diagnosing the electron density. Nevertheless, for illustration, the values of individual ratios derived from the background-subtracted observations (Sect. \ref{Sect:4.1}) are plotted in Fig.~\ref{Fig:Diag_ne} together with their uncertainties.

The densities diagnosed using the \ion{Fe}{11} and \ion{Fe}{13} ratios are consistent within their respective errors (Fig.~\ref{Fig:Diag_ne}, Table \ref{Table:2}). However, the densities diagnosed using \ion{Fe}{12} are higher by about 0.5 dex compared to those diagnosed using \ion{Fe}{13}. I.e., given the densities diagnosed using \ion{Fe}{11} and \ion{Fe}{13}, the calibrated intensities of the \ion{Fe}{12} 186.887\AA~selfblend seem to be too high compared to the intensities of the \ion{Fe}{12} 195.119\AA~sbl by about 20--30\%. The reason for this discrepancy is unknown. \citet{DelZanna12a} showed that the new atomic model for \ion{Fe}{12} provides significantly lower densities than the previous model, by about 0.4 dex. The previous model consistently provided values much higher than those obtained from \ion{Fe}{11} and \ion{Fe}{13} \citep[see, e.g.,][]{Young09}. The densities obtained for the background-subtracted intensities from \ion{Fe}{12} are much reduced using the new atomic data of \citet{DelZanna12a}, but still higher by about 0.5 dex than those obtained here from \ion{Fe}{11} and \ion{Fe}{13}. The discrepancy is lower when background-subtraction is ignored. The reason is unclear despite the fact that we use here the same atomic data and calibration as \citet{DelZanna12a} and \citet{DelZanna13a}.
Nevertheless, the intensity of the 186.887\AA~line is somewhat suspect since the shape of the EIS effective area curve in the \citet{DelZanna13a} calibration differs from that of the ground calibration (see Fig. 8 \textit{top} therein), or from that of  \citet[][Fig. 7 therein]{Warren14}. The differences are of the order of several tens of per cent, which is comparable to the 20\% calibration uncertainty. We note that changing the \ion{Fe}{12} 186.887\AA\,sbl /195.119\AA sbl ratio by about 20\% would bring the diagnosed densities into consistency with the \ion{Fe}{11} and \ion{Fe}{13} ones (Fig. \ref{Fig:Diag_ne}). Furthermore, if the 20\% calibration uncertainties are taken into account, the observed \ion{Fe}{12} ratio still yields densities consistent with the other ratios, since the lower violet line in Fig. \ref{Fig:Diag_ne} yields densities of log($n_\mathrm{e}/\mathrm{cm}^{-3}$)\,=\,9.2--9.4.

Additionally, because of the strong overlap of the \ion{Fe}{11}--\ion{Fe}{13} relative ion abundances (Fig. \ref{Fig:Ioneq_AIA}) at temperatures corresponding to the peaks of the background-subtracted DEMs for all $\kappa$ (Fig. \ref{Fig:AIA_DEM}), it is unlikely that the diagnosed loop is a multi-density one. If it were, the \ion{Fe}{13} density-sensitive ratios, which have have stronger density sensitivity at log($n_\mathrm{e}/\mathrm{cm}^{-3}$)\,=\,9.5, should be biased towards higher densities more than the \ion{Fe}{12} ratio, whose increase with $n_\mathrm{e}$ is weaker.

For these reasons, we adopt the densities diagnosed using \ion{Fe}{11} and \ion{Fe}{13} for further analysis. For the average loop spectrum, the diagnosed value is log$(n_\mathrm{e}/\mathrm{cm}^{-3})$\,=\,9.0--9.5, while for the loop spectrum averaged through pixels $y$\,=\,300--309 we obtain the value of log$(n_\mathrm{e}/\mathrm{cm}^{-3})$\,=\,8.8--9.6. We note that these values are conservative, as they are chosen to contain all the densities diagnosed using \ion{Fe}{11} and \ion{Fe}{13} together with their respective errors (Table \ref{Table:2}).

\subsection{Diagnostics of $T$ and $\kappa$}
\label{Sect:4.3}

\subsubsection{Ratio-Ratio Diagrams}
\label{Sect:4.3.1}

Having obtained constraints on the electron density, we can next diagnose $\kappa$. A principal constraint is that the diagnostics of $\kappa$ have to be performed using lines that are also sensitive to temperature. This comes from the nature of the task, as both the excitation as well as ionization and recombination rates are a function of both $T$ and $\kappa$. Typically, line ratios sensitive to $\kappa$ involve lines populated by different parts of the distribution \citep{Dzifcakova06}, i.e., either lines with different behaviour of the excitation cross-section with $E$; or lines with different excitation thresholds $\Delta E_{ji}$, i.e., lines formed at different wavelengths; or both. Such line ratios will always be sensitive to temperature as well. Line ratios sensitive to $\kappa$ are then be combined with the line ratios involving lines from neighbouring ions, that are strongly sensitive to $T$, but also sensitive to $\kappa$.

Furthermore, if the relative level population depends on density, $\kappa$-sensitive line ratios belonging to the same ion typically have smaller sensitivity to $\kappa$ than to $n_\mathrm{e}$, although exceptions occur \citep{Dzifcakova06,Dzifcakova10,Mackovjak13,Dudik14b}. The diagnosed electron density is then used to contrain the  diagnostics using the ratio-ratio diagrams \citep{Dzifcakova10,Mackovjak13}.

This is the approach we chose here. From the lines listed in Table \ref{Table:1}, \ion{Fe}{11} is the only ion offering strongly temperature-sensitive line ratios. There are four temperature-sensitive ratios of \ion{Fe}{11} lines, each involving a line from the EIS short-wavelength channel (182.167\AA~or 188.216\AA) and a line from the EIS long-wavelength channel (257.554\AA~sbl or 257.772\AA~sbl, see Table \ref{Table:1}). These \ion{Fe}{11} line ratios are combined with ratios involving a \ion{Fe}{12} line. Both the selfblends at 186.887\AA~and 195.119\AA~are used. This is because of the possible calibration problems involving the 186.887\AA~sbl (Sect. \ref{Sect:4.2}). The various combinations are shown in Figs. \ref{Fig:Diag_TK1}, \ref{Fig:Diag_TK2} and \ref{Fig:Diag_TK3}. The observed line ratios together with their uncertainties are shown as large crosses where the azure and violet errorbars are calculated using the $\sigma_{10\%}$ and $\sigma_{20\%}$ uncertainties of the individual lines involved, respectively (see Table \ref{Table:1}). The calibration uncertainty is included since various ratios involve lines observed in both EIS channels. 

The ratio-ratio diagrams involving the 257.554\AA~sbl (Fig.~\ref{Fig:Diag_TK1}) show strong sensitivity to $T$ and sensitivity to $\kappa$ that is of the order of $\approx$20\%. However, the large calibration uncertainties together with the uncertainty in the diagnosed log$(n_\mathrm{e}/\mathrm{cm}^{-3})$ prevent positive diagnostics using some of these diagrams for both loop segments. This is because the 1-$\sigma_{20\%}$ errorbar intersects all curves for different $\kappa$ for at least one of the diagnosed limits on log$(n_\mathrm{e}/\mathrm{cm}^{-3})$. The situation is worse for the loop (300:309), where the observed ratios are closer and intersect more curves for different $\kappa$ on the ratio-ratio diagrams (Fig.~\ref{Fig:Diag_TK1}, \textit{left}) than in the case of the average loop spectrum (loop (288:317), Fig.~\ref{Fig:Diag_TK1}, \textit{right}). For the average loop spectrum, we can diagnose $\kappa$\,$\lesssim$\,5 using the ratios involving \ion{Fe}{11} 188.216\AA\,/\,(\ion{Fe}{11} 257.554\AA~sbl). This is because for all ratios in Fig.~\ref{Fig:Diag_TK1}, \textit{right}, the red and orange ratio-ratio curves corresponding to $\kappa$\,=\,2 and 3 intersect all the violet errorbars, while others do not. Nevertheless, the green ratio-ratio curve ($\kappa$\,=\,5) is in close vicinity of the errorbar in the \textit{top right} panel.

The situation is much better when using ratio-ratio diagrams involving the \ion{Fe}{11} 257.772\AA~selfblend (Fig.~\ref{Fig:Diag_TK2}). Compared to the spread of the curves for individual $\kappa$ and log$(n_\mathrm{e}/\mathrm{cm}^{-3})$, the uncertainty of the observed ratios is smaller, and the observed ratios are located further away from the curves. This permits diagnostics of $\kappa$\,$\lesssim$\,3, a strongly non-Maxwellian distribution. We note that this result is the same whether the \ion{Fe}{12} 186.887\AA~selfblend or the 195.119\AA~selfblend is used. 

To confirm this diagnostic, in Fig.~\ref{Fig:Diag_TK3} we substitute the \ion{Fe}{11} 257.772\AA~sbl by other \ion{Fe}{11} lines in the \ion{Fe}{12}\,/\,\ion{Fe}{11} ratio. Again, $\kappa$\,$\lesssim$\,3 is found independently of the combination of lines used (Fig.~\ref{Fig:Diag_TK3}), with majority of the ratio-ratio diagrams indicating $\kappa$\,$\lesssim$\,2.

We note that the results of the diagnostics of $\kappa$ are not dependent on which of the two \ion{Fe}{12} lines is used. Both the \ion{Fe}{12} 186.887\AA~and 195.119\AA\,selfblends yield similar results in terms of $\kappa$ (Figs. \ref{Fig:Diag_TK1}--\ref{Fig:Diag_TK3}). This allows us to establish confidence in using the \ion{Fe}{12} lines to diagnose $\kappa$ despite the possible calibration problems mentioned in Sect. \ref{Sect:4.2}.

Finally, we point out that the loop evolution as observed by AIA (Sect. \ref{Sect:3.2}) is unlikely to strongly affect the diagnostics of $\kappa$. This is because the AIA 193\AA~intensities \citep[dominated by \ion{Fe}{12};][]{DelZanna13b} do not change by more than 1--2\% during the EIS observations at 13:19\,UT (Fig. \ref{Fig:AIA_Lightcurves}, \textit{top}).

\subsubsection{Influence of the DEM on Diagnostics of $\kappa$}
\label{Sect:4.3.2}

Originally, the ratio-ratio diagrams were developed for simultaneous diagnostics of $T$ and $\kappa$ based on the assumption that the observed structure is isothermal \citep{Dzifcakova10,Mackovjak13}. The results presented for the transient loop in Sect. \ref{Sect:3.3} however suggest that the loop is multi-thermal independently of the value of $\kappa$. Therefore, we investigated the influence of DEM on the diagnostics of $\kappa$.

Unfortunately, the EIS raster obtained during HOP 226 does not contain enough strong lines for determination and constraining the DEM, especially throughout the entire log$(T/$K)\,=\,5.5 -- 7.0 range. Therefore, we use the AIA data to perform the DEM diagnostics and note that \citet{DelZanna13b} showed that the AIA observations can be used to predict the EIS radiances to within the calibration uncertainties of both instruments. Although these authors used preferentially a DEM reconstruction technique different from the \citet{Hannah12} one that is used here (Sect. \ref{Sect:3.3}), the \citet{Hannah12} technique was also tested by \citet{DelZanna13b} and a reasonable agreement was found.

First, we produce an AIA pseudo-raster for each AIA band similar to that shown in Fig.~\ref{Fig:EIS_raster}, \textit{left}. Subsequently, the DEM is derived using the technique of \citet{Hannah12,Hannah13} similarly as in Sect. \ref{Sect:3.3} for each pixel of these AIA pseudo-rasters. We perform the averaging of the DEMs the same way as for the EIS raster (i.e., over pixels at position 9 corresponding to the loop, $y$\,=\,288--317 and 300--309), subtract the background at position 5, and then use these background-subtracted DEMs to predict the intensities of individual EIS \ion{Fe}{11}--\ion{Fe}{13} lines. 

The predicted ratios of individual lines are shown as a function of $\kappa$ on each of the ratio-ratio diagrams in Figs. \ref{Fig:Diag_TK1}--\ref{Fig:Diag_TK3}. Diamonds are used for the diagnosed lower limits on log$(n_\mathrm{e}/\mathrm{cm}^{-3})$ (Sect. \ref{Sect:4.2}), asterisks for the upper limits. Generally, the predicted ratios for each $\kappa$ are located close to the corresponding curves, indicating that these curves can still be used to indicate the value of $\kappa$ in the observed plasma even if this plasma is multi-thermal. Furthermore, we find that \textit{with decreasing $\kappa$, the predicted ratios converge on the observed values in all cases,} indicating that the plasma is strongly non-Maxwellian with $\kappa$\,$\lesssim$\,2. This is so even for the cases when the isothermal ratio-ratio diagrams cannot be used to constrain the value of $\kappa$, such as the \ion{Fe}{11} 188.216\AA\,/\,(\ion{Fe}{11} 257.554\AA~sbl) -- (\ion{Fe}{12} 186.887\AA~sbl)\,/\,(\ion{Fe}{11} 257.554\AA~sbl) in Fig.~\ref{Fig:Diag_TK1}, \textit{middle row}, or the similar combination with \ion{Fe}{12} 195.119\AA~sbl in Fig.~\ref{Fig:Diag_TK1}, \textit{bottom row}.



\section{Atomic Data Uncertainties}
\label{Sect:5}

The atomic datasets for astrophysical spectroscopy are always incomplete, since they contain only a finite number of energy levels and the corresponding transitions. Therefore, we investigated the influence of the atomic data uncertainties on the analysis presented here. To do that, we repeated the analysis presented in Sect. \ref{Sect:4} using older atomic data that are available in the CHIANTI, version 7.1 \citep{Dere97,Landi13}. The calculations for the Maxwellian distribution were performed using CHIANTI v7.1 and the corresponding calculations for the $\kappa$-distributions were performed using the KAPPA package \citep{Dzifcakova15}.

The density diagnostics performed in Sect. \ref{Sect:4.2} using \ion{Fe}{11} and \ion{Fe}{13} ratios remain valid, since these density-sensitive ratios do not change appreciably \citep[see Figs. 5 and 7 in][]{Dudik14b} even if the individual line intensities change by up to $\approx$20\%.  Therefore, the conservative densities adopted (Table \ref{Table:2}) are kept.

Examples of the diagnostics of $\kappa$, including the effect of DEM using the older atomic data, are presented in Fig.~\ref{Fig:Diag_TK2_old}, which shows ratios involving the \ion{Fe}{11} 257.772\AA~selfblend. These diagrams correspond to those shown in Fig.~\ref{Fig:Diag_TK2} for the newest atomic data. From Fig.~\ref{Fig:Diag_TK2_old} we see that the diagnostics of $\kappa$\,$\lesssim$\,2 remain valid even if older atomic data are used.

We note that the newest atomic data of \citet{DelZanna10c}, \citet{DelZanna11}, \citet{DelZanna12a}, and \citet{DelZanna13} used in Sects. \ref{Sect:2}--\ref{Sect:4} represent significant improvement over the previous ones, especially in the case of \ion{Fe}{12}, where differences of up to 60\% in intensities of key lines were found \citep{DelZanna12a}, including the 195.119\AA~selfblend. We find similar increases found for all $\kappa$. The \ion{Fe}{11} line intensities are found to be increased for all $\kappa$ by about 10--20\%, although details depend on $\kappa$ and the particular line. These increases in line intensities are due to the increased contribution from resonances as well as cascading from the $n$\,=\,4 level \citep{DelZanna12e,DelZanna12a,DelZanna13} and cause the change in the theoretical ratio-ratio diagrams (compare Fig.~\ref{Fig:Diag_TK2} with Fig.~\ref{Fig:Diag_TK2_old}).

Finally, we note that the newest atomic data used in Sects. \ref{Sect:2}--\ref{Sect:4} do not include contributions from the $n$\,$\geqq$\,5 levels. For \ion{Fe}{11}, the contributions of cascading from $n$\,$\geqq$\,5 to \ion{Fe}{11} line intensities are about 10-20\% \citep{DelZanna13}. For the transitions from the 3s$^2$\,3p$^2$\,3d configuration in \ion{Fe}{12} (i.e., the 186.887\AA~and 195.119\AA~lines), the missing contributions from cascading is smaller, of the order of 5\% only \citep{DelZanna12a}. Including these contributions would move the diagnostic diagrams (Fig.~\ref{Fig:Diag_TK1}--\ref{Fig:Diag_TK3}) down in the $y$-direction. Therefore, more complete atomic data are unlikely to change the results of the diagnostics of $\kappa$.

%
\section{Summary}
\label{Sect:6}

We reported on imaging and spectroscopic observations of a transient coronal loop observed within the core of AR 11704 on 2013 March 30. The loop reappeared in the same location as the already faded flare arcade of the B8.9-class microflare peaking about three hours earlier. The transient loop persisted for nearly two hours, during which it evolved into a series of individual threads. By examining the AIA data, we found no signatures of hot, flare-like emission being associated with the loop. These results were confirmed using the DEM reconstruction by the regularized inversion method of \citet{Hannah12} and \citet{Hannah13}, showing the relative absence of hot plasma except a spurious peak that is also present in the background. This method was employed in conjunction with the AIA responses emission calculated for the $\kappa$-distributions by \citet{Dzifcakova15}.
We found that the loop is multi-thermal for all $\kappa$ considered, with the DEMs peaking at log$(T/$K)\,=\,6.3 for the Maxwellian distribution, and at 6.5 for $\kappa$\,=\,2. The spurious high-temperature peak becomes less prominent with decreasing $\kappa$, and it disappears for $\kappa$\,=\,2.

We analyzed the spectroscopic data obtained by \textit{Hinode}/EIS in order to perform diagnostics of density and $\kappa$. Several density-sensitive ratios of \ion{Fe}{11}, \ion{Fe}{12} and \ion{Fe}{13} were used in conjunction with the latest EIS calibration and the latest available atomic data. We find consistency between the densities diagnosed from the background-subtracted \ion{Fe}{11} and \ion{Fe}{13} intensities, with the \ion{Fe}{12} intensities yielding higher densities. However, the density-diagnostics using \ion{Fe}{12} may still be consistent with the \ion{Fe}{11} and \ion{Fe}{13} ones if the EIS calibration uncertainty is taken into account. Using \ion{Fe}{11} and \ion{Fe}{13} yields a conservative estimate of electron density of the order of log$(n_\mathrm{e}/\mathrm{cm}^{-3})$\,=\,8.8--9.6 for a loop segment 10$\arcsec$ long, and log$(n_\mathrm{e}/\mathrm{cm}^{-3})$\,=\,9.0--9.5 for the average spectrum of the loop along position 9 of the EIS slit.

This diagnosed electron density is then used to constrain the diagnostic of $\kappa$. To do that, we use the temperature-sensitive ratios involving \ion{Fe}{11} lines from both EIS detectors in combination with the ratios involving \ion{Fe}{12} and \ion{Fe}{11} lines. We found that ratios involving the \ion{Fe}{11} 257.554\AA~selfblend can preclude a succesful diagnostic due to the calibration uncertainty being larger than the spread of the curves for individual $\kappa$. However, all ratios involving the \ion{Fe}{11} 257.772\AA~selfblend together with other \ion{Fe}{11} and \ion{Fe}{12} lines consistently yield $\kappa$\,$\lesssim$\,2, i.e., an extremely non-Maxwellian situation. 

We next studied the influence of the plasma multithermality on the diagnostics of $\kappa$. Since the EIS data are insufficient to obtain the DEMs, we obtained the DEMs from the AIA data. These DEMs were used to predict the EIS line intensities as a function of $\kappa$. We found that, with decreasing $\kappa$, all ratios of the predicted Fe line intensities converge on the observed values. These results confirm the diagnosed value of $\kappa$\,=\,2 and provide first quantitative description of the non-Maxwellian distribution of electron energies in a coronal loop.

We note that the transient loop studied here is not a typical coronal loop. This is due to its rather low temperature compared to typical active region cores, the ongoing magnetic reconnection as evidenced by the brightenings and jetting activity near its right footpoint, as well as its appearance in the same spatial location as the previous B8.9-class microflare. Nevertheless, the results presented here demonstrate the viability of the methods for diagnostics of $\kappa$. Such methods should be applied in the future on the analysis of typical, well-defined coronal loops to search for possible signatures of impulsive heating.

Finally, we note that the 20\% calibration uncertainty, typical of the EUV spectroscopic instrumentation, can represent a severe limitation to the diagnostics of non-Maxwellian distribution. Such large calibration ucertainty contributes to the possible mis-interpretation of the observations, as it is always possible to obtain \textit{some} temperatures and DEMs, if a diagnostic of $\kappa$ is not performed. Decreasing the calibration uncertainty to about 10\% would significantly contribute to enabling the diagnostics of non-Maxwellian distributions from EUV spectra.

\acknowledgements
The authors thank the referee for comments leading to the improvement of the manuscript. The authors are also grateful to Dr. Iain Hannah for the availability of the regularized inversion method for reconstructing the DEMs. JD acknowledges support from the Royal Society via the Newton Fellowships Programme. JD, EDZ, \v{S}M, PK, and FF acknowledges the support by Grant Agency of the Czech Republic, Grant No. P209/12/1652. \v{S}M also acknowledges Comenius University Grant No. UK/620/2014. GDZ and HEM acknowledge STFC. The atomic data were calculated within the UK APAP network, funded by STFC. The authors also acknowledge the support from the International Space Science Institute through its International Teams program. Hinode is a Japanese mission developed and launched by ISAS/JAXA, with NAOJ as domestic partner and NASA and STFC (UK) as international partners. It is operated by these agencies in cooperation with ESA and NSC (Norway). AIA data are courtesy of \textit{NASA}/SDO and the AIA science team. CHIANTI is a collaborative project involving the NRL (USA), RAL (UK), MSSL (UK), the Universities of Florence (Italy) and Cambridge (UK), and George Mason University (USA).

\bibliographystyle{apj}
\bibliography{Loop_Kappa}



\end{document}